\documentclass{aa}  
\usepackage{graphicx}
\usepackage{natbib,twoopt}
\usepackage[breaklinks=true]{hyperref} %% to avoid \citeads line fills
\usepackage{xcolor}
\usepackage{color}
\usepackage{amsmath}    
\usepackage{txfonts}
\usepackage{hyperref}
\usepackage{comment}
\usepackage{tablefootnote}
\usepackage{booktabs, multirow}
\usepackage[normalem]{ulem}

\usepackage{hyperref}
\hypersetup{colorlinks=true,linkcolor=[rgb]{1.,0.2,0.2},citecolor=[rgb]{0.1,0.4,1.},filecolor=[rgb]{0.7,0.2,0.2},urlcolor=[rgb]{0.7,0.2,0.2}}
\def\kms{\mbox{km~s$^{-1}$}}
\newcommand{\nic}[1]{{\color{black}{#1}}}
\newcommand{\massi}[1]{{\color{black}{#1}}}
\newcommand{\chiara}[1]{{\color{black}{#1}}}

\begin{document} 
\title{COSTA: the COld STream finder Algorithm.}   
\subtitle{Searching for kinematical substructures in the phase space of discrete tracers}

   \author{M.~Gatto\inst{1,4}
          \and N.R.~Napolitano\inst{2}\thanks{E-mail: napolitano@mail.sysu.edu.cn}
          \and          C.~Spiniello\inst{3,4}
          \and 
          G.~Longo\inst{1}
          \and
          M.~Paolillo\inst{1}
          }

   \institute{University of Naples Federico II, C.U. Monte SantAngelo,
    Via Cinthia, 80126, Naples, Italy\\
    \and
    School of Physics and Astronomy, Sun Yat-sen University Zhuhai Campus, 2 Daxue Road, Tangjia, Zhuhai, Guangdong 519082, P.R. China\\
    \and 
    Sub-Department of Astrophysics, Department of Physics, University of Oxford, Denys Wilkinson Building, Keble Road, Oxford OX1 3RH, UK\\
    \and
    INAF-Osservatorio Astronomico di Capodimonte, Via Moiariello  16, 80131, Naples, Italy.
}

\date{Received 24 July 2020}

% \abstract{}{}{}{}{} 
% 5 {} token are mandatory
 
  \abstract
  % context heading (optional)
  % {} leave it empty if necessary  
{We present COSTA (COld STream finder Algorithm), a novel algorithm to search for cold kinematical substructures in the phase space of planetary nebulae (PNe) and globular clusters (GCs) in the halo of massive galaxies and intracluster regions.}
{COSTA aims at detecting small sized, low velocity dispersion streams, as the ones produced in recent interactions of dwarf galaxies with the halo of more massive galaxies, \nic{including the ones sitting in the central region of rich galaxy clusters.}}
%\st{, including bright cluster galaxies.}
{COSTA is based on a deep friend-of-friend procedure that isolates groups of $N$ particles with small velocity dispersion (between 10 \kms and $\sim100$ \kms), using an iterative {($n$)} sigma-clipping over a defined number of {($k$)} neighbor particles. The algorithm has three parameters {($k-n-N$)}, plus a velocity dispersion cut-off, %,$\sigma_{\rm cut}$, 
which defines the ``coldness'' of the stream, that are set using Montecarlo realizations of the sample under exam. }
{\massi{In this paper, we show}
%\st{We demonstrate} 
the ability of COSTA to %how COSTA can 
recover simulated streams on mock data-sets of discrete kinematical tracers with different sizes and measurement errors, from publicly available hydrodynamical simulations.  We also 
%demonstrate how to 
show the best algorithm set-up for a realistic case of stream finding in the core of the Fornax cluster, for future applications of COSTA to real populations of PNe and GCs.}
%\massi{which will be presented in a companion paper.} %\st{However,}
%\massi{Moreover, }
{COSTA can be generalized to all problems of finding small substructures in the phase space of a limited sample of discrete tracers, provided that the algorithm is trained on realistic mock observations reproducing the specific dataset under exam.} 


   \maketitle
%
%-------------------------------------------------------------------

%%%%%%%%%%%%%%%%%%%%%%%%%%%%%%%%%%%%%%%%%%%%%%%%%%

%%%%%%%%%%%%%%%%% BODY OF PAPER %%%%%%%%%%%%%%%%%%

\section{Introduction}
\label{sec:introduction}
In the hierarchical formation scenario, massive structures grow in a bottom-up manner assembling mass by merging of small systems \citep{white-1978}). 
This is a process that is still ongoing, as demonstrated by cosmological simulations \citep{naab-2007,naab-2009, cooper-2010, oser-2010, oser-2012}.

In the Local Group, the accretion of smaller building
blocks has been observed in the last decades \citep{ibata1994,ibata1995,Yanny2000} under the form of dwarf
debris \citep{ibata2001d,majewsky-2003}. 
In denser environments, like galaxy groups or clusters, 
this mechanism is enhanced because of the whirl of encounters and collisions, leading to the formation of an extended halo around the central galaxy \citep[e.g.][]{cooper-2010}. % of a group or a cluster  
%NRN: add this reference and more recent simulations from Cooper?
%MG: Cooper et al. performed simulations of Galactic stellar haloes formed by the tidal disruption of accreted dwarf galaxies, so is not an example of dense environment, isn't it?
%NRN2: but he shows the build up of the halo around the central galaxy of group/clusters
%, \chiara{which are often very massive (stellar masses $M_{\star}\geq 10^{12}M_{\odot}$)} 
Central dominant (cD) galaxies \chiara{in the innermost region of} %near the center of 
the richest clusters are the archetypes of this scenario, with their exceptional merging histories \citep[e.g.][]{Ruszkowski&springel-2009,Weinzirl-2014,iodice-2016}. 
The remarkable \chiara{large} stellar masses of cDs ($M_{\star}\geq 10^{12}M_{\odot}$) are well explained in the hierarchical scenario, as the mass assembly is expected to happen either through tidal stripping of stars and globular clusters from their satellite or dwarfs \citep{gallagher&ostriker-1972,moore-1996,gregg&west-1998,willman-2004,read-2006}, or through major mergers with other bright galaxies or minor ones in which the cD 'eats' smaller systems \citep{ostriker&tremaine-1975,white-1976,malumuth&richstone-1984,merritt-1985,Liu_F.S.-2015,Nipoti-2017}. 
In particular, numerical simulations and semi-analytic models have demonstrated that the bulk of their accreted mass and extended halos was built up in the last few Gyr, especially through minor merger events (e.g. \citealt{delucia&blaizot-2007,amorisco-2019}). 
%, leading to extreme masses and luminosities.
As all these processes are expected to be still in action, one can search for observational signatures of such events in the cluster core, either with the deep photometry \citep[see e.g.][]{mihos2005, iodice-2017,mihos2017} or in the kinematics of stars and other kinematical tracers as planetary nebulae and globular clusters \citep[e.g.][]{napolitano-2003,romanowsky-2012,longobardi-2015,spiniello-2018,pota-2018,amorisco-2019}.
Hence, stellar substructures in galaxy haloes (and beyond, i.e. in the intracluster regions), under the form of debris of past or recent merger events, are invaluable pieces of information to study the mechanisms supplying mass in the assembly history of galaxies \chiara{in dense environments}. 

In recent years, the study of the signature of minor mergers in the local universe, like halo shells and ripples, tidal streams or other stellar substructures, has become an important tool to probe the assembly histories of galaxies \citep[e.g.][]{helmi-1999,ibata-2001,Belokurov-2006,tal-2009,Martinez-Delgado-2010,cooper-2011,Mouhcine-2011,xue-2011,bate-2014}.
%NRN: maybe you can merge the list above with the list below?
%GL I agree
The number of newly discovered stellar streams and other substructures in the halos of nearby galaxies has dramatically increased, showing that remnants of merger events could be almost ubiquitous. 
Among the most important examples, we can count the Sagittarius Stream in the Milky Way  \citep{ibata-1997,ibata-2001b,majewsky-2003} and many other substructures detected around M31 \citep[Andromeda galaxy; e.g.][]{mcconnachie-2009,ibata-2001a}.

Early investigations of such substructures were based on photometric observations. However, this approach is challenging due to the faint surface brightness of the remnants, typically below $\mu \sim 27$ mag/arcsec$^2$. This implies that only the brightest substructures are generally detected, while most of the accreted mass provided by the fainter events, having generally a surface brightness of the order of 30 mag/arcsec$^2$ or below \citep{cooper-2010}, remain hidden in the central galaxy background.

In the last few years, deeper and more accurate spectroscopy has allowed to include kinematic information of the debris, in order to go beyond the purely photometric studies and look into the phase-space (projected positions and line-of-sight velocities) to search for the typical signatures expected in these interactions \citep[e.g. ][]{johnston-2008,romanowsky-2012}.
Within the Local Group, these substructures can be studied using individual stars  \citep[e.g.][]{koch-2008,gilbert-2009,starkenburg-2009,xue-2011,Belokurov2016}.
%, while at larger distances, where stars are not resolved, one \chiara{needs to }%can 
Outside the Local Group stars cannot be resolved and other kinematical tracers have to be used. 
Planetary nebulae (PNe) and globular clusters (GCs) are suitable tracers of this kind as they are observable out to large distances from the galaxy centers \citep{durrell-2003,merrett-2003,douglas-2007,shih&mendez-2010,cortesi-2011,richtler-2011} and their velocity can be measured with good precision in nearby galaxies and galaxy clusters. They represent a viable alternative to study the outskirts of galaxies where it is very 
%\st{hard to measure the absorption line from spectroscopy} 
\massi{hard to measure stellar absorption lines and thus obtain kinematical information from the integrated light}
(PNe: \citealt{hui-1995,napolitano-2002,romanowsky-2003,delorenzi-2009,coccato-2009,napolitano-2009,richtler-2011,forbes-2011,pota-2013,longobardi-2015,hartke-2017,spiniello-2018}; GCs: \citealt{cote-2003,romanowsky-2009,schuberth-2010,woodley&harris-2011,romanowsky-2012,foster-2014,veljanoski-2016,pota-2018,longobardi-2018}).

The combined information of position and velocity of tracers in the halo regions of galaxies allows us to study substructures in the tracer phase-space, where they have not yet fully mixed due to the long dynamical times \citep{napolitano-2003,arnaboldi-2004,bullock-johnston-2005,arnaboldi-2012,coccato-2013,longobardi-2015}.
Historically, the methods adopted to search for streams have been 
%CS: I have changed in this way to make the transition with the next paragraph smoother. 
very empirical and 
lacking well encoded (objective) criteria 
to systematize the search of streams in the full phase-space.
Only recently, there has been a large effort to develop stream finding algorithms suitable for different data sets. 
For the Milky Way, \citet{Malhan&Ibata2018}, implemented STREAMFINDER, with the aim of unveil dynamically cold structures in the 6D phase-space, by taking advantage of the Gaia space mission data. This code looks for a handful number of particles (as few as ~ 15 members) which lie along a similar orbit, allowing to detect tiny and ultra-faint streams in the Galactic Halo. 
Other algorithms have been focused on automatic search of tidal structures, like shells or ridges, in deep images \citep[e.g.][]{Kado-Fong2018,Hendel2019}. Such approaches are more directed to large sample of galaxies to build statistically significant samples of stream features, but they do not rely on kinematics. As stated before, this is not ideal to look for low-surface brightness tidal features, as expected to be those originating from minor mergers \citep[see][]{cooper-2010}.

In this context, we present COSTA, the COld STream finder Algorithm, a new method to search for candidate cold substructures which can be interpreted as signatures of recent or past interaction between a main galaxy and the dwarf galaxies surrounding it. 
COSTA aims to fill the gap, left by the above algorithms, introducing a method that relies on kinematics (namely a reduced 3D phase-space of projected positions and line-of-sight velocities) which can reveal streams even beyond the Local Group and that can be still applied to large galaxy samples but below the detection limits imposed by the photometry.

We introduce the basic statistical methods that allow the identification of cold kinematical substructures made of few tens of particles, compatible with what is expected for faint streams around galaxies.  The method is based on a k-nearest-neighbors approach (KNN), which groups nearby particle in 2D positions and in velocity to find coherent kinematic substructures. 

%CS: MOVED IN SEC. 2 (this was too much details for the intro. In Sec.2 instead, there was this (up) more generic sentence. So I SWAP THEM. 
%The method relies on a pseudo-k-nearest neighbors algorithm (KNN) method which is based on a deep \textit{friend-of-friend} algorithm that isolates groups of ($N$) particles with a small velocity dispersion ($\sigma_{\rm cut}$, chosen between 10 \kms and $\sim$ 100 \kms), using an iterative ($n$) sigma-clipping in a defined number ($k$) of neighbor particles. Thus, the method has the three parameters ($k-n-N$) for any given (upper) dispersion threshold, $\sigma_{\rm cut}$, that need to be set upon Montecarlo realizations of the specific sample under exam.
%mixes a \textit{friend-of-friend} algorithm in the right ascension (RA), declination (DEC) and a \textit{$\sigma$-clipping} procedure in radial velocity, in order to search for groups of tracers (particles) correlated both in position and velocity. 

The algorithm is general and can be applied to any nearby stellar system, either galaxies or galaxy clusters cores (where large galaxy haloes and intracluster light concentration reside). 

As a template case of this latter kind to show the potential of the method, we discuss here the specific case of the Fornax cluster core.  The Fornax cluster is particular suitable for such a test as different studies have provided evidences of recent galaxy interactions \citep[e.g.][]{dabrusco-2016,iodice-2017,spiniello-2018,sheardown-2018}. 
This complexity represents a challenging test bench for the algorithm.

%indeed a very representative case since %quite representative of the problem because 
%\chiara{already} 
%\chiara{REFERENCES MISSING} 
%, due to its complexity \glo{Complessità di chi? algoritmo o ammasso? Ovviamente dell'ammasso ma andrebbe specificato meglio}.

Here, in this paper, we use mock observation of the Fornax Cluster to assess the reliability of the method and to demonstrate how to set-up the best parameters in a real case. 
In a companion paper (in preparation), we will then apply COSTA to identify real streams of GCs and PNe from the Fornax VST Spectroscopic Survey (FVSS,  \citealt{pota-2018, spiniello-2018}, hereafter P+18 and S+18 respectively). 

The paper is structured as follows: \massi{in Section \S\ref{sec:An algorithm to detect cold substructures} we present a brief description of the algorithm. In Sections \S\ref{sec:galmer simulations} and \S\ref{sec:fornax} we test it on hydrodynamical simulations of pair interacting galaxies and on Montecarlo simulations of the Fornax cluster core, respectively. } \chiara{Finally, in Section \S\ref{sec:conclusion} we draw our conclusions}. 

\section{COSTA: The COld STream finder Algorithm}
\label{sec:An algorithm to detect cold substructures}
In this section we introduce the COld STream finder Algorithm (COSTA) to detect cold substructures in the reduced phase space (position on the sky and radial velocity) of discrete tracers. 
%\st{We first check whether the algorithm is able to find known streams in simulations of galaxy encounters between a massive galaxy and a dwarf system, then, we describe the procedure to optimize the parameter choice in the specific case of the search for streams in the center of the Fornax cluster.} 
%\subsection{The COld STream Algorithm (COSTA) definition}

\begin{table*}
%	\begin{minipage}{120mm}
	\centering
    \caption{Parameters of the selected galaxies in the galmer simulation. The values of the velocity dispersion $\sigma_{\rm{v}}$, listed in the last column, have been measured in the configuration $N_{\rm{giant}} = 2000$ - $N_{\rm{dwarf}} = 150$}
	\label{tab:galmer parameters}
    \begin{tabular}{c c c c c c c c c c}
		\hline
		& M\textsubscript{B} &
		M\textsubscript{H} &
		M\textsubscript{disk} &
		r\textsubscript{B} &
		r\textsubscript{H} &
		N\textsubscript{stars} &
		N\textsubscript{DM} &
		N\textsubscript{gas} &
		$\sigma_{\rm{v}}$\\
		& 
		[$2.3 \times 10^9 M_{\odot}$] &
		[$2.3 \times 10^9 M_{\odot}$] &
		[$2.3 \times 10^9 M_{\odot}$] &
        [Kpc] & [Kpc] & & & & [$kms^{-1}$]\\
		\hline
		gE0 & 70 & 30 & 0 & 4 & 7 & 320000 & 160000 & 0 & 145\\
		dE0 & 7 & 3 & 0 & 1.3 & 2.2 & 32000 & 16000 & 0 & 77\\
		gSa & 10 & 50 & 40 & 2 & 10 & 240000 & 160000 & 80000 & 162\\
		dS0 & 1 & 5 & 4 & 0.6 & 3.2 & 32000 & 16000 & 0 & 149\\
		\hline
	\end{tabular}
%    \end{minipage}
\end{table*}

In order to find cold substructures that are correlated both in position and in velocity, we implemented an algorithm looking for points close both in the RA/DEC position-space and in the reduced phase-space (velocity vs. radius). 
The method relies on a pseudo-KNN method which is based on a deep \textit{friend-of-friend} algorithm that isolates groups of ($N$) particles with a small velocity dispersion ($\sigma_{\rm cut}$, chosen between 10 \kms and $\sim$ 100 \kms). 
%, using an iterative ($n$) sigma-clipping in a defined number ($k$) of neighbor particles. 

%CS: MOVED TO SEC1 %The method is based on a k-nearest-neighbors approach (KNN), which groups nearby particle in 2D positions and in velocity to find coherent kinematic substructures. 

The main difficulty is to efficiently detect particles belonging to the stream, which should preserve the low velocity dispersion of the dwarf progenitor while they are moving in regions where the potential of the cluster rules and the local velocity dispersion is the one of the cluster (i.e. up to 50 times larger than typical dwarf-like velocity dispersions). 
To do that, for each particle, the algorithm starts performing an iterated sigma clipping on a number ($k$) of %neighbor 
neighbors. In particular, it removes all the particles with a velocity outside the interval [$\bar{v} - n \times \sigma$, $\bar{v} - n \times \sigma$], where $\bar{v}$ and $\sigma$ are the mean velocity and the velocity dispersion %(for which we use the standard deviation of the individual velocities as a proxy) 
of the $k$ particles and $n$ is the sigma clipping value. 
As a proxy for the velocity dispersion, we use the standard deviation of the individual velocities (see \S\ref{sec:COSTA_GALMER}). The algorithm iterates the procedure, with the mean velocity and velocity dispersion of the remaining particles, until there are no outliers to be clipped. Once the procedure is over, the algorithm  %keeps
 selects all structures in the position and velocity space with a minimal number ($N_{\rm min}$) of particles.
%, coherent both in position and in velocity spaces (near by particles with low velocity dispersion).

%\st{Roughly speaking, for each particle the algorithm performs first an iterated ($n-$) sigma clipping on a number ($k$) of neighbor neighbours, by removing particles which are moving following the potential of the cluster, finally keeping a structure with a minimal number ($N$) of particles, coherent both in position and in velocity spaces, and having a low velocity dispersion.}\glo{QUESTA FRASE \'E ABBASTANZA OSCURA} 

\nic{To define the maximum velocity dispersion acceptable for a given substructure to be considered cold, COSTA uses another parameter, the cut-off velocity dispersion, $\sigma_{\rm cut}$.}
%, also a parameter of the algorithm, defines the maximum velocity dispersion acceptable for a given substructure to be considered cold. 
%Moreover, 
We fine-tune our algorithm to find cold streams originating from the interaction of dwarf galaxies with the cluster.  
In fact, we expect that dwarf disruption is the
%most common event still acting in the local universe and 
main mechanism contributing to the later formed intracluster stellar population and the assembly of large stellar halos around galaxies. Hence, we allow for
%\nic{Therefore we use for 
$\sigma_{\rm cut}$ values ranging from 10 to $\sim$ 100 kms$^{-1}$, based on the typical dwarf-like dispersion values found in the Coma cluster \citep{coma_dw_FJ}.

The final COSTA output is then a list of substructures with low velocity dispersion, below the fixed threshold, $\sigma_{\rm cut}$.

%In fact, the remnants of disrupted dwarf galaxies events preserve their kinematic properties for a longer time than those of massive satellites (see \S\ref{sec:introduction}). Furthermore, more massive galaxies produce more diffuse substructures due to a higher velocity dispersion and larger sizes, and these are harder to be ``filtered'' in the phase space as they might look more mixed to the warm halo environment. 
%Thus, 
%and 
We note here that more massive galaxies would produce more diffuse substructures, due to a higher velocity dispersion and larger sizes. These would be harder to be ``filtered'' in the phase space as they would be more mixed to the warm halo environment.  

Thus, to summarize, the COSTA algorithm has, in total, three parameters (\emph{k, n, N\textsubscript{\rm min}} for any given (upper) dispersion threshold, $\sigma_{\rm cut}$, that 
need to be properly chosen  
to maximize the number of real cold substructures (completeness) and minimize the number of spurious detections
(purity), caused by the intrinsic stochastic nature of the 
velocity field of hot systems. \chiara{To this purpose, one can use Montecarlo realizations of the specific sample under exam.}

%%This cut-off velocity dispersion, $\sigma_{\rm cut}$, is also a parameter of the algorithm, which defines the maximum velocity dispersion acceptable for a given substructure to be considered cold. 
%As we expect that dwarf disruption is the most common event still acting in the local universe to contribute to the intracluster stellar population and the assembly of large stellar halos around galaxies, we have set a series of cut-off values to account for a variety of dwarf masses, i.e. from 10 to 80 kms$^{-1}$ 
%%as a maximum $\sigma_{\rm cut}$
%, e.g. according to typical dwarf-like dispersion values found in the Coma cluster \citep{coma_dw_FJ}. 
%%, but we have tested the impact of even smaller values (see \S\ref{sec:galmer simulations}).
%The remnants of these events preserve their kinematic properties for a longer time than those of massive satellites (see \S\ref{sec:introduction}). On the other hand, more massive galaxies produce more diffuse substructures due to a higher velocity dispersion and larger sizes, so they are harder to be ``filtered'' in the phase space as they might look more mixed to the warm halo environment (see \S\ref{sec:introduction}).
%This 

Our approach has the advantage of being able to refine the selection of coherent spatial and velocity substructures, but it has the disadvantage to be biased toward round geometries.  In fact, the algorithm is based on a simple metric which uses the distances from every single particle. This reduces the chance to identify chain-like structures, which are likely expected in elongated streams. 
To remove this bias, we add a second stage in COSTA, where we verify if some of the groups do belong to a single structure. 
\massi{In particular, we define two or more groups belonging to a single structure if they show at least one common particle and their velocity dispersion values differ by less than their uncertainties.}
%\st{this is verified when there are common points among two o more groups showing similar velocity dispersion (i.e. consistent within the individual group uncertainties)}. 
%\st{and the overall ``grouped'' substructure has a velocity dispersion below a given threshold (between 10 and 80 kms$^{-1}$) to optimize the selection of streams of different size and morphology} (see \S\ref{sec: selection of the parameters}).
%In synthesis the algorithm can be divided into two steps: in the first step it defines as "friends" all points that are close enough in the space of configurations and have a similar velocity; then it inspects all common points belonging to a different group of friends in order to check if these groups belong to the same substructure; this allows finding also structures that have a chain-like pattern.
%Finally, it takes just the structures with a velocity dispersion below a certain threshold.

To demonstrate that it is possible to identify regions in this parameter space that can reliably detect streams with an acceptable fraction of false positives, we  first test the algorithm over a simulated sample from the publicly available hydrodynamical simulations Galmer \citep{chilingarian-2010}, and then train the algorithm to search for stellar streams in the Fornax cluster core. \chiara{The results of these tests are presented in the next Sections.}
%We \chiara{expect }%aspect 
%that on $k$ neighbours, there could be some with a similar velocity by randomness, hence the chance to gather a real stream (i.e. the purity) increases by increasing the minimum number $N_{min}$ of points belonging to the group.
%\st{that on $k$ neighbours, most of them might belong to the cluster background, but the chance to gather more stream members increases by increasing the number $N$ of points belonging to the group, hence having a better chance to make the sigma clipping doing the job of iteratively cutting particles belonging to the cluster and finally select a real structure.}
%Values of the parameters must be chosen in order 
%To check the effectiveness of the scheme presented above, we

\section{Testing COSTA on hydrodynamical simulations}
\label{sec:galmer simulations}

We use a suite of publicly available simulations, the Galmer database \citep{chilingarian-2010}, to test the ability of our algorithm to recover streams originating from a dwarf during a close passage to a giant galaxy. The simulated data-cubes are needed to test the algorithm self-consistently. 

We first define the series of (\emph{k, n, N\textsubscript{\rm min},$\sigma_{\rm cut}$}) set-ups that minimize the false detections, and then apply them to find the stream. Finally, we check how meaningful the recovered properties (e.g. mean velocity, local velocity dispersion, fraction of particles) are with respect to the intrinsic property of the stream. 
At this point, we are interested to verify whether for a given stream there is a series of parameter set-ups that allow COSTA to find it and how these might change as a function of the observational conditions (i.e. measurement errors and total number of particles).
%NRN5: Massimiliano vedi se questa cosa ti va bene

\subsection{The Galmer Simulations}
\begin{figure*}
    \centering
    \includegraphics[scale=0.6]{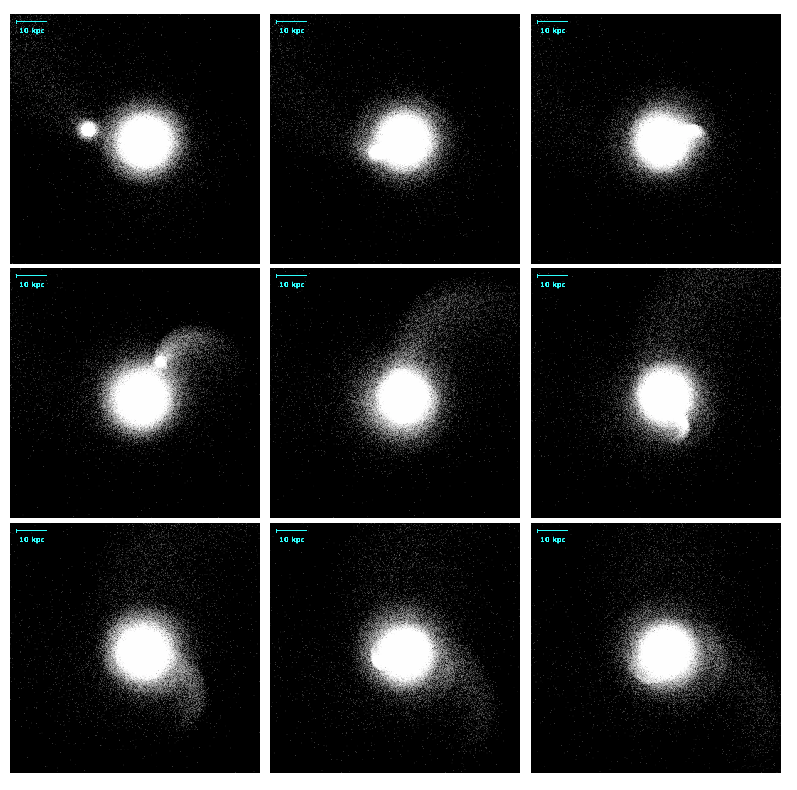}
    \caption{Snapshots of the gE0-dE0 encounter, from 1850 Myr (top left) up to 2250 Myr (bottom right) after the begin of the simulation, and separated by steps of 50 Myr. To test COSTA, we use the configuration at the center of the image, temporarily located at 2050 Myr after the start of the encounter.}
    \label{fig:gE0-dE0 interaction}
\end{figure*}{}

\begin{figure*}
    \centering
    \includegraphics[scale=0.6]{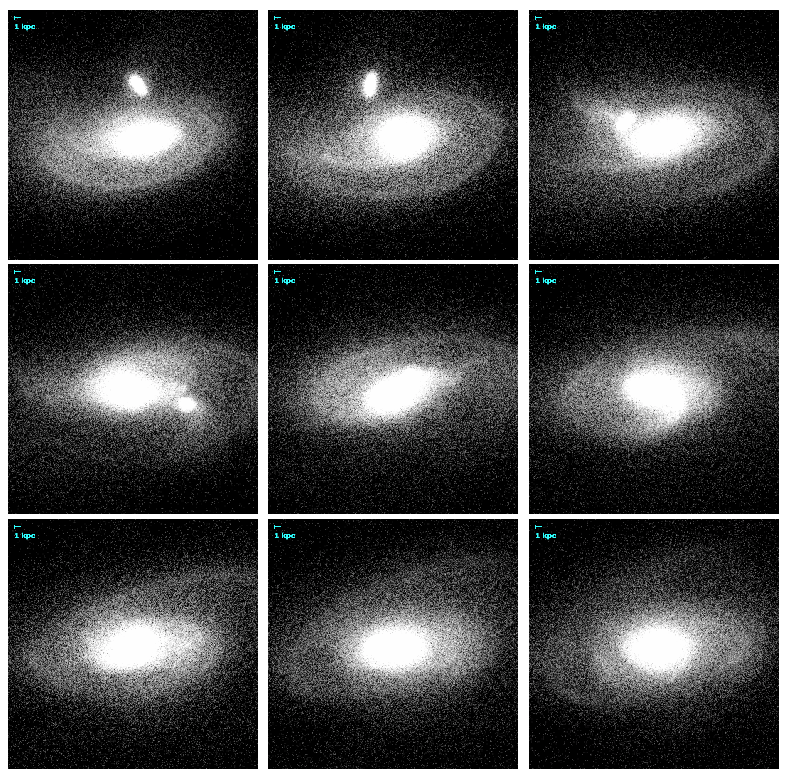}
    \caption{Same as Fig. \ref{fig:gE0-dE0 interaction} but in the case of the gSa-dS0 interaction. These snapshots correspond to a time interval comprised between 1650 Myr and 2050 Myr after the begin of the simulation, with our test configuration (1850 Myr) at the center of the image.}
    \label{fig:gSa-dS0 interaction}
\end{figure*}{}

%NRN5: we need to give more details about the technical information: dark matetr particle mass, stellar mass...
The Galmer simulations are based on a Tree-Smoothed Particle Hydrodynamics (SPH) code, in which gravitational forces are calculated using a hierarchical tree method \citep{barnes-1986} and gas evolution is followed by means of SPH  \citep[][]{lucy-1977,gingold-1982}. 
Dark matter particles 
and baryon particles have both masses of $\sim10^5 M_{\odot}$, while the softening lengths is $\epsilon = 280$ pc for giant-giant interactions, and $\epsilon = 200$ pc for giant-intermediate and giant-dwarf runs. 
This gives an appropriate mass and spatial resolution to trace low mass and low surface brightness substructures. 
The typical mass residing in stellar streams stripped by the dwarf during his interaction with the large galaxy is of the order of 10\% of its mass, i.e. given the typical GC and PN specific number densities (the number of particle per unit of luminosity) the stream is fairly sampled with a few tens and up to hundred tracers (e.g. GCs and PNe together). 
We will simulate different depth of our observational set-up by assuming different \massi{numbers} %\st{sampling of the complete sample} 
of stellar particles expected to populate the stream. 

The advantage of testing the algorithm on simulations is that we can separate the particles belonging to the dwarf galaxy from the ones belonging to the target system. We can, therefore,  characterize the phase-space of both galaxy target and streams produced in the fly-by of the dwarf galaxy through the central galaxy halo.  

%NRN5: Massimiliano vedi se questa cosa ti va bene
\massi{The Galmer database\footnote{http://galmer.obspm.fr} provides about 1019 simulations of colliding galaxies and more than 70000 snapshots showing the development of these interactions up to 3 Gyr from the beginning of the encounter with a bin interval, for each snapshot, of 50 Myr.}

%From this database, 
From the whole  database, we focus on two encounter configurations: 1) the one between a giant Elliptical galaxy, gE0, and a dwarf E0 galaxy, dE0, with a mass ratio of 1:10 (the minimum found in the database for all simulations) and 2) the one between a giant Sa galaxy, gSa, and a dwarf S0, dS0, galaxy, also with a mass ratio 1:10. \massi{Table \ref{tab:galmer parameters} shows the parameters of the four galaxies}.
We need to point-out that a mass ratio of 1:10 is not optimal to generalize the results, as a wider population  of systems, also with lower mass ratio, exist in real cases.  %population of lower mass ratio systems in real cases. 
As we will show later, though, this is a conservative starting point as our algorithm is more efficient in detecting group in phase-space which are generally much colder than the surrounding environment. Hence, the difference in velocity dispersion which characterize the Galmer systems are representative of extreme cases: if COSTA is able to detect substructures in these systems, then   
%there is a good chance that 
it will be %to be 
even more successful in cases involving lower mass satellites.

%Old version: From this database, we selected the simulated interaction between a gE0 and a dE0 with a mass aspect ratio of 1:10 (see also Table \ref{tab:galmer parameters}), the maximum value present in the database.
%Although the bulk of the dwarfs in the core of Fornax is made of dE0, we used a dS0 because this had the mass and effective radius more similar to the dwarfs in the inner regions of Fornax.

\massi{We select the gE0-dE0 and gSa-dS0 cases}
%\st{We picked these cases} 
as realistic representations of the dynamics of a dwarf-giant encounters. In particular the case of the gE0-dE0 is fairly representative of a typical encounter between a hot, high dispersion system and a colder satellite, 
%which is reasonably reproducing typical events 
like the one happening in large galactic halos \citep[see e.g.][]{cooper-2010,iodice-2016}. 
In both configurations, we choose an encounter with the satellite starting 100 kpc away, and falling toward the larger galaxy in a prograde orbit with an inclination of 33 degrees and a pericentral distance of 16 kpc. We initially used a prograde orbit because this is expected to exchange a lower amount of energy and therefore to minimize the scatter of dwarf particles into warmer tails. However, since in the case of the gSa-dS0 encounter the dwarf and the giant stars were too mixed, for this latter case we also consider a retrograde encounter. 
%\st{The parameters of the simulated galaxies are listed in the table .}

In Fig. \ref{fig:gE0-dE0 interaction} and Fig. \ref{fig:gSa-dS0 interaction}, \massi{we show a few snapshots of the gE0-dE0 and gSa-dS0 encounters, respectively. The final configuration we adopt for our tests is shown in the central panel (encounter), while the other panels show different time-snaps, each one spaced in time by 50 Myr, with the top left corner temporarily located 200 Myr before the chosen configuration.} %from 1850 Myr to 2250 Myr after the begin of the simulation.} 
%\st{we show a few snapshots of the gE0-dE0 and gSa-dS0 encounters, respectively, centered around the final configuration we adopted for our tests, which then resides in the central panel.}
\massi{
%We have selected two 
The choice of the central configurations 
%where the dwarf galaxy is at a minimum 
is motivated by the fact that there the distance between the intruder and the giant galaxy is the shortest (as evident also from figures \ref{fig:gE0-dE0 interaction} and \ref{fig:gSa-dS0 interaction}).} \chiara{This allows us to have a sufficient spatial mix of the two systems, and thus stress as much as possible the ability of COSTA to recover stream particles well embedded in high density regions.} 
%\st{We have selected two configurations where there is a clear mix of the target and intruder systems, in order to stress as much as possible the ability of COSTA to recover stream particles.} 
%NRN6: Massimiliano occorre fare questa figura

%In the second case, the satellite merges into a colder system with a low intrinsic dispersion but a higher rotation. 

From the figures, it is also clear that the encounters start producing a stream-like structure since the first passage at a few tens of kpc. Particles belonging to the original stream become mixed after a few hundreds Myr, but subsequent close passages produce even brighter streams. These latter remain visible and well separated %quite better distinct 
from the background galaxy %again 
for hundreds of Myr. Then, later on in time, they diffuse and mix with galaxy halo particles. This time scale is set by the specific dynamical time of the system under exam and, for  hotter central systems and lower mass ratios, this can be larger. Unfortunately, the Galmer database does not provide \massi{lower mass ratios than the ones adopted here. Nevertheless, these examples allow us to test the ability of COSTA in finding such cold streams as a function of a few observational parameters.}  
%\st{such specific combinations of galaxy encounters, but this example that can serve our purpose to show the ability of COSTA to find such cold streams as a function of a few observational parameters.} 

\subsection{Running COSTA on Galmer simulations}
\label{sec:COSTA_GALMER}
\nic{In order to apply COSTA to Galmer simulations, we }
%\st{To do that,} 
first need to extract from the simulated 6D datacubes
%from the Galmer simulation 
a velocity field (i.e. RA and DEC, and a radial velocity) that mimics a typical observational situation. Then, COSTA can be applied \massi{to the mock velocity field} to recover the cold substructures, together with their intrinsic kinematical parameters. 
Here, we intend to test the capability of identifying streams made of a few particles in velocity fields of different sizes. In particular we %\st{will} 
test the case of $N_{\rm part}=2000, 1000, 500$ extracted from the giant galaxy. These are typical number of test particles found in external galaxies, like planetary nebulae (Fornax cluster: $\sim1000$, $\sim1500$ PNe: \citealt{spiniello-2018} and references therein; M31, $\sim$2000 PNe: \citealt{merrett-2006}; NGC 5128, $\sim1100$ PNe: \citealt{peng-2004}; NGC 4374, $\sim500$ PNe: \citealt{napolitano-2011}) or globular clusters (Fornax cluster: $\sim1000$ GCs: \citealt{pota-2018} and references therein; M87, $\sim500$:  \citealt{romanowsky-2012}). 
%MG: The sentence below is not correct (is the old version): actually, I extracted from the dwarf a number of particles = 150, 75, 38. 
%and for 
For the dwarfs, we consider instead $N_{\rm part}=150, 75, 38$, respectively \citep[e.g.][]{Fahrion-2020}.  
These number of particles are choose to match with the expected particles observable from streams of surface brightness of the order of 28-30 mag arcsec$^{-2}$ (see discussion below). 

Finally, to test different observational conditions, we adopt, for each of the three different selected encounters (i.e. gE0-dE0 and gSa-dS0 prograde/retrograde),  three orders of measurement errors, $\Delta_v=10,~20,~40$ \kms\ by re-sampling the particle velocities with a Gaussian distribution centered on the particle velocity and having $\sigma=\Delta_v$ ($v_{\rm obs}$ hereafter).  
These values are comparable to what typically reached %in 
with mid and low spectral resolution.  
%\massi{For each of the three different selected encounters (i.e. gE0-dE0 and gSa-dS0 prograde/retrograde),}   we produce mock observations with $\Delta_v=10, 20, 40$ kms$^{-1}$. 
\massi{Measurement errors have the effect to dilute the observed velocity distribution of the cold substructure by increasing the observed squared velocity dispersion, 
%. The increment is roughly the square of the measurement error
i.e. $\sigma_{\rm obs}^2$=$\sigma_I^2$+$\Delta_v^2$, where we indicate with $\sigma_I$ the intrinsic velocity dispersion of the stream and with $\sigma_{\rm obs}$ the observed velocity dispersion.}

In the following, we define the mean velocity and velocity dispersion of the detected substructures using some standard statistical definition (see also P+18):

\begin{equation}
v_{\rm mean} = \frac{1}{N}  \sum v_{\rm obs,i},~~~ 
\sigma_{I}^2 = \frac{1}{N-1}  \sum (v_{\rm obs,i} - v_{\rm mean})^2 - (\Delta_v)^2.
\label{eq:std}
\end{equation}
%NRN6 fino a qui ore 23:18

Hence, the larger the $\Delta_v$ is, the larger the chance is that a cold structure become warm enough to skip the cold criterion on $\sigma_{\rm{cut}}$, or that some of the particles are discarded 
%of the stream have measured velocities large enough to be selected-off 
by the sigma-clipping part of the algorithm. This would then leave too few particles to meet the minimum particle number ($N_{\rm min}$) limit, hence making COSTA loosing good candidate streams. 

%CS: I moved a bit earlier the beginning of the new section, because you are already talking about reliability.
\subsection{Setting the reliability of COSTA}
\label{sec:reliability}
Before running COSTA to search for streams, we need to check whether and how often COSTA returns spurious detections. In the case of simulations, this is easily performed by running COSTA on the central galaxy particles only, which represents the smooth warm background, over which streams have to be found when the intruder is added. 

For our analysis, we define the following datasets:
\begin{itemize}
    \item {\it white noise sample (WNS):} RA, DEC and $v_{\rm obs}$ of the giant galaxy or cluster regions without any artificial stream added;
    \item {\it detection sample (DS):} RA, DEC and $v_{\rm obs}$ of the full system including the WNS and the particles of the stream.
\end{itemize}

We use the WNS to select those set-ups (i.e. combination of \emph{k, n, N\textsubscript{\rm min}} and $\sigma_{cut \rm}$) that have a reasonably low probability to find artificial detection and to be used to look for streams in the DS. 
A given set-up that finds no spurious in the WNS has maximum ``reliability'', which means that if it detects a stream in the DS then this is likely to be real. On the other hand, a set-up that finds many spurious detections is highly unreliable and has to be discarded. 

%\chiara{MAYBE MOVE THIS FOLLOWING PARAGR. ELSEWHERE? SO THAT HERE YOU ONLY FOCUS ON RELIABILITY? MAYBE IT CAN GO BEFORE THE BEGINNING OF SEC.3.3? OR LATER ON WHEN YOU DEFINE CF?} 
%It is likely that a detected stream contains, in addition to the real stream particles, also some contaminant background particles, given the statistical nature of COSTA.  This ``contamination'' is a critical parameter to evaluate because contaminants will alter the inferred stream properties.

%\nic{Once a stream is detected, given the statistical nature of COSTA,
%within a detected stream, COSTA can %this is expected to include also some contaminant background particles on top of the real stream particles.} %\st{Of course when the stream is detected (given the statistical nature of the detection) COSTA can pick both real stream particles and some contaminant background particles.} 

%quantitative and 
In order to have a statistical definition of the reliability of the set-ups in the \emph{k, n, N\textsubscript{\rm min}} and $\sigma_{\rm cut}$ space, we use 100 different mock datasets randomly extracted from all 
particles in the simulations. \massi{We use different combinations of number of particles $N_{\rm part}$ and velocity errors $\Delta_v$ for each of the three encounters, and present here some representative cases.  %For sake of space and clarity of the discussion, we present here \chiara{only} some representative cases.
Specifically, we discuss the cases where we randomly extract 2000 particles with errors $\Delta_v=$10 and 40 \kms; 1000 particles with $\Delta_v=$40 and 500 particles with $\Delta_v=20$ \kms).} 
%\st{for five over the nine possible combination of number of particles velocity errors for each encounter configuration (specifically 2000 particle with errors $\Delta v=$10 and 40 \kms\, 1000 particles with $\Delta v=$40 and 500 particles with $\Delta v=20$ \kms\).} 
%\massi{COMMENT: Actually I run COSTA in less than 9 cases}.
For each case, we uniformly sample the \emph{k, n, N\textsubscript{\rm min}} parameters space, for different $\sigma_{\rm{cut}}$ 
%NRN6: forse dovremmo dire i limiti qui?
and run COSTA with all the possible combinations of the free parameters selected in the following ranges:
\begin{itemize}
    \item \emph{k}: from 10 to 30 with steps of 5
    \item \emph{n}: from 1.3 to 3 with steps of 0.2-0.3
    \item \emph{N\textsubscript{\rm min}}: all values from 5 up to \emph{k}
    \item $\sigma_{\rm cut}$: from 10 to 80 \kms{} with steps of 5 \kms{}
\end{itemize}{}

\noindent
For each combination of these parameters %of these sets 
we define the reliability over the 100 random extractions as
\begin{equation}
    \texttt{Rel} = 100 - N_{\rm spu}~ \%
    %\texttt{\# sim. with spurious}~ \%.
\end{equation}
\noindent
\massi{where 
%``\texttt{\# sim. with spurious}'' 
$N_{\rm spu}$ is the number of times we obtain at least one spurious detection from COSTA.}

\massi{We use 70\% as threshold to define a set-up reliable.}
This threshold is somehow arbitrary, as it might depend on the risk one is willing to take in considering a group of particles as a stream.
%\st{In order to consider a set-up reliable enough to qualify for detection, we have fixed a threshold for the reliability to be 70\%.} 
 %whats to take to select a bunch of particles as a stream.
In principle, one should set the reliability toward 100\%, to be sure that none of the detection is spurious. However, this could result in a too conservative choice that might cut all streams that are statistically closer to the white noise given by the background particles. For instance, the properties of streams with a small number of particles and/or too close to the $\sigma_{\rm{cut}}$ may be very close to the properties of the spurious detections, and thus would be filtered out by too conservative thresholds. For this reason we are motivated to choose a lower threshold which might provide a larger completeness but a lower purity, due to the increased chance 
%presence of 
to find some spurious detections. 
Since the main scope of COSTA is to %basically 
provide stream candidates that shall be confirmed with deeper observations, then a fair amount of false detections are acceptable. 
%Any lower reliability threshold would be too close to a random bet. \massi{COMMENT: Maybe we should remove the previous sentence since we choose a threshold of 50\% in Fornax. Anyway I run COSTA on galmer simulations also with different threshold 
We will discuss the impact of the threshold in \S\ref{sec:different thresholds}).

Here below we discuss the results for \massi{the gE0-dE0 and the gSa-dS0 encounters separately and in details.} 
%\st{the three encounters in details.}

%should be used for the research of streams.
%Indeed such a configuration represents relaxed just particles in the phase-space, since no stream are present, hence each cold substructure detected in this way is a false one, allowing us to give a measure of a reliability of each different set-up by counting the number of times COSTA does not find any spurious structure in every set-up. 
%We used 100 different mock datasets for each of the nine cases described above (three different number of particles and three different velocity errors), sampling the whole space of parameters, or running COSTA with all the possible combinations of the three free parameters; thus the reliability for a set-up is defined as
%\begin{equation}
%    \texttt{Rel} = 100 - \texttt{\# sim. with spurious structures}
%\end{equation}

%\medskip
%\massi{In the next subsections we report the results for both interactions that we selected from Galmer simulations.}
%NRN6: fino a qui

\begin{figure*}
\centering
%\vspace{-0.2cm}
%\hspace{-0.8cm}
    \includegraphics[scale = 0.28]{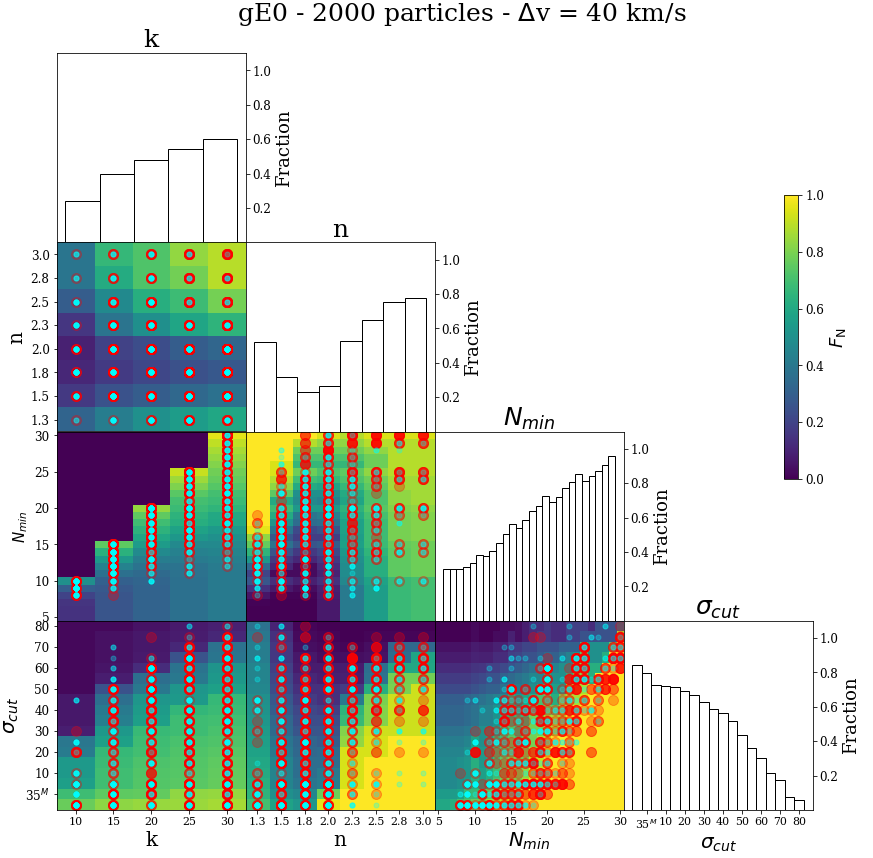}
    %\includegraphics[scale=0.4]{colorbar.png}
%\vspace{-0.2cm}
%\hspace{-0.2cm}
   \includegraphics[scale = 0.28]{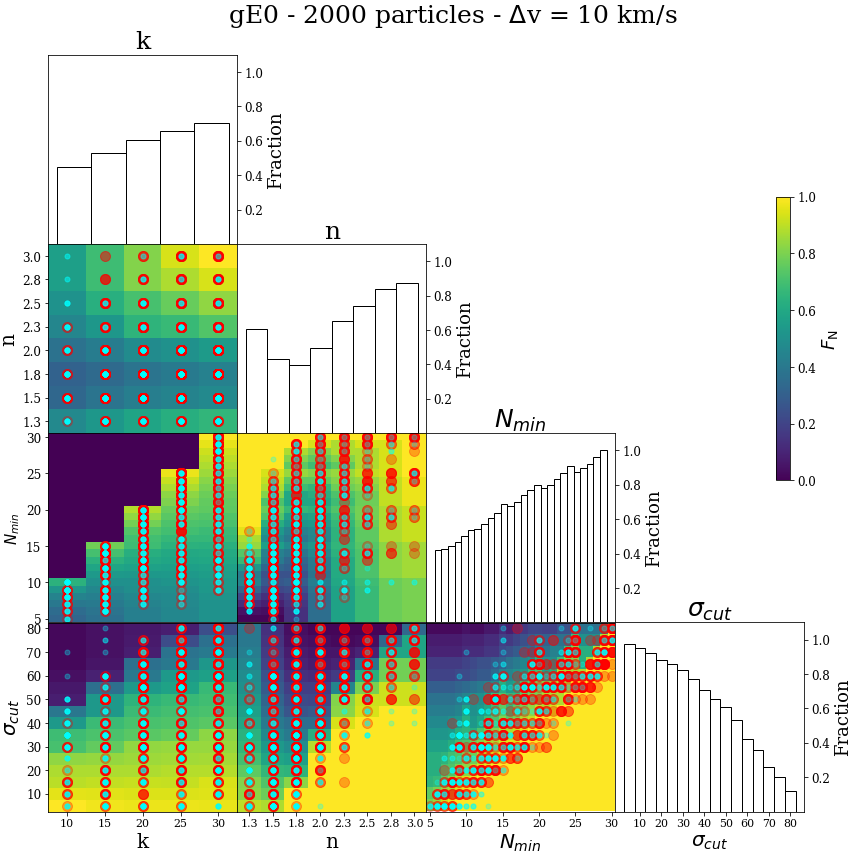}\\
   %\includegraphics[scale=0.4]{colorbar.png}\\
%\vspace{-0.2cm}
%\hspace{-0.8cm}
    \includegraphics[scale = 0.28]{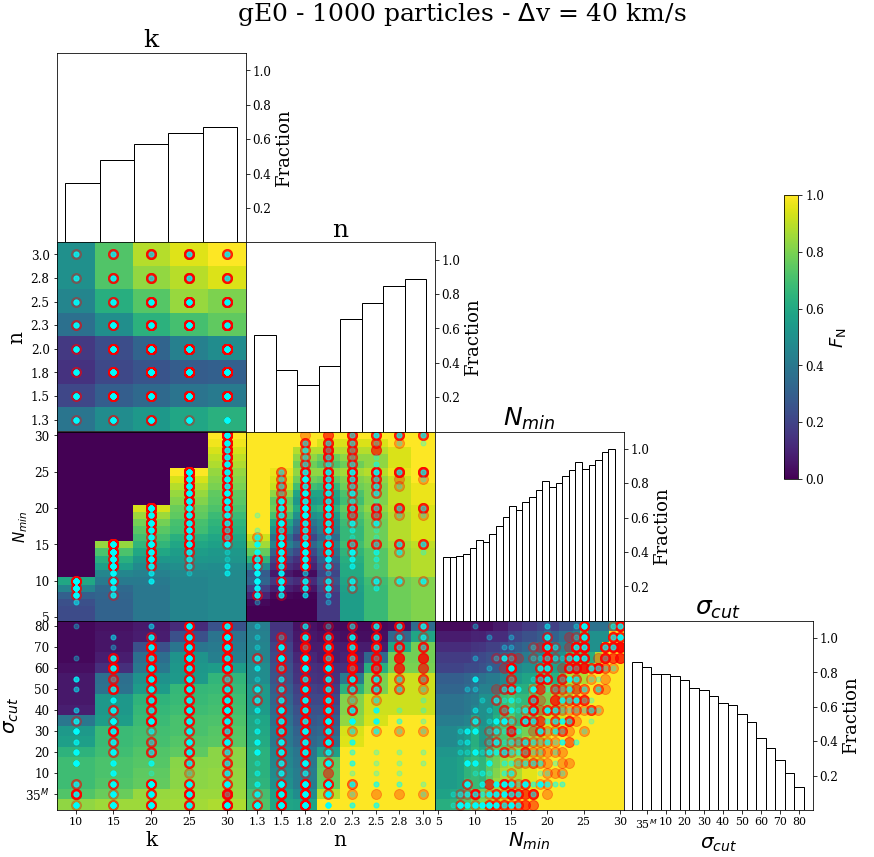}
    %\includegraphics[scale=0.4]{colorbar.png}
%\vspace{-0.2cm}
%\hspace{-0.2cm}
   \includegraphics[scale = 0.28]{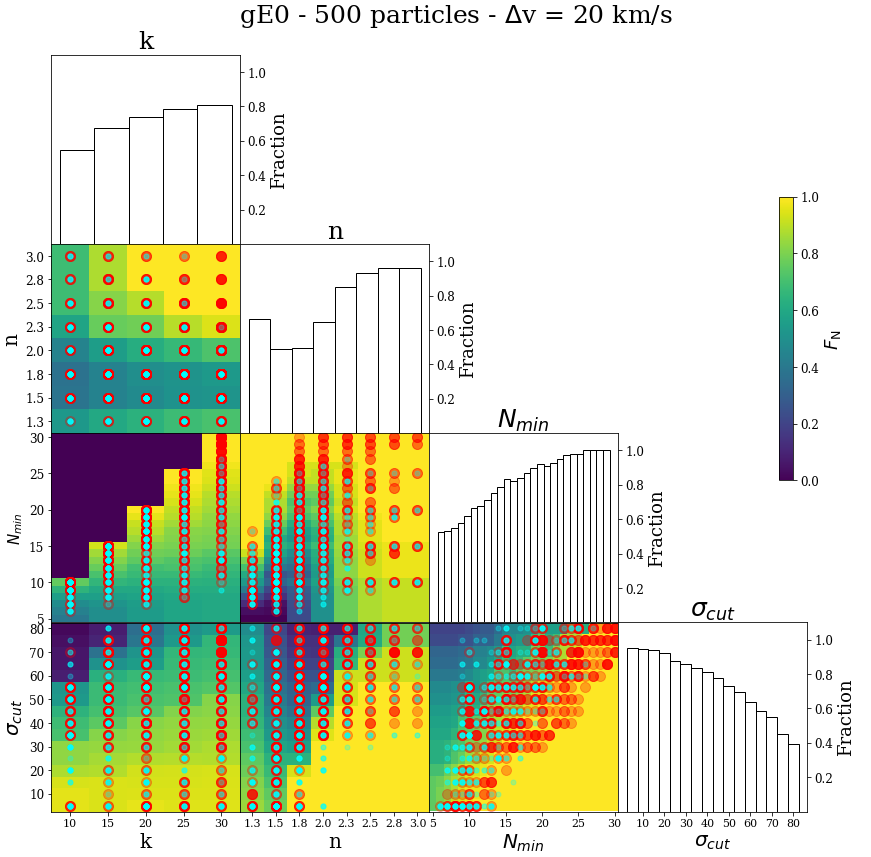}\\
   \caption{Scatter plot between all the possible free parameter pairs, color-coded by the fraction of times ($F_{\rm N})$ a given couple of parameters has a reliability greater than 70\% over the total number of possible configurations, along with the distribution of each parameter. \massi{Red points (cyan points) indicate the combination of parameters where COSTA detected a real (spurious) stream in ten random extraction of the giant+dwarf system.}}
    \label{fig:gE0_parameter_space}
\end{figure*}

\subsection{The case of gE0-dE0 encounter}
\label{sec:gE0-dE0}
%\st{In Table ... we report the tests performed on a number of configurations of the COSTA parameters that allowed to recover the stream particles in different type of encounters and different observation conditions.}
%, i.e. the galaxy particles without the dwarf/stream particles, 
We first test COSTA parameter combinations on the WNS
to check which configurations produce spurious detections over 100 re-extractions of the same catalog, re-sampling the velocity errors for each particle. We exclude the particles in the central 1 kpc of the main galaxy, since these regions are usually highly incomplete in discrete tracers' %recovery rate in observations 
detection \citep[see e.g.][]{napolitano-2001} and any attempt to look for streams would produce very uncertain results.

Then, we collect all configurations which return at least 70\% of the re-simulated field COSTA analysis with no spurious detection (e.g. $\texttt{Rel} \geq 70)$.

\subsubsection{Reliability as a function of the COSTA parameters}
In Fig. \ref{fig:gE0_parameter_space} we show, for the four different combinations of $N_{\rm part}-\Delta_v$, the scatter plot between all the possible free parameter pairs, color-coded by the fraction of times a given pair overcomes the chosen threshold on the reliability, marginalized over the other two free parameters. This cumulative fraction, related to each parameter pair, is hence defined as

\begin{equation*}
    F_{\rm N} =\frac{\texttt{\# parameter-pair with Rel. $\geq 70$\%}}{\texttt{\# total tests for a fixed parameter-pair}}.
\end{equation*}

The latter quantity is expected to be higher for the combinations of parameters that have a lower chance to produce spurious detection for any other choice of the other parameters, and, as such, it represents a quality flag for a given configuration. \massi{Indeed, when a parameter configuration with a low $F_{\rm N}$ finds a stream, the chance that this is a spurious one is higher.} %\st{Indeed even if a parameter configuration with a low \texttt{Fract} finds a stream, it is higher the chance that this is a spurious.} 
In the following, we will use the high $F_{\rm N}$ regions in the parameter space to label the detection as higher-quality (see also below), as they represent the regions where all parameters give a high reliability.

\nic{In Fig. \ref{fig:gE0_parameter_space} the regions of the parameter space that reach the maximum level of reliability  ($>70$\%  for all possible parameter configurations) are shown in yellow, while the quality of the configuration degrades toward the blue as the fraction of >70\% reliability decreases over the total combination including that particular pair.}
%\st{The figure clearly shows that some regions of the parameter space reach the lower acceptable reliability (70\% in this case) for all possible parameter configurations (yellow color in the figure);} 
For the yellow area, this means that, having fixed two of the four free parameters, the threshold for the reliability is reached regardless the values of the other two. 
%A potential stream found within these regions should be much more credible, since these ones are indicative of a lower number of spurious structures.

Looking at the results for the different $N_{\rm part}-\Delta_v$ cases, we can see that regions with $F_{\rm N}\sim 1$ are located %\st{localized} 
at the upper right corner of the $k-n$,  \chiara{on the diagonal of the $k-N_{\rm min}$,} at the bottom right corner of the $k-\sigma_{\rm cut}$, $n-\sigma_{\rm cut}$ and of the $$N_{\rm min}-\sigma_{\rm cut}$$ panels,   
at the upper left corner \chiara{and right side} of the $n-N_{\rm min}$ projection. 
This is valid for all $N_{\rm part}-\Delta_v$, but with different extensions.
Moreover, many other regions in the parameter space have a $ F_{\rm N} > 70$\% (light green color),   %very close to the unity, 
indicating that there are numerous combinations of the four parameters yielding a high reliability, and hence producing very reliable streams with little or no chance to be spurious. 

The same figure also shows the one dimensional distribution of all the four COSTA parameters, corresponding to {\tt Rel}$>70$\%, showing which are the peak values in the parameter space that allow the largest chance of producing little or no spurious streams.

%\begin{figure*}
%    \centering
%\hspace{-0.8cm}
%    \includegraphics[scale = %0.35]{sigma_vs_Nmin_gE02000_err40_cut0.png}
%    \includegraphics[scale = %0.35]{sigma_vs_Nmin_gE02000_err10_cut0.png}\\
%\hspace{-0.8cm}
%    \includegraphics[scale = %0.35]{sigma_vs_Nmin_gE01000_err40_cut0.png}
%    \includegraphics[scale = %0.35]{sigma_vs_Nmin_gE0500_err20_cut0.png}\\
%    
%    \caption{Density plot of the number of set-ups with a %reliability above the selected threshold. Data have been smoothed %with a gaussian function.}
%    \label{fig:disp_vs_sigma_cut_noFrac}
%\end{figure*}{}
%E.g., 
\nic{
In particular, the distribution of $N_{\rm min}$ shows that this is a critical parameter to avoid spurious detections, as the probability of finding spurious structures (i.e. groups of particles with similar velocities) is larger for small $N_{\rm min}$ and monotonically decreases as $N_{\rm min}$ increases, producing a higher overall reliability at larger $N_{\rm min}$.
Indeed, too small $N_{\rm min}$ would increase the chance that a bunch of particles in $k$ neighbors have close velocities by chances, thus returning a spurious detection. On the other hand, as $N_{\rm min}$ defines the minimal mass of the stream that COSTA would detect, too large $N_{\rm min}$ will produce high reliability but also high incompleteness in the final list of stream candidates (as small streams would be filtered out).
%In particular, the distribution of $N_{\rm min}$ shows, as expected (see the discussion above about the effect of this parameter), that the probability of finding spurious structures (i.e. groups of particles with similar velocities) is larger for small $N_{\rm min}$ and monotonically decreases as $N_{\rm min}$ increases, hence producing a higher overall reliability at larger $N_{\rm min}$.
%Thus, the larger is the number of particles that define a stream the better one can assess that this is a distinct substructure, and viceversa, the smaller the $N_{\rm min}$, the higher is the probability that one can find a spuriously cold substructure.} 

The distribution of $\sigma_{\rm cut}$, instead, justifies our choice to avail with many cut-offs, as larger values allow us to find warmer streams with larger $N_{\rm min}$ ($F_{\rm N}$ is close to 1 in the bottom right of the $N_{\rm min}-\sigma_{\rm cut}$ panel), while lower $\sigma_{\rm cut}$ minimize the number of spurious structures (from the 1D distribution of the $\sigma_{\rm cut}$ Fraction). 
About the $\sigma_{\rm cut}$ parameter, it is worth mentioning that we adopt cut-offs in some case lower than the nominal instrumental errors. This is because a stream with an intrinsic velocity dispersion smaller than the instrumental error would give an observed value which can be any random number smaller than $\Delta_v$. Hence, using $\sigma_{\rm cut}\ge\Delta_v$ would  exclude real stream colder than $\Delta_v$. Using smaller cuts, we expect to detect such streams, although we cannot evaluate their intrinsic kinematics. In this cases, we will consider the $\sigma_{\rm mea}\sim\sigma_{\rm obs}$, i.e. without subtracting $\Delta_v$ in quadrature, and mark this latter with an apex.}
%since the measured velocity dispersion, $\sigma_{\rm mea}$ is generally obtained by subtracting in quadrature the measurement errors, $\Delta_{v}$, from the observed velocity dispersion, $\sigma_{\rm obs}$, (i.e. $\sigma_{\rm mea}=\sqrt{\sigma_{\rm obs}^2-\Delta_{v}^2}$), in these cases, we will consider the $\sigma_{\rm mea}\sim\sigma_{\rm obs}$  (and mark this latter with an apex). Indeed, these measurements represent an upper limit of the true structure velocity dispersion, but since we are mainly interested in the detection of streams and not on the estimate of the true kinematics, it is useful to probe also this region of the parameter space.

Looking at the \emph{n} distribution, the set-ups with the highest reliability fractions are located at high \emph{n}, since a shallow sigma-clipping remove less outliers. Hence only structures with an initial low velocity dispersion value fall below a given threshold, unless one sets a higher $N_{\rm min}$, in which case there is little chance of finding a spurious structure with small \emph{n} (see e.g. the central panel in the top left of Fig.~\ref{fig:gE0_parameter_space}).

Finally, the \emph{k} distribution increases monotonically.
Since the higher is \emph{k} the larger are the possible values of $N_{\rm min}$ ($N_{\rm min}$ is varied from 5 up to \emph{k}), it is expected that this distribution mimics the trend of the $N_{\rm min}$ one. \\

As all the previous considerations may be dataset dependent, it is crucial to explore the behavior of the reliability in the parameter space as a function of the number of particles, the velocity errors and also the minimal reliability threshold (e.g. changing this to a lower or higher threshold than 70\%).
In this section we take care of the first two quantities (number of particles and the velocity errors), while we will discuss the reliability threshold in \S\ref{sec:different thresholds}.

%In appendix \ref{appendix: galmer other configurations} we will show similar diagnostics as in  Fig.\ref{fig:gE0_2000p_err40} but for the cases of 2000 particles with $\Delta_v = 10$ \kms and 1000-500 particles with $\Delta_v = 40$ \kms.
%NRN7: devi muovere questa figura sotto dall'appendice a qui

%We start with
%can now have a close look to the effect of the 
%measurement errors. 
By comparing Fig. \ref{fig:gE0_parameter_space} top--right panel, which shows the case of a lower $\Delta_v=10$ \kms\, with the top--left panel, showing $\Delta_v=40$ \kms\,
%one sees that the fraction of reliable set-ups as a function of the parameter pairs is higher in a wider parameter space. \massi{More precisely, 
the number of possible combinations which have \texttt{Rel} $\geq$70\% increases by more than 10\%
when considering smaller errors. In fact, the smaller velocity error values allow COSTA to recognize more efficiently the absence of streams without spurious detections. This is quite encouraging, as it shows that there is little room for spurious cold structures to be produced by the white noise of the background velocity field. Also, this shows that the velocity measurements are crucial to increase the purity of stream detection.

%Also, this test draws attention on the importance of a follow-up on the kinematic tracers in order to improve their spectroscopic measure, since this greatly improve the purity of COSTA.
The bottom left and right panels of Fig.~\ref{fig:gE0_parameter_space} show the configurations with 1000 and 500 particles with $\Delta_v = 40$ \kms and $\Delta_v = 20$ \kms, respectively.
These situations exhibit a wider parameter space with high $F_{\rm N}$ of reliable set-up, in a way similar to the case of 2000 particles and $\Delta_v = 10$ \kms. This is likely because the smaller number of particles reduce more the effect of the noise for a given $\Delta_v$, and consequently the probability that COSTA finds a spurious structure becomes lower. 
However, the smaller number of particles also decreases the sampling of the stream and its signal, overall decreasing the signal-to-noise ratio by roughly $\sqrt{\rm N_{\rm part}}$. This has the consequence that COSTA might not detect the stream with the same efficiency than for higher numbers of particles. Thus, it is essential to test also the detection ability of COSTA as a function of $N_{\rm part}$, when a stream is present in the detection sample.

\begin{table}
    \centering
    \caption{\emph{Column 1}: adopted configuration. \emph{Column 2}: percentage of set-ups where the stream has been recovered with respect to the total set-ups in which COSTA detected at least a cold substructure averaged on ten simulations. \emph{Column 3}: the contaminant fraction (\texttt{CF}, see definition in the text).}
    \begin{tabular}{l|c|c}
    \hline
    gE0 - dE0 & $F_N$ & \texttt{CF} \\
    \hline
    2000 part - 40 \kms & 0.54 $\pm$ 0.18 & 0.71 $\pm$ 0.12\\
    $^a$2000 part - 40 \kms & 0.59 $\pm$ 0.24 & 0.73 $\pm$ 0.11\\
    2000 part - 10 \kms & 0.54 $\pm$ 0.08 & 0.67 $\pm$ 0.13\\
    $^a$2000 part - 10 \kms & 0.67 $\pm$ 0.10 & 0.71 $\pm$ 0.12\\
    1000 part - 40 \kms  & 0.40 $\pm$ 0.22 & 0.70 $\pm$ 0.13\\
    $^a$1000 part - 40 \kms & 0.47 $\pm$ 0.25 & 0.71 $\pm$ 0.12\\
    500 part - 20 \kms & 0.51 $\pm$ 0.20 & 0.64 $\pm$ 0.17\\
    $^a$500 part - 20 \kms & 0.56 $\pm$ 0.21 & 0.66 $\pm$ 0.17\\
    \hline
    \end{tabular}
    \begin{minipage}{128mm}
    \textbf{a:} in these configurations we ruled out set-ups with $F_{\rm N} < 50\%$ \\
    in the $n-N_{\rm min}$ space as described in the text.
    \end{minipage}
%\\
%    \textbf{a:} in these configurations we ruled out set-ups with \texttt{Fract} $< 50\%$ in the $N_{min}-n$ space as described in the text.}
    \label{tab:galmer_statistic}
\end{table}{}

\subsubsection{Stream detection}
\label{sec:stream_detection}
%To this purpose, 
We now run COSTA using all set-up configurations with {\tt Rel}$\ge$70\% over the DS made of the gE0 and dwarf/stream particles, 
%(i.e. 150 randomly extracted particles from the whole dwarf simulated particles for the case of the $N=$ 2000), 
to test the ability of the algorithm to detect cold structures embedded in the hot environment of the central galaxy.
We repeat this procedure ten times in order to take into account statistical fluctuations due to a random extraction of the detection sample particles. 
Furthermore, in order to reproduce a lower limit for the surface brightness of the extracted stream, we impose a minimum number of ten particles to be picked up in an area of about 40 kpc located in the tail of the dwarf.  These numbers correspond to a stream with a surface brightness of the order of 28-30 mag arcsec$^2$ (see discussion in \S\ref{sec:simulated_streams})\footnote{We stress here that this condition has been imposed regardless the $N_{\rm part}$, which might bias the detection toward intrinsically denser streams for lower $N_{\rm part}$ and, as such, increase the detection power of COSTA for these cases. As we are interested to cover a variety of observational conditions, we keep this condition, however we will take into account the detection efficiency as a function of $N_{\rm part}$ when drawing conclusions.}

%NRN: I suggest to add here comment about the fact 
We note that COSTA does not only detect streams in the proximity of the dwarf, but it also correctly identifies other groups of stream particles 
%which are identified as stream members
, including portions of the stream that are far from the dwarf body. However, these latter detections are fairly occasional because particles that are far from the dwarf have spent more time in the halo of the host galaxies and have started to mix in the phase space of the host halos to be detected as part of a decoupled stream (see discussion in \S\ref{sec:kinem_gE_dE}).
%\chiara{MAYBE MOVE THIS FOLLOWING PARAGR. ELSEWHERE? SO THAT HERE YOU ONLY FOCUS ON RELIABILITY? MAYBE IT CAN GO BEFORE THE BEGINNING OF SEC.3.3? OR LATER ON WHEN YOU DEFINE CF?} 
%It is likely that a detected stream contains, in addition to the real stream particles, also some contaminant background particles, given the statistical nature of COSTA.  
\nic{It is likely that streams detected from COSTA contain, along with the actual dwarf particles, also some contaminants, i.e. particles close to the stream that accidentally also have a similar velocity to that of the particles of the dwarf.
This ``contamination'' is a critical parameter 
to evaluate because contaminants will alter the inferred stream properties.
Since we know, from the simulation, which of the system particles belong to, we will use this information to estimate the contamination fraction (see \S\ref{sec:stream_contam}). }
%\st{If the stream is detected we expect that COSTA identifies groups of particles belonging to the dwarf and possibly also some particle that accidentally have a velocity close enough to the ones of the dwarf to be also grouped as part of the stream.} 
Regardless of the mix of the dwarf/stream particles with the background main galaxy particles, we expect that the stream particles closer to the dwarf body are the ones that most keep their kinematics clearly decoupled by the hot background (see also \S\ref{sec:kinem_gE_dE}).
%Since we are mainly interested on stream detection, %leaving the analysis of an accurate measure of the stream properties in a follow-up, 
Among all candidate streams that COSTA recognizes on the DS, we consider as true %positive 
detections the ones where COSTA correctly identifies at least 4 particles of the stream/dwarf or where at least one third of the total particles (dwarf + contaminants from the main galaxy)%belonging to the identified group (generally made of dwarf particles plus some contaminants from the main galaxy) 
are from the stream/dwarf.

The final results of COSTA true positive detections are shown also in Fig.~\ref{fig:gE0_parameter_space}. Here we overplot on the density plots, showing the $F_{\rm N}$, the combination of parameters where COSTA found the stream (i.e. true positives, red points) and spurious groups (i.e. false positives, cyan points), in the ten repeated DS extractions. 
\chiara{In many panels, real (red points) and spurious (cyan) streams cluster in different regions, even though, visually it is not always simple to see this. 
The most evident case is the $N_{\rm min}-\sigma_{\rm cut}$ plot, where red points are slightly shifted towards the right corner, where $F_{\rm N}$ is higher. More quantitatively, in the $N_{\rm min}-\sigma_{\rm cut}$ projection of the case with 2000 particles and $\Delta_v=40$ \kms, the median $F_{\rm N}$ for true positive equals 0.88 while that of spurious streams is only 0.73. }

%CS: I Have rewritten everything but the original version is still here. 

%Starting from the top left panel, which represents the case of 2000 particles and $\Delta_v=40$ \kms, {\color{red} one can notice two things: 1) true positive tend to agglomerate in the high \texxtt{Frac} regions while false positive are more abundant in the low \texxtt{Frac}; 2) there are regions where true positive come with no false positive and these are always in the high \texxtt{Frac}, while there are almost no true positive in the regions with \texxtt{Frac}$\leq0.2$ in almost all projections. 
%it seems, at a first glance, that cyan points are in the exact same region as the red ones. Anyway, in some of the projection in the parameter space, the streams correctly detected tend to agglomerate towards higher \texxtt{Frac} values with respect the false positives. 
%In particular, in the $N_{\rm min}-\sigma_{\rm cut}$ plot, red points are slightly shifted towards the right. 
%\chiara{Quantitatively,}in the $N_{\rm min}-\sigma_{\rm cut}$, the median \texxtt{Frac} for true positive equals 0.88 while that of spurious streams is only 0.73. 
Another useful projection which slightly separates real stream from spurious is the $n-N_{\rm min}$ in the middle of each corner plot. This panel shows the compromise between how strong the sigma clipping can be depending on the minimal number of particles expected in the stream.  Indeed a closer inspection of the $n-N_{\rm min}$ plot reveals that many spurious structures have been detected in the bottom left region, while red points tend to cluster on the upper right. Being more quantitative, the median $F_{\rm N}$ of red and cyan points in the $n-N_{\rm min}$ panel are 0.58 and 0.49, respectively. %In the $N_{\rm min}-\sigma_{\rm cut}$ the difference between true and false positives is even more striking, being the median \texxtt{Frac} equals to 0.88 and 0.73 for streams and spurious, respectively.

Thus, in order to minimize the chance of over-collecting spurious streams we adopt a threshold on the $F_{\rm N}$ in the $n-N_{\rm min}$ panel. In particular, setting a minimum value of $F_{\rm N}$ = 0.5, 
%in the $n-N_{\rm min}$ panel 
we remove about 50\% of the spurious structures. 
\massi{We note that, despite the separation is clearer in the $\sigma_{\rm cut}-N_{\rm min}$ panel, we prefer to set a threshold in a perpendicular direction of the parameter space, with respect to $\sigma_{\rm cut}$ in order to reduce the chances to bias the final selection in a projection which is strictly related to stream physical properties.  
%\texttt{Fract} only in the $n-N_{\rm min}$ panel.
%Such a choice prevents that we introduce some bias in the inferred stream properties. 
In fact, a further clean involving  $\sigma_{\rm cut}$ might alter the estimated stream kinematics. Since some of the very low $F_{\rm N}$ regions lie at high $\sigma_{\rm cut}$ values, removing such regions would rule out all the combination of parameters having $\sigma_{\rm cut}$ close to the actual dwarf velocity dispersion (77 \kms,  see also Table~\ref{tab:galmer parameters}).

In the following, we use this threshold as further condition over the detected structures to clean out our list of candidate streams. \chiara{The effectiveness of this choice will become} more clear in \S\ref{sec:kinem_gE_dE}}.

In Table~\ref{tab:galmer_statistic} we report the fractions of set-ups that reveal the stream tail without any false positive, averaged over ten simulations returning at least one detection (i.e. either a true or false positive), \chiara{with and without applying the threshold of $F_{\rm N}$ = 0.5. We also report the contaminant fraction, which will be defined in \S\ref{sec:stream_contam}. 

Generally, the threshold in $n-N_{\rm min}$ increases the number of set-ups where the stream is recovered. This is particularly evident for the bast case with $N_{\rm part}=2000$ and $\Delta_v=10$ \kms\, where the fraction of setups returning streams with no spurious is $\sim$ 67\% when applying the the threshold of $F_{\rm N}$ = 0.5 in the $n-N_{\rm min}$ plane and goes down to $0.54$ without it. 
%in the case we operate the \texxtt{Fract} = 0.5 or not. 
%For $N_{\rm part}=2000$ and $\Delta_v=40$ \kms\, we can see in the former case the fraction of setups returning streams with no spurious is $\sim$ 59\%, in the latter case this decreases to $\sim$ 55\%. 
Given the uncertainties, however, this makes very little difference. 
The same can be said for the impact of changing the number of particles and adopting different velocity uncertainties. 
Going from 2000 to 1000, keeping $\Delta_v$ fix to $40$ \kms\, $F_{\rm N}$  goes down from $0.54$ to $0.40$, but it is always consistent within one-$\sigma$ errors. 
%The larger changes are due to the adoption of different velocity errors, although also in this case the numbers are always consistent within their uncertainties. 
} 

\chiara{Lower velocity errors tend to shift detected streams towards ``more reliable" regions of the parameter space. This is also visible directly from Fig.~\ref{fig:gE0_parameter_space}, comparing the top-left and top-right panels and using again the the $n-N_{\rm min}-$ and the $N_{\rm min}-\sigma_{\rm cut}$ panels to discriminate between real streams and spurious. Yellow region are more extended in all panels for $\Delta_v=10$ \kms. }

\massi{The bottom-left panel of Fig.~\ref{fig:gE0_parameter_space} shows the results of the case 1000-75 giant-dwarf particles and $\Delta_v=40$ \kms.}
%is displayed in Fig.\ref{fig:gE01000_err40_stream}; 
Here COSTA is still able to detect the stream, even though 
\chiara{the ratio of the number of set-ups where the stream has been recovered over the total number of set-ups is the lowest (see column 2 of Table~\ref{tab:galmer_statistic}), with and without the threshold.}

%\st{the mean contaminants (i.e. main galaxy particles associated to the stream)
%NRN7 questi numeri non li abbiamo dati sopra per le altre configurazioni?
%is increased (from 56\% to 66\%)}.
%\st{the fraction of set-ups in which the stream is recovered is smaller %are %considerably 
%less in number 
%(see column 2 of Table ).}

%, likely due to a lower signal respect to the previous case. 

Finally, we consider the case with 500-38 giant-dwarf particles. Here we show the result for $\Delta_v=20$ \kms in the bottom-right plot of the same figure. This is, in fact, %an average 
the precision one can obtain with typical mid resolution spectroscopy. In this case also, COSTA is well able to catch the stream in a quite ample range of configurations in the parameter space ($\sim 50\%$). %\st{even if the signal is significantly reduced because of the smaller number of particles tracing the stream.}
%\st{It is interesting to notice that the fraction of set-ups in which the stream is recovered are larger than the case with $N_{\rm part}=1000$ (see Table , column 2), mainly because of the smaller velocity errors.} \chiara{I AM NOT CONVINCED THIS IS THE REASON. WHY GOING FROM 2000-40 TO 2000-10 WE DO NOT SEE AN INCREASE IN THE FRACTIONS?}

%\st{Hence here we can further conclude that  in planning observations to look for streams, velocity accuracy should be the primary driver also at cost of smaller samples (i.e. survey depth).}

\chiara{In conclusion, for all the different configurations we test, changing the number of particles and the velocity accuracy, COSTA is able to recover the stream in a relatively broad space of parameters (ranging between 40 and 67\%). We note that a fifty percent of success is } acceptable in blind stream searches, if one wants to find a list of candidates to follow-up, and \chiara{represents} a fair compromise between purity (no false positives) and completeness (i.e. find as many real stream as possible), see also \S\ref{sec:stream_contam}. 

\begin{figure*}
    \centering
    \includegraphics[scale = 0.5]{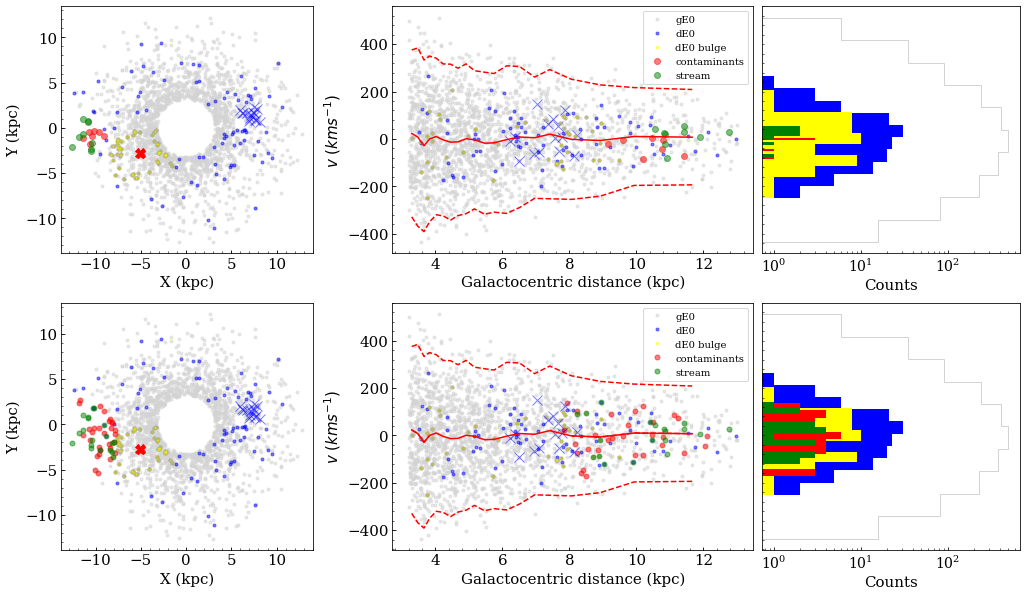}
    \caption{\emph{Top:} relative positions (left panel) and reduced phase-space (right panel) in the case of the stream recovered with $N_{min}$ = 15. \emph{Bottom:} same as above but with the stream recovered with $N_{min}$ = 30. Light gray points are gE0 particles, while blue ones are those belonging to the dE0. Yellow points represent dwarf particle within 3 effective radii from the dwarf center, while the recovered stream is colored in green (real stream particles) and in red (contaminants).}
    \label{fig:galmer_found}
\end{figure*}{}

\begin{figure*}
    \centering
\hspace{-0.8cm}
    \includegraphics[scale = 0.33]{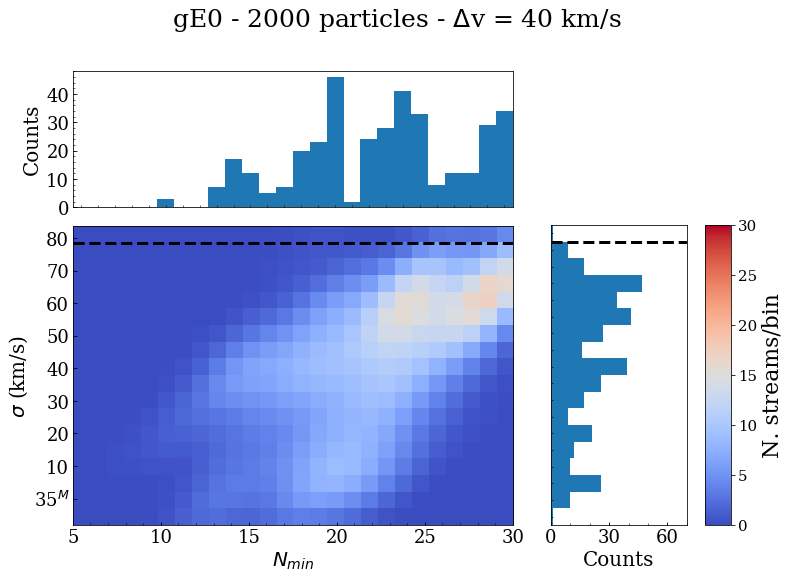}
    \includegraphics[scale = 0.33]{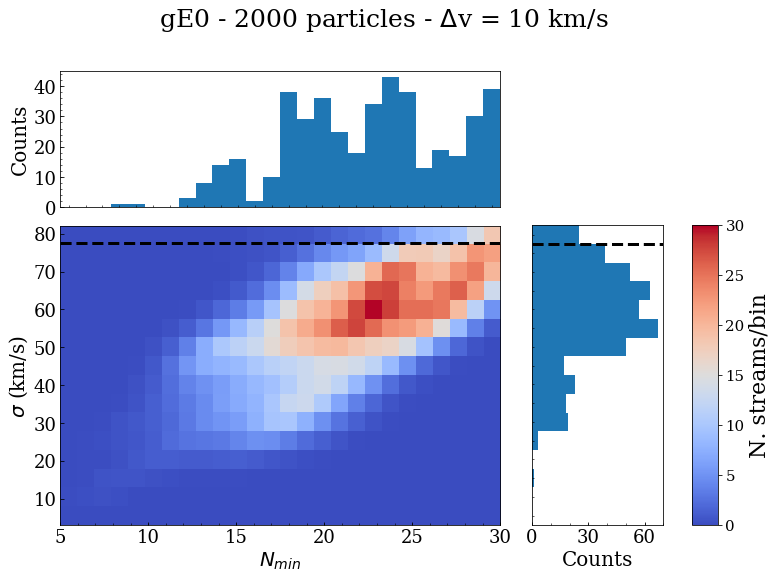}\\
\hspace{-0.8cm}
    \includegraphics[scale = 0.33]{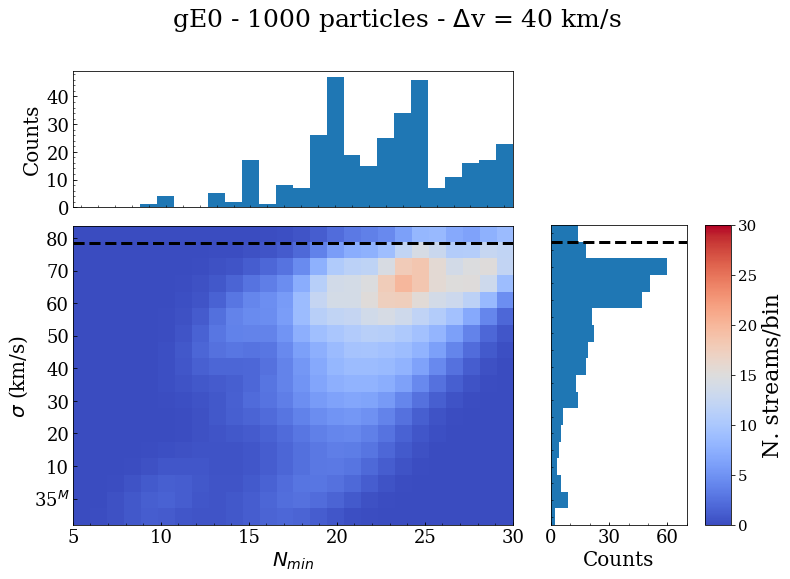}
    \includegraphics[scale = 0.33]{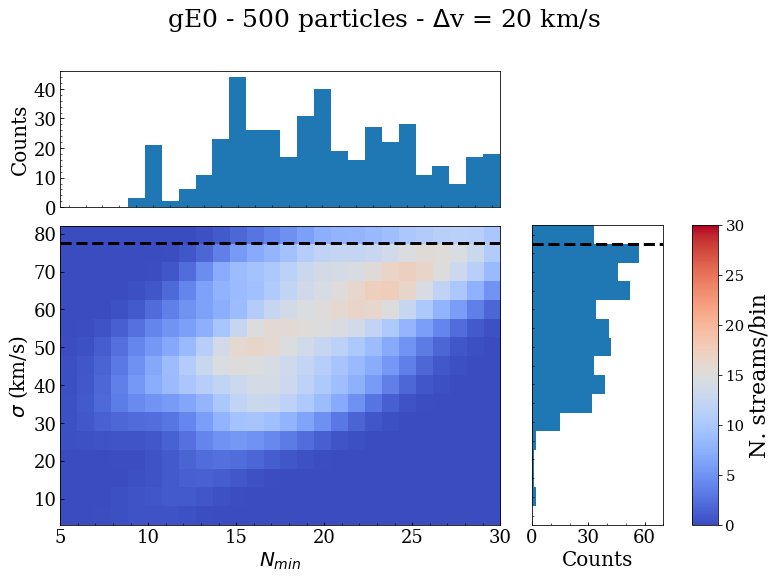}\\
    
    \caption{Density plot of the number of set-ups with a reliability above the selected threshold and with $F_{\rm N}$ ($\geq0.5$) in the $n-N_{\rm min}$ space. Data have been smoothed with a Gaussian kernel having a bandwidth equal to 3 bins.}
    \label{fig:disp_vs_sigma_cut}
\end{figure*}{}

\subsubsection{Completeness and contamination}
\label{sec:stream_contam}
%As anticipated, 
We \chiara{can now better describe and quantify }%can now proceed to assess closer 
the stream properties as returned by the different set-ups. So far, we have identified the set-up that give the true positives, but every set-up produces different groups of particles, including real stream particles. In particular we can check what is the degree of contamination introduced by the different set-ups \chiara{with the purpose of finding} %and finally maybe find 
a method to define the best set-up, e.g. the one optimizing the ratio between number of real particles \chiara{and contaminators}. %included and contamination.
To do that, we define the observed completeness (\texttt{OC}), as 
\begin{equation*}
\texttt{OC} = \frac{\texttt{\# recovered stream particles}}{\texttt{total recovered (stream and non-stream)}}.
\end{equation*} 
%also contains the derived contamination of the detected streams. 

This parameter is clearly complementary to the contaminant fraction (\texttt{CF}) of the stream (i.e. \texttt{CF} = 1 - \texttt{OC}): %, or
\begin{equation*}
\texttt{CF} = \frac{\texttt{\# recovered non-stream particles}}{\texttt{total recovered (stream and non-stream)}}
\end{equation*}
The mean \texttt{CF} derived over the totality of the set-ups producing no false positive (3 column of Table~\ref{tab:galmer_statistic})
%, \st{as in Table ~ column 2, are reported in column 3 of the same Table: as we can see these} 
are always $\sim65-70\%$, almost independently from the sample size and velocity accuracy, which, by definition correspond to $\sim 35-30\%$ of \texttt{OC}. This high fraction of contaminants can significantly affect the conclusion about physical properties of the stream (see e.g. \ref{sec:kinem_gE_dE}). However, we stress that these quantities are %first 
an average over many set-ups and, in principle, one can define the optimal set-up that maximize the \texttt{OC}. We will enter in more details about this optimization in \S\ref{sec:fornax_contam}. 
We also remark here that the contaminant fraction does not impact the detection of the stream that still remain a good candidate for subsequent follow-ups.  These are %meant to be 
needed in any cases to obtain the physical properties of the stream (luminosity, colors, surface brightness, kinematics, etc.).

\begin{figure*}
%\vspace{-0.2cm}
\hspace{-1.2cm}
    \includegraphics[scale = 0.155]{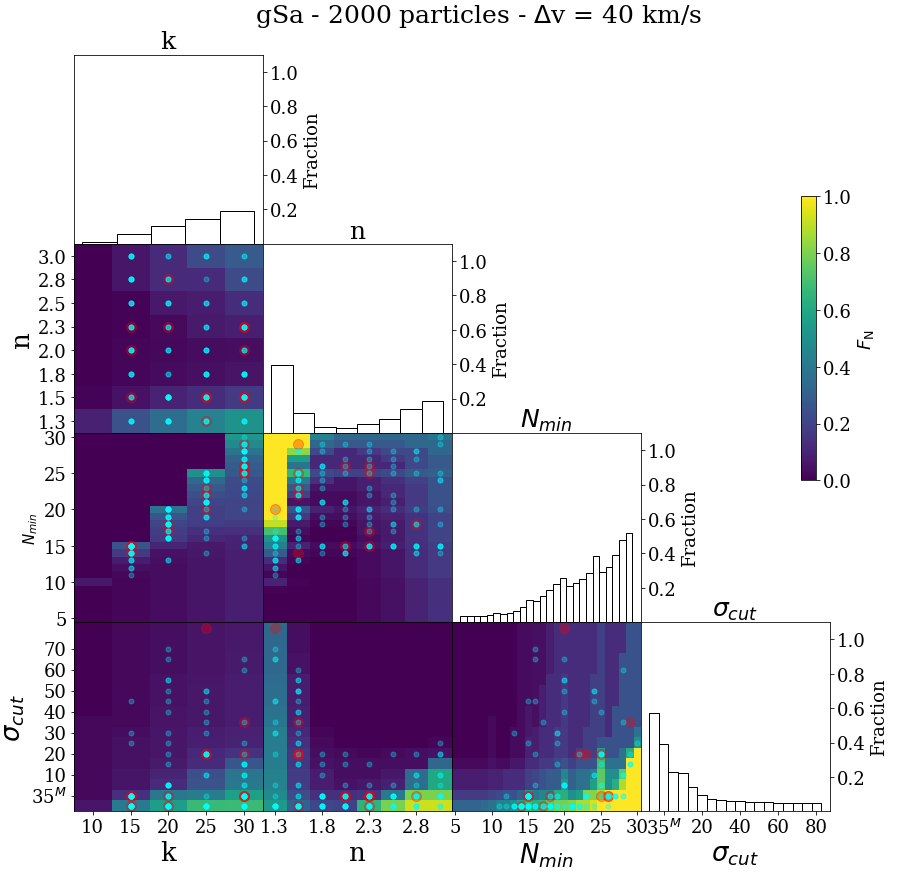}
    %\includegraphics[scale = 0.15]{colorbar.png}
%\vspace{-0.2cm}
\hspace{-0.23cm}
    \includegraphics[scale = 0.155]{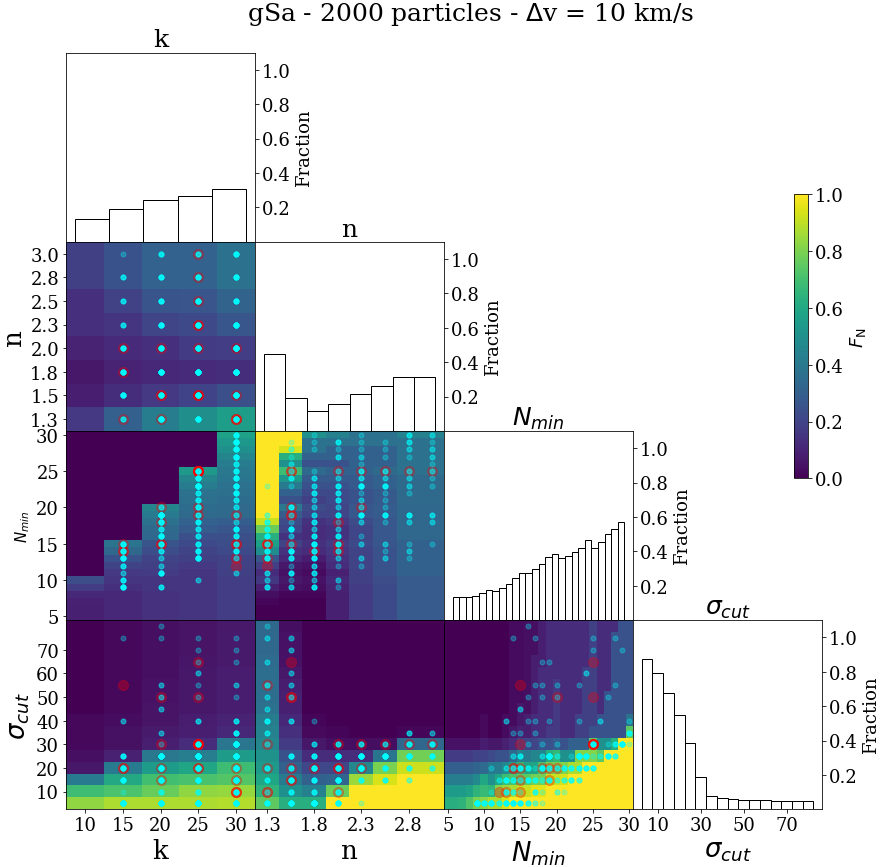}
    %\includegraphics[scale = 0.15]{colorbar.png}
%\vspace{-0.2cm}
\hspace{-0.23cm}
    \includegraphics[scale = 0.155]{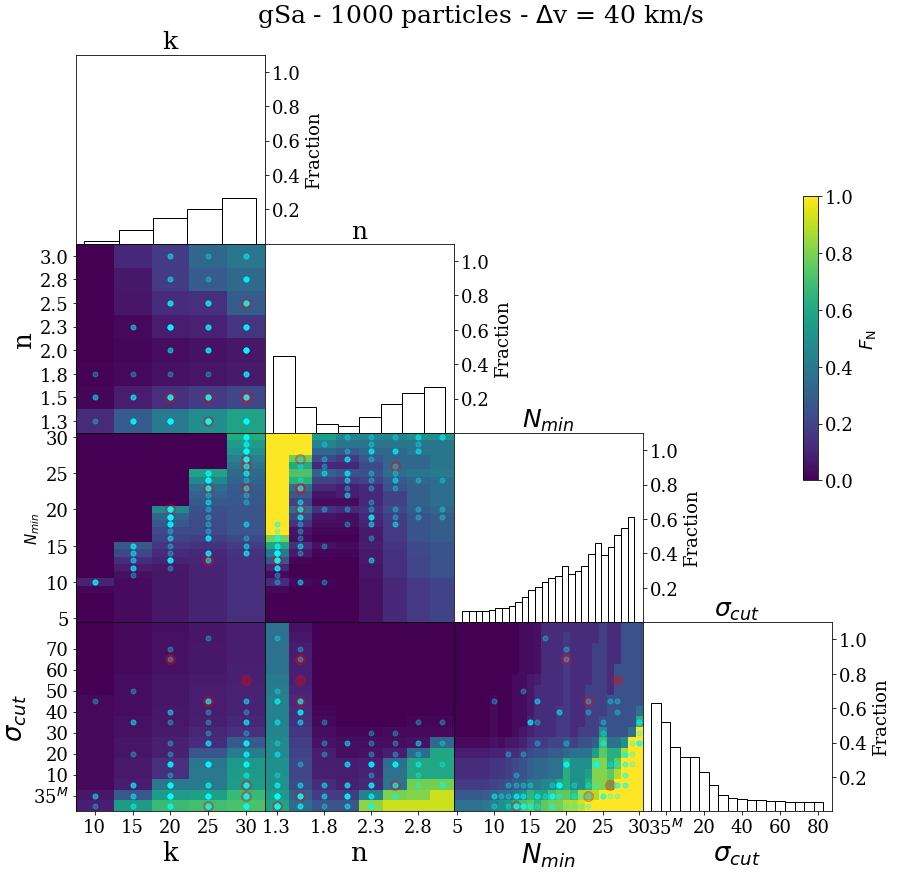}
    %\includegraphics[scale = 0.15]{colorbar.png}
%\vspace{-0.2cm}
\hspace{-0.23cm}
    \includegraphics[scale = 0.155]{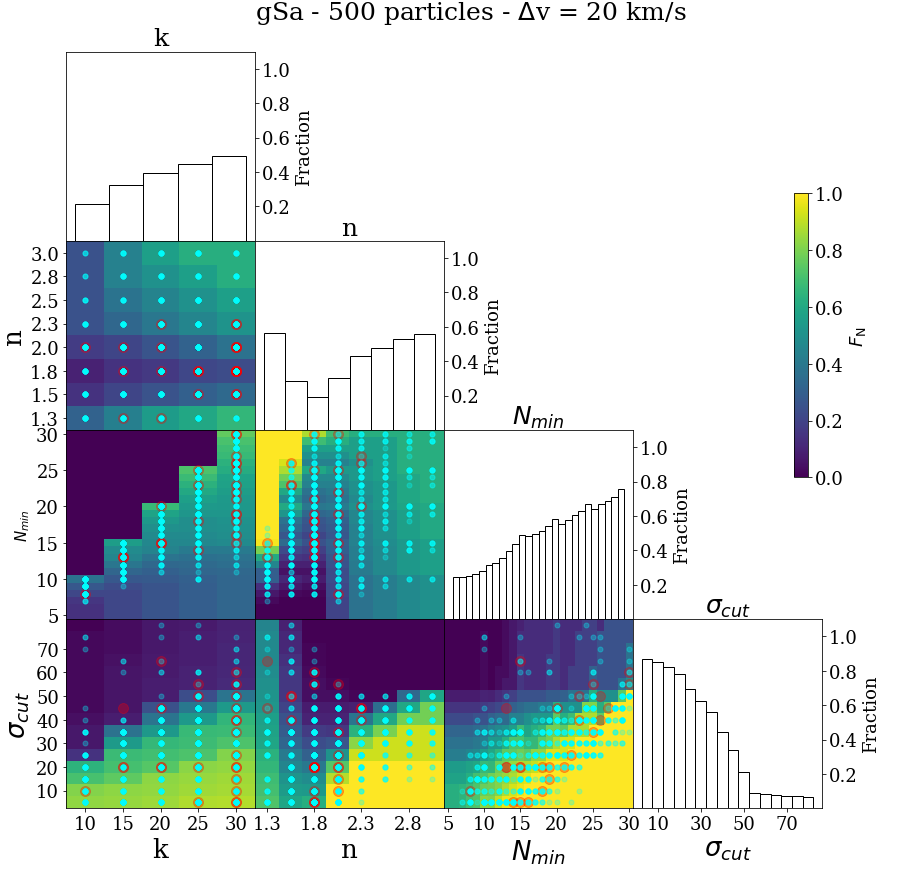}\\
    %\includegraphics[scale = 0.2]{colorbar.png}\\

%\vspace{-0.2cm}
\hspace{-1.2cm}
    \includegraphics[scale = 0.155]{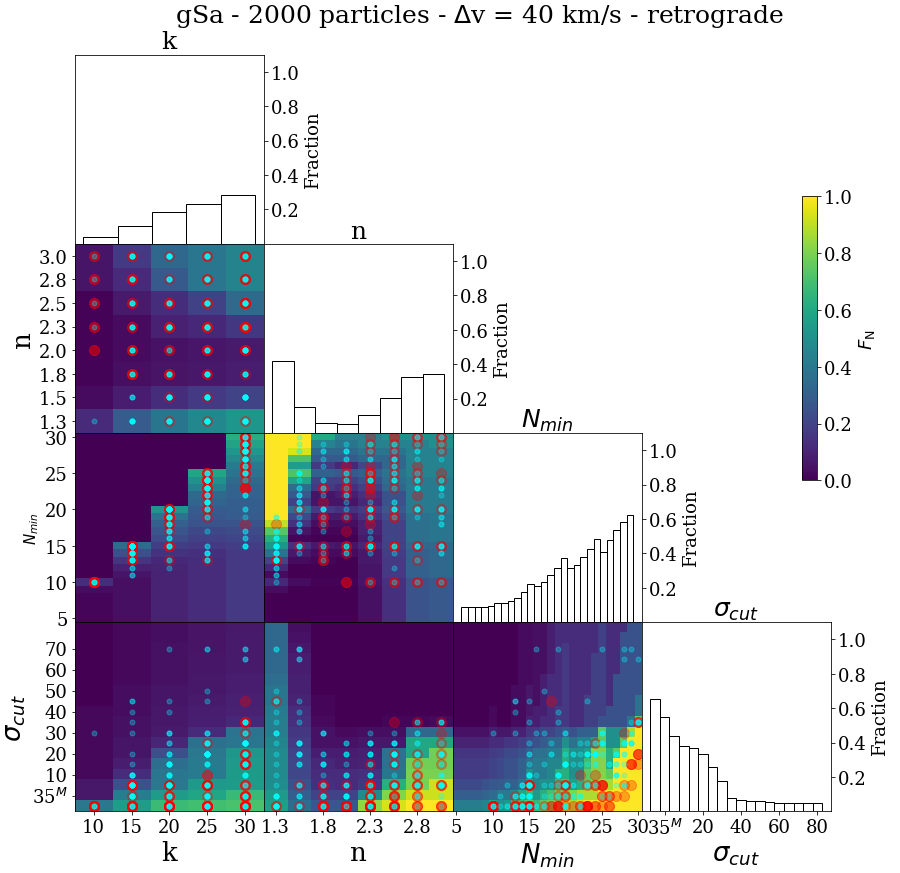}
%\vspace{-0.2cm}
\hspace{-0.23cm}
    \includegraphics[scale = 0.155]{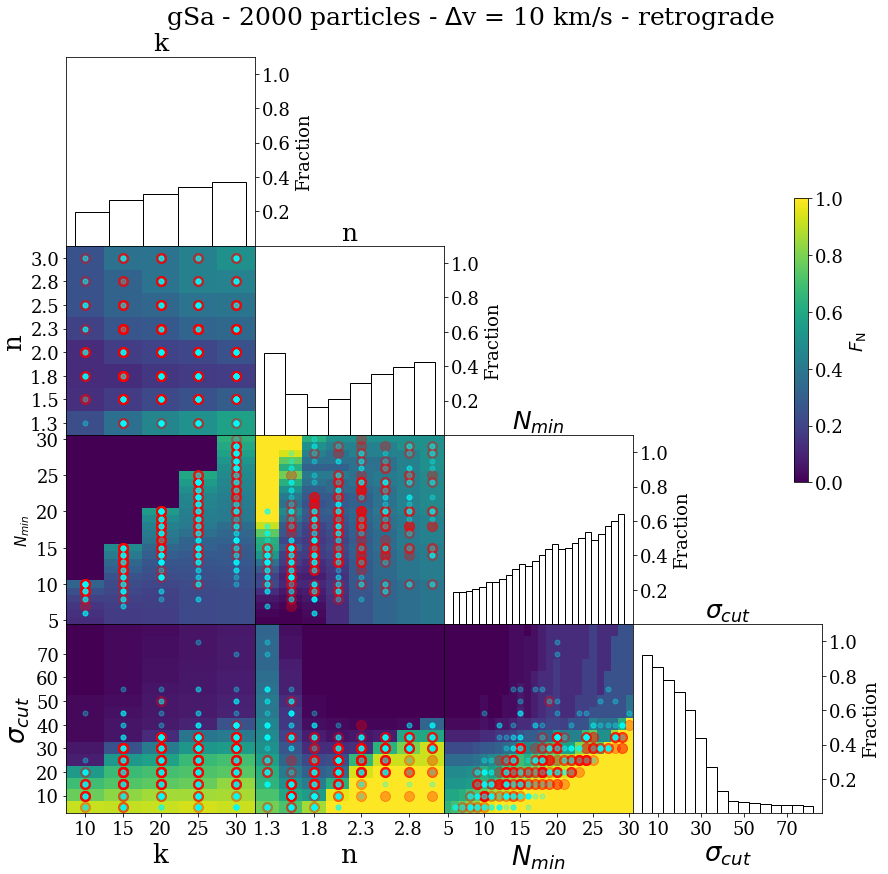}
%\vspace{-0.2cm}
\hspace{-0.23cm}
    \includegraphics[scale = 0.155]{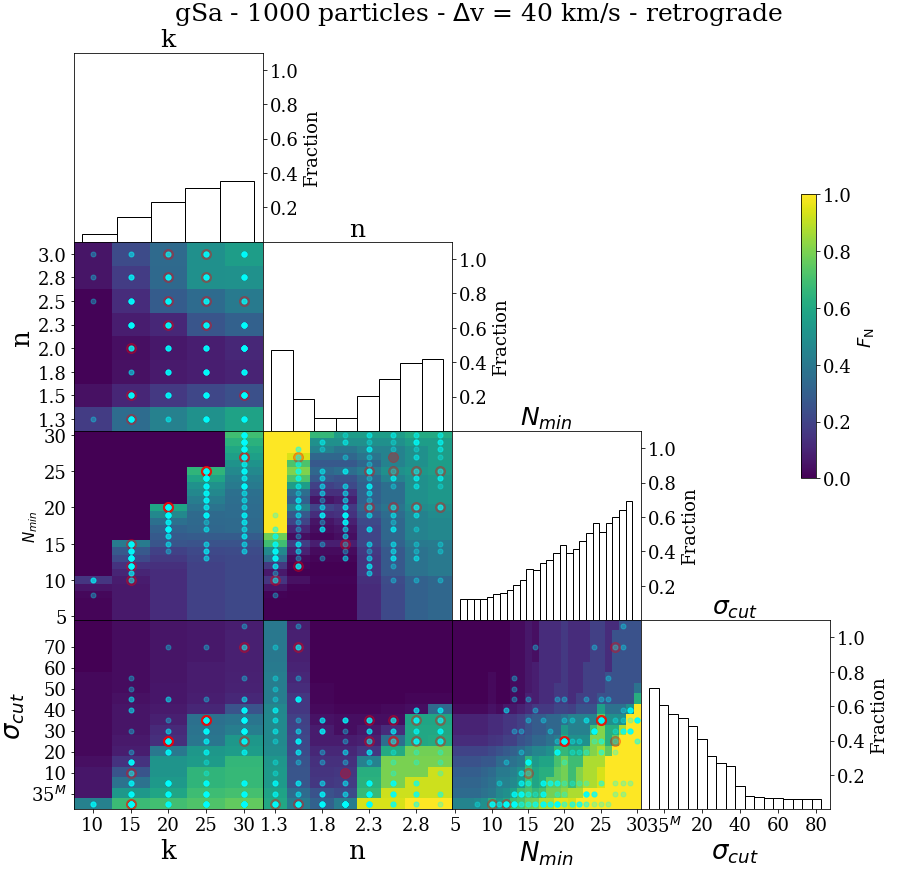}
%\vspace{-0.2cm}
\hspace{-0.23cm}
    \includegraphics[scale = 0.155]{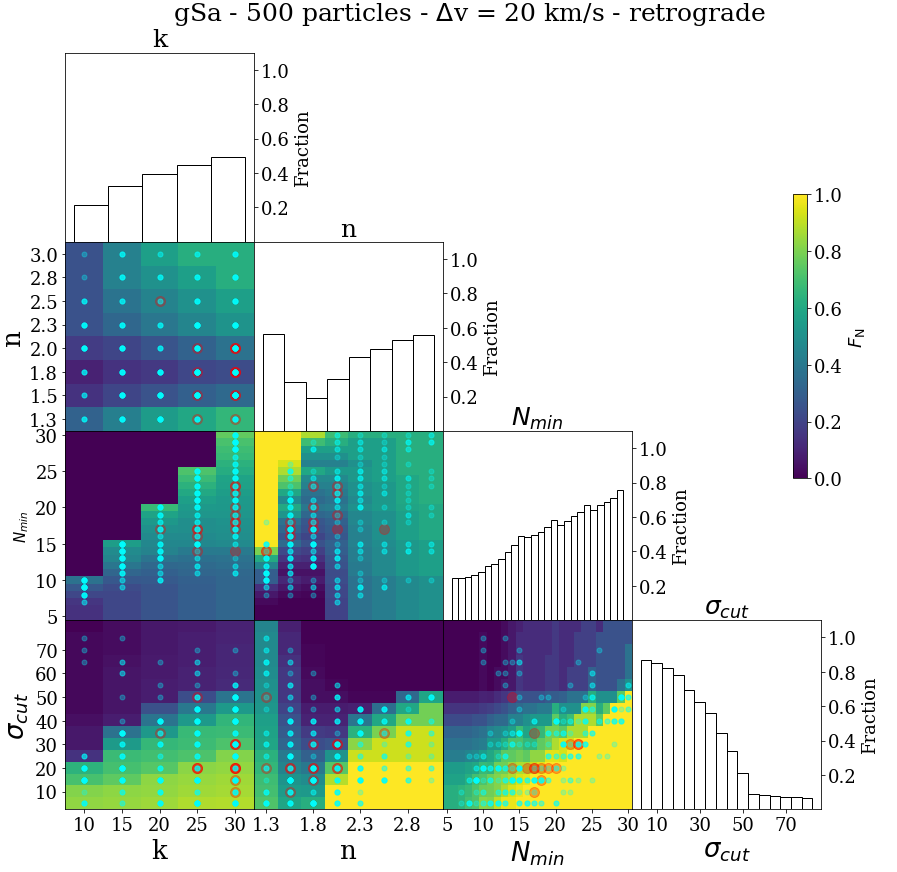}\\

    \caption{Same of Fig.\ref{fig:gE0_parameter_space} but for the gSa in the prograde and retrograde encounter.}
    \label{fig:gSa_prograde_parameter_space}
\end{figure*}

\subsubsection{Stream kinematics}
\label{sec:kinem_gE_dE}
Once having demonstrated that COSTA is able to detect a stream, if any, we are interested to extract some physical properties of the recovered stream. 
In particular, we are interested in deriving some kinematical information of the stream from the velocities of the tracers collected as part of it. Hence, we want to find some rule of thumb to apply to the many configurations that find the stream,  to identify the set-ups which better characterize its kinematical properties (e.g. its velocity dispersion). %of the stream. 

%To establish a kinematical connection between the stream and the parent galaxy, we need to compare the inferred velocity dispersion from the stream candidate particles and the velocity dispersion of the dwarf. 
%\nic{Unfortunately, it is not easy, even in simulated samples like the ones from Galmer, to unequivocally define the velocity dispersion of the stream. This is because, technically,  the stream is made of all particles left behind by the disrupting dwarf, which have a different degree of mixing, depending on the time they have become unbound. On the other hand, one can expect that the latest stripped particles are the ones easier to detect as they are more spatially concentrated and they keep memory of the kinematics they had when still bound to the system, hence looking cold to COSTA\footnote{Dynamically speaking this means that both the energy transfer from orbital momentum to the dwarf GCs \citep[e.g. via tidal approximation, see][]{napolitano-2002} and the dynamical friction ($\propto \rho/\sigma_{\rm env}^2$, where the density, $\rho$ and the inverse of the velocity dispersion $1/\sigma_{\rm env}^2$ of the environment are small), have a minor impact on the stripped particles.}.}

Note that a dynamical definition of the stream velocity dispersion is not straightforward, even in simulated samples like Galmer. 
%, it is not easy, even in simulated samples like the ones from Galmer, to unequivocally define the velocity dispersion of the stream. This is because, 
Technically, the stream is made of all particles left behind by the disrupting dwarf, which have a different degree of mixing, depending on the time they have become unbound. In Fig.~\ref{fig:galmer_found} we compare the position and velocity distribution of particles belonging to the galaxy background (in grey), the ones belonging to the outskirt of the dwarf galaxy (in yellow, which center is shown as a red cross), and the stream particles (in blue) for one of the run discussed in \S\ref{sec:gE0-dE0} and two different $N_{\rm min}$ values ($N_{\rm min}$=15 top, $N_{\rm min}$=30 bottom). In the same figure we also plot the true stream particles detected by COSTA in this run (in green), and contaminants that COSTA selected but are instead not part of the true streams (in red). 
From this figure we can see that the stream particles (blue) overall have %the overall blue particles have
a wider distribution with respect to both the dwarf body particles (yellow) and the ones that COSTA detects in the proximity of the dwarf (red and green), while they are not as dispersed as the gray particles of the central galaxy halo. As such, they are both unbound to the parent dwarf and unmixed to the host halo, hence their velocity dispersion does not have a dynamical meaning because hydrostatic equilibrium does not hold. On the other hand, the ``youngest'' regions of the stream (green particles) show a distribution which is similar to the ones of the dwarf particles (yellow) that are at the equilibrium.
Thus, the particles recently lost in the tail (and more likely detected from COSTA) keep the record of the kinematics of the parent galaxy\footnote{Dynamically speaking this means that both the energy transfer from orbital momentum to the dwarf GCs \citep[e.g. via tidal approximation, see][]{napolitano-2002} and the dynamical friction ($\propto \rho/\sigma_{\rm env}^2$, where the density, $\rho$ and the inverse of the velocity dispersion $1/\sigma_{\rm env}^2$ of the environment are small), have a minor impact on the stripped particles.}.
This means that these latter particles have not yet fully dynamically decoupled from their progenitor and
%hope to derive information about the Faber-Jackson of the progenitors and use this to infer the global properties of the  disrupted/disrupting dwarf population. 
%Thus, 
we are dynamically motivated to compare their velocity dispersion 
%of stream particles that might likely be detected from COSTA in proximity of the parent dwarf galaxy 
with the dwarf velocity dispersion (i.e. 77 \kms, see table \ref{tab:galmer parameters}). This is useful for two main reasons: 1) algorithm-wise this is the best way to identify set-ups that better describe some (dynamical motivated) kinematical properties of a detected stream; 2) dynamically-wise we %\textcolor{magenta}{believe }%postulate 
postulate that the stream velocity dispersion should follow the Faber-Jackson \citep{Faber&Jackson1976} of the parent dwarfs, i.e. the velocity dispersion should correlate with the luminosity of the progenitor, if any.

To illustrate how this works on data, looking at Fig.~\ref{fig:galmer_found} again we clearly see a typical situation of a stream detection where stream particles (including some contaminants) are close to the bulk of the parent galaxy.
%we compare the position and velocity distribution of particles belonging to the galaxy background (in grey), the ones belonging to the outskirt of the dwarf galaxy (in yellow, which center is shown as a red cross),} and the stream particles (blue) for one of the run discussed in \S\ref{sec:gE0-dE0} and two different $N_{\rm min}$ values ($N_{\rm min}$=15 top, $N_{\rm min}$=30 bottom). 
%The true stream particles detected by COSTA are highlighted in green, while contaminants that COSTA selected but are instead not part of the true streams are color-coded in red. 

%\chiara{I AM NOT SURE I UNDERSTAND. BLUE = ALL PARTICLES OF THE STREAM NOT DETECTED. GREEN DETECTED ONES, RED CONTAMINANTS THAT COSTA THINKS ARE FROM THE STREAM BUT I NREALITY ARE NOT? }
For both the case with $N_{\rm min}$=15 and that with $N_{\rm min}$=30, COSTA selects only a limited fraction of particles and very close to the dwarf (tail). The ratio of the red particles over all red+green gives the \texttt{OC}, which 
%\st{stays similar in the two $N_{\rm min}$ cases.} 
\massi{decreases towards higher $N_{\rm min}$ (e.g. 0.56 vs 0.40).}
On the contrary, the overall velocity dispersion increases from the $N_{\rm min}=15$ to 30 (as seen in both the phase-space diagram -- the velocity-radius plot in the middle panels -- and the velocity histogram in the right panels) and, in the latter case, it becomes closer to the one of the dwarf (i.e. 77 \kms). 
Green and red particles have rather similar velocity dispersion values, i.e. (40 and 74 \kms - green, 35 and 83 \kms - red for the recovered stream respectively with $N_{\rm min}$ = 15, top row, and with $N_{\rm min}$ = 30, bottom row). This suggests that the contaminants only slightly alter the true velocity dispersion of the stream. 

%Interestingly, all the previously stripped particles (in blue) show a similar velocity dispersion, 
%hence suggesting that in principle COSTA might also be able to find clumps of particles formerly stripped and currently being far from the dwarf center. 
Only in a very few runs, COSTA also detects %(as second detection) 
groups of particles in the tail of the stream that are further away from the dwarf main body, on the opposite side of the central galaxy (see e.g., blue crosses in Figure~\ref{fig:galmer_found}). This shows that COSTA can, in principle, identify also portions of the debris of a stream in absence of a close dwarf (e.g. at the pericenter/apocenter of stream orbits where lost particles tend to accumulate around zero systemic velocity in the reference frame of the central galaxy). This is due to the fact that, being stream (blue) particles still unmixed to the halo, they are also recognized as cold substructures.
%, located at $\sim$ (8,3) Kpc from the center. 
%The stream particles are further divided in all stream particles (blue) and detected particles from COSTA where we have the true stream particles (in green) and contaminants (red).

The fact that the majority of the detections occur in the regions close to the galaxy depends strongly on the fact that these particles are fully unmixed. For a detection at larger distances to occur, one needs an \textit{ad-hoc} combination of poor mixing and occasional overdensity, which is more difficult to happen.
%However, this example clearly show that COSTA favor to select particles in the tail of the intruder and that our assumption that the stream is kinematically connected to the parent galaxy is correct.
It remains that the velocity dispersion of these latter detections cannot be dynamically connected to the parent dwarf (e.g. via Faber-Jackson relation). The only case one is motivated to dynamically interpret the velocity dispersion of an isolated group of particles that has no clear dwarf association is in the case there are evidence that the parent dwarf has been recently disrupted and the remaining particles are the latest lost. 

%We have already seen that the group of particles selected as part of a stream contains a number of contaminants that depends on the ratio of the number of stream particles vs. galaxy particles. 
%Hence, we can expect that whatever kinematical measurements we derive from the selected stream particles, these have to be taken with precaution and just considered indicative of the real kinematics. 
%\st{However, the availability of some ground truth from the simulations allows us%here 
%to check whether we can answer the question: what is the configuration that best recover the kinematics of the stream?}
%\massi{However, in case some ground truth is available from the simulations, we can ask ourselves what is the configuration that best recover the kinematics of the stream.}
%depends on $N_{\rm min}$From both the phase-space (the velocity--radius plot) and the velocity histogram, one can see that 

%We have also seen that, if one \chiara{selects }%picks 
%the closest particles to the dwarf, these do retain the overall kinematics of the dwarf galaxy. Hence, 
Finally, since we have postulated a connection between the kinematics of the stream and the one of the parent dwarf (e.g. a sort of stream Faber-Jackson), and given that
%Since 
COSTA can detect the same stream with different configurations, we are interested to 
check whether we can identify configurations that %closer 
reproduces \chiara{as close as possible the real } internal dispersion of the dwarf. 
%(i.e. 77 \kms, see table \ref{tab:galmer parameters}).
%NRN7: Massi, puoi aggiungere il valore piu' corretto? (forse c'e' prima ma controlla). Occorre fare anche questa figura sotto
\massi{In Fig.~\ref{fig:disp_vs_sigma_cut} we show the density plot} of the velocity dispersion estimates of the selected stream particles as a function of the most sensitive parameter discussed in this paragraph, $N_{\rm min}$. %, for the four cases shown in Fig. \ref{fig:gE0_parameter_space}.
In particular, we show the values obtained using a threshold on $F_{\rm N}$ = 0.5 in the $n-N_{\rm min}$ space as described in the \S\ref{sec:stream_detection}. %meaning that we excluded from the following considerations all set-ups falling in a region of the \emph{N\textsubscript{min}-n space} below the above quoted threshold}. 
%Using this {\it clean} selection of set-ups, {\color{red} we can see, by making a quick comparison with red points in the four plots of figure \ref{fig:gE0_parameter_space}, that most of the stream detection we remove are those ones underestimate the actual velocity dispersion. This is particularly evident in the two cases with a smaller velocity uncertainty, i.e. $\Delta v\leq$20 \kms (top right and bottom rigth plot), since almost all detection with $\sigma \leq$30 \kms have been removed.}
%\st{we can see that the estimated velocity dispersion become closer to the dwarf value for the higher $N_{\rm min}$. In the real case the true velocity dispersion this is not known so this diagnostic turns to be very effective to obtain the closest estimate to the true velocity dispersion of the stream candidate.} 

The four plots correspond to the measured velocity dispersion, $\sigma_{\rm mea}$, from the streams selected according to the \massi{four different}
%\st{different observational} 
cases as in Fig.~\ref{fig:gE0_parameter_space}.  Overall, we notice that the velocity dispersion estimates tend \massi{to cluster around the true value of the dwarf (77 \kms)}, \chiara{with tails towards lower values.}
%\st{to peak within $15-20$\kms from the true value of the dwarf (77 \kms, as marked by the horizontal dashed line).} 
This happens regardless of the sample size and velocity errors, although the higher velocity accuracy (e.g. $\Delta_v\leq$ 10 \kms, top right panel, and $\Delta_v\leq$ 20 \kms, bottom left panel) give \massi{less pronounced tails} toward low $\sigma_{\rm mea}$ in the velocity  distribution. This is particularly evident when comparing the 2000-particles cases (top row of Fig. \ref{fig:disp_vs_sigma_cut}). %: the small velocity errors (right panel) show a narrower distribution.

%for the smaller velocity dispersion errors ($\Delta v\leq$20 \kms, see bottom left panel, and $\Delta v\leq$10 \kms, top right panel), the estimates tend to accumulate closer to the true value for $N_{min}>20$, and there are almost no stream candidates that have $\sigma<40$ \kms. In these cases, \nic{the peak values of the $\sigma_{\rm mea}$ distribution ($\sim67-70$\kms) 
%of the estimated stream velocity dispersion in the upper 25\% quantile 
%differ by $\sim$10 \kms from} the true value of the dwarf (77 \kms). 
%\st{is enclosed in the upper 25\% quantile of the $\sigma_{\rm mea}$ distribution}.
%NRN-8: controlla il quantile qui sopra
%The distribution of the estimated $\sigma_{\rm mea}$ is quite spreader in the case of larger errors ($\Delta v=$40 \kms) and while for the 1000 particles the true dispersion is still 
%in the upper 25\% quantile
%within $\sim 10$ \kms\ from the peak of the $\sigma_{\rm mea}$ distribution ($\sim70$\kms), for the 2000 particle case the situation is less clear. Overall, in all cases 
%of the if averaging the $\sigma_{\rm meas}$ in the upper 25\% quantile, 
Being more quantitative, using the median of the distribution as probe of the peak, we obtain the following stream velocity dispersions: 43 $\pm$ 23 \kms, 57 $\pm$ 12 \kms, 57 $\pm$ 15 \kms, 57 $\pm$ 18 \kms, for the 2000 (40), 2000 (10), 1000 (40) and 500 (20) cases respectively. 

%\chiara{CHECK THESE NUMBERS. THEY INVALIDATE WHAT YOU SAID ABOVE. 43 $\pm$ 23 FOR 2000-40 AND 57 $\pm$ 12 FOR 2000-10. FOR LOWER $\Delta_v$ THE TAIL IS *MORE* PRONOUNCED AND HAS A LARGER SCATTER!}\chiara{ALSO, IF IT WOULD HAVE BEEN A MATTER OF $\Delta V$,THEN THE TWO LEFT PLOTS, UP AND DOWN SHOULD HAVE HAD THE SAME STREAM VELOCITY DISPERSION, BECAUSE THEY BOTH HAVE THE SAME $\DeltaV$. THE CASE 1000-40 GIVES 57kms JUST LIKE THE 500-20 and 2000-10. HOW CAN WE JUSTIFY THIS?}
Of course, the final dispersion of the stream in all cases is affected by the contaminants from the main galaxy, but overall 
\chiara{the median values are always consistent within 1-$\sigma$ uncertainties and all close, although slightly lower, to the true velocity dispersion of the dwarf. This }
%the final dispersion estimates are close to the intrinsic dwarf dispersion, which 
implies that the %only
contaminant selected %picked 
by COSTA as part of the stream are almost statistically  indistinguishable from the stream particles as they hold a similar overall kinematics (see Fig.~\ref{fig:galmer_found}). 

%\massi{COMMENTO: Piu' che il quantile credo che la moda della distribuzione indichi meglio la dispersione di velocita' dello stream. Vedi ad esempio gli stream artificiali messi in Fornax (sec. 4.3 e Fig.\ref{fig:disp_vs_sigma_cut_artificial_streams}), in quel caso la regola del quantile non si avvicina alla vera dispersione di vel. dello stream (indicata dalla dashed line)}
%\nic{ok: vuoi implemenrare la moda allora?}
%NRN-8 stream (: aggiungi i valori qui sopra l. delloe ne discutiamo,  la regola del 25\% va calibrata se serve.

%for which there is a peak of the number of configurations. I.e., the configurations returning the more realistic stream selection tend to accumulate in the parameter space.
%NRN-8: inoltre occorre dire a Vrad (media delle velocita') corrispondono questi stream e come si confrontano con la Vrad della dwarf: Massi aggiungi questo commento 

%NRN-8: Massi aggiungi il valore della dispersione dello stream come linea tratteggiata orizzontale (sono tutti valori error subtracted?). Inoltre aggiungi anche gli altri casi di Fig. 4.

%\begin{figure*}
%    \centering
%    \includegraphics[scale=0.6]{gE02000_err40_stream.png}
%    \caption{Same of the figure \ref{fig:gE0_2000p_err40}, along with position on the parameter space where COSTA found the stream (red points) and spurious structures (cyan points)}
 %   \label{fig:gE0_2000p_err40_stream}
%\end{figure*}

\subsection{The case of gSa-dS0 encounter: testing COSTA on a cold system}
\label{sec:gSa-dS0}

Having demonstrated that COSTA is able to find cold streams embedded in the halo of hot early-type systems, we now need to test the case of late-type galaxies. 
We select a gSa-dS0 encounter, \chiara{and test both a prograde and a retrograde motion for the dwarf, }%both with the dwarf having a prograde and a retrograde motion, 
because the stronger rotation of the galaxy disc might have a different impact in the two cases. 
We follow the same steps as in the gE0-dE0 case, \chiara{and we highlight the results in the following sections}. 

\subsubsection{Reliability}
\label{sec:reliab_gSa_dS0}

First we run COSTA over the 
%central galaxy without dwarf+stream particles 
WNS using all parameter combinations to determine the reliability distribution in the parameter space.
In Fig.~\ref{fig:gSa_prograde_parameter_space} we show the reliability maps for the prograde and retrograde cases, on two separate rows. 
\chiara{Also in this case, we show the density plot obtained with different number of particles for the giant and the dwarf and different values of $\Delta v$. }
%There are no significant differences between the two cases in the `white noise' sample because the main galaxy is not dynamically changed, while the difference in the retrograde case is the direction of the dwarf impact.

%Starting with the prograde case (top row), the \emph{top left} panel of Fig.~\ref{fig:gSa_prograde_parameter_space} shows the case of 2000 particles and velocity errors $\Delta v=$40 \kms, color coded by $F_{\rm N}$.
%\st{Being the gSa colder than the gE0, the differences with this latter case are sensitive, since COSTA finds easily combination of galaxy particles that locally have a velocity dispersion close to the $\sigma_{\rm cut}$ (i.e. there is a smaller contrast), and pick them as possible (spurious) substructures.}

\massi{Since the gSa is colder than the gE0, it is much easier for COSTA to find combinations of galaxy particles with a local velocity dispersion close to that of the $\sigma_{\rm cut}$ (i.e. there is a smaller contrast). } \chiara{It is much easier that COSTA find spurious %  hence picking them as possible (although spurious) 
substructures and consequently   %.} 
%As a consequence, it is 
harder to find set-ups with high reliability (i.e. with more than 70\% of no-spurious detections). As a result, } %Hence, 
the regions of the parameter space with high $F_{\rm N}$ (yellow) are considerably reduced with respect to the gE0 case, and there is generally a higher chance to find some false positive. 

%A similar effect is found for a lower number of tracers and high velocity errors (3-rd  panel on the top row in the same figure).
As for gE0, the adoption of smaller velocity errors produces a slightly  wider number of good set-ups, especially in the retrograde case,  %(see 2-nd panel top row), 
and the $F_{\rm N}$ increases over a relatively wider area, also for the smallest number of tracers tested in our simulations (4-th panel).

\chiara{The results for the prograde (top) and retrograde (bottom) cases are }
%\massi{The second row of Fig \ref{fig:gSa_prograde_parameter_space} shows the retrograde case results, which are 
\massi{very similar, % to the prograde one, 
substantially because in the two cases the WNS does not change dynamically in a significant way, despite the different interaction with the intruder might have introduced different perturbations.
%, while the difference in the retrograde case is the direction of the dwarf impact.
}

\begin{table}
    \centering
    \caption{Same of Tab.~\ref{tab:galmer_statistic} but in the case of the spirals.}
    \begin{tabular}{l|c|c}
    \hline
     gSa-dS0 Prograde & $F_N$ &  \texttt{CF} \\
    \hline
    %\multicolumn{5}{c}{Prograde (gSa-dS0)}\\
    2000 part - 40 \kms & 0.10 $\pm$ 0.13 & 0.61 $\pm$ 0.14\\
    $^a$2000 part - 40 \kms & 0.14 $\pm$ 0.20 & 0.74 $\pm$ 0.15\\
    2000 part - 10 \kms & 0.08 $\pm$ 0.15 & 0.57 $\pm$ 0.15\\
    $^a$2000 part - 10 \kms & 0.11 $\pm$ 0.19 & 0.67 $\pm$ 0.19\\
    1000 part - 40 \kms  & 0.04 $\pm$ 0.06 & 0.73 $\pm$ 0.13\\
    $^a$1000 part - 40 \kms & 0.04 $\pm$ 0.12 & 0.89 $\pm$ 0.11\\
    500 part - 20 \kms & 0.03 $\pm$ 0.04 & 0.71 $\pm$ 0.12\\
    $^a$500 part - 20 \kms & 0.04 $\pm$ 0.05 & 0.73 $\pm$ 0.16\\
    \hline
    %\multicolumn{5}{c}{Retrograde (gSa-dS0)}\\
     gSa-dS0 Retrograde & $F_N$ &  \texttt{CF} \\\hline
    2000 part - 40 \kms & 0.14 $\pm$ 0.22 & 0.13 $\pm$ 0.14\\
    $^a$2000 part - 40 \kms & 0.11 $\pm$ 0.19 & 0.26 $\pm$ 0.17\\
    2000 part - 10 \kms & 0.29 $\pm$ 0.34 & 0.16 $\pm$ 0.15\\
    $^a$2000 part - 10 \kms & 0.14 $\pm$ 0.21 & 0.26 $\pm$ 0.12\\
    1000 part - 40 \kms  & 0.04 $\pm$ 0.12 & 0.36 $\pm$ 0.14\\
    $^a$1000 part - 40 \kms & 0.03 $\pm$ 0.09 & 0.44 $\pm$ 0.19\\
    500 part - 20 \kms & 0.04 $\pm$ 0.05 & 0.57 $\pm$ 0.10\\
    $^a$500 part - 20 \kms & 0.03 $\pm$ 0.04 & 0.50 $\pm$ 0.13\\
    \hline
   \end{tabular}
    \begin{minipage}{128mm}
    \textbf{a:} in these configurations we ruled out set-ups with $F_{\rm N}$ < 50\% \\
    in the $n-N_{\rm min}$ space as described in the text.
    \label{tab:galmer_statistic_spiral}
    \end{minipage}
\end{table}{}

%\begin{figure*}
%%\vspace{-0.2cm}
%\hspace{-0.8cm}
%    \includegraphics[scale = %0.171]{gSa2000_retro_err40_stream_10sim.png}
%%\vspace{-0.2cm}
%\hspace{-0.2cm}
%    \includegraphics[scale = %0.171]{gSa2000_retro_err10_stream_10sim.png}
%%\vspace{-0.2cm}
%\hspace{-0.8cm}
%    \includegraphics[scale = %0.171]{gSa1000_retro_err40_stream_10sim.png}
%%\vspace{-0.2cm}
%\hspace{-0.2cm}
%    \includegraphics[scale = %0.171]{gSa500_retro_err20_stream_10sim.png}\\
%    \caption{Same of the fig.\ref{fig:gE0_parameter_space} but for %the gSa in the case of the retrograde encounter.}
%    \label{fig:gSa_retrograde_parameter_space}
%\end{figure*}

\begin{figure*}
%\vspace{-0.2cm}
%\hspace{-1cm}
    \includegraphics[scale=0.2]{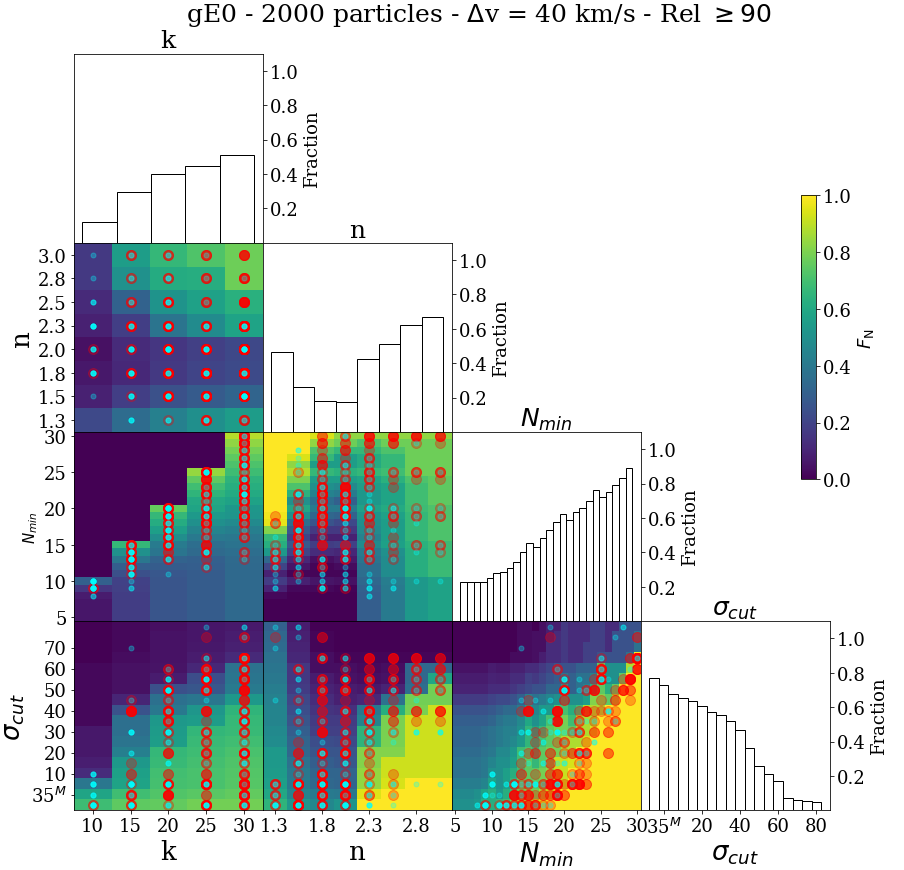}
    \includegraphics[scale=0.2]{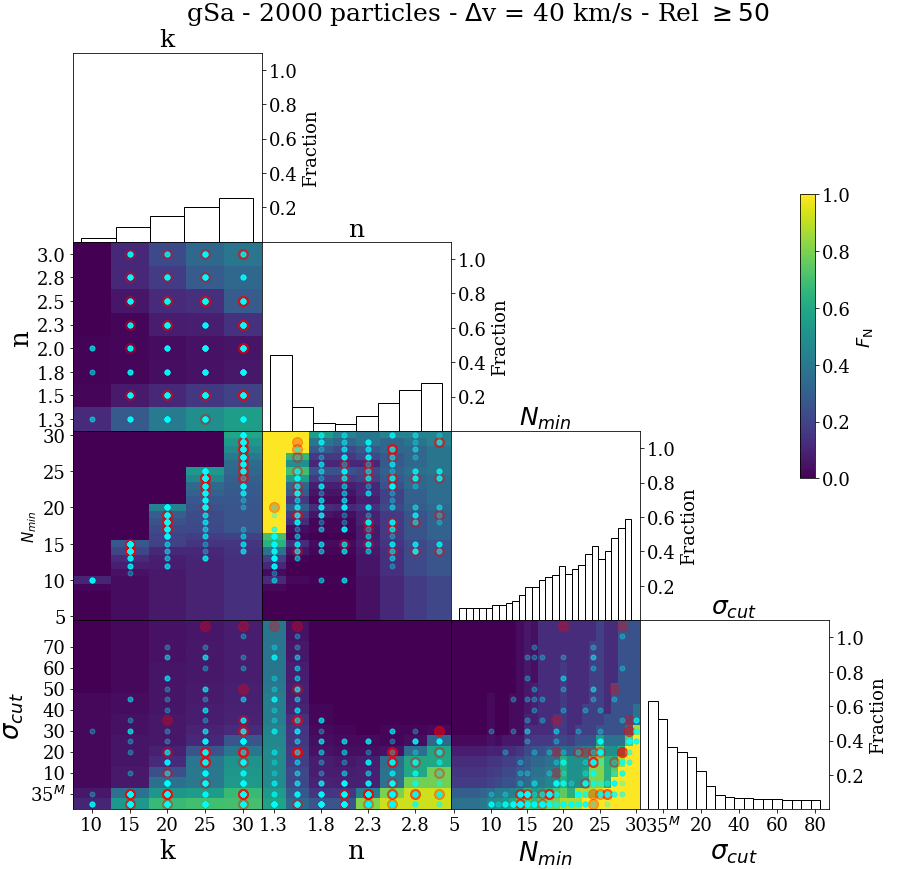}
    \includegraphics[scale=0.2]{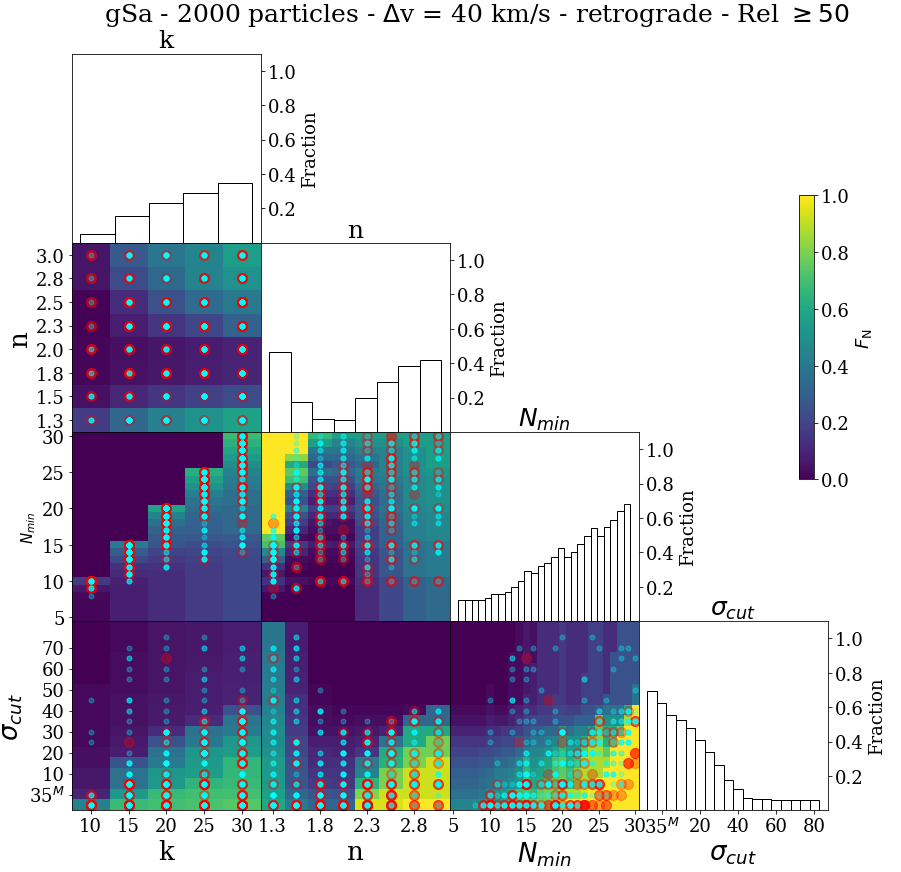}
    \caption{Parameter space overlapped with stream and spurious detected by COSTA for both early and late type galaxies, \chiara{in the standard configurations} (2000 particles and $\Delta_v$= 40 \kms) for the three cases \chiara{(gE0, gSa prograde and gSa retrograde)}, but using different reliability thresholds.}
    \label{fig:parameter_space_different_threshold}
\end{figure*}

\subsubsection{Stream detection}
\label{sec:detect_gSa_dS0}

The second step is to run COSTA on the DS, made by the main galaxy and dwarf particles, to recover the stream particles. As done for the gE0+dE0 encounter, we perform ten random extractions of the giant+dwarf system, imposing a limit on the lower surface brightness of the stream.
The results of this test are listed in Tab.~\ref{tab:galmer_statistic_spiral}, both for the prograde and the retrogade encounters. 
\chiara{In general, COSTA detects the stream only in few set-ups with reliability {\tt Rel}$\ge$70\%$\geq$ 70\% both for the prograde and retrograde motions. Furthermore, in the gSa-dSO case, most of the set-ups which correctly found the stream particles, also detect spurious substructures (see bottom left panel of Figure~\ref{fig:gSa_prograde_parameter_space} or Table \ref{tab:galmer_statistic_spiral}). 

Overall, for the retrograde encounter COSTA performs better, with a much lower \texttt{CF} (e.g., $\sim$ 15\% versus $\sim$ 60\% for the 2000 particles case with $\Delta_v=40$ \kms). The number of set-ups in which COSTA detects a stream is also higher in the retrograde case, at least for the best possible configuration, i.e. 2000 part - 10 \kms (11\% for the prograde and 29\% for the retrograde, assuming a threshold of 50\% in the $n-N_{\rm min}$ panel, although with a large uncertainty)}.  
In this configuration, and partially also in that with 2000 particles and larger velocity error, the stream has been detected in regions that tend to accumulate toward the high $F_{\rm N}$ (yellow areas) of high-reliability in the parameter space. This is especially visible in the $N_{\rm min}-\sigma_{\rm cut}$ plot, as we have seen for the gE0 system. \chiara{In particular, high-reliability configurations favor }%They favor high 
%\st{intermediate}  %$N_{\rm min}$ and 
smaller $\sigma_{\rm cut}$ ($\le40$\kms). %This is particularly true for the set-ups \chiara{with  $F_{\rm N}>0.5$ in the $n-N_{\rm min}$, which we use as a further quality threshold as already done in \S\ref{sec:gE0-dE0}.} 
However, this is not always true for the prograde cases and the retrograde cases with a smaller number of particles, at least not for all the projections. In these cases, the stream has been detected only few times and they are very sparse in the region of the single plots of the parameter pairs. 
%\nic{For all other cases 
%We cannot achieve a similar conclusion in the other cases since 

%\massi{We notice that, for the 2000 particles case with $\Delta_v=40$ \kms, COSTA has detected the stream in a very few set-ups with reliability $\geq$ 70\% both in the prograde and retrograde motions, but in the retrograde encounter \texttt{CF} is much lower ($\sim$ 15\%)}.
%CS: I rephrased because this was true also for other cases.

%\st{In the former case (top panel of Fig. ), for the 2000 particles case with $\Delta_v=40$ \kms, COSTA has detected the stream in a very few set-ups with reliability $\geq$ 70\%, while in the retrograde case (bottom panel of Fig. ) it has detected stream in more configurations and with a very low percentage of contaminants (below 15\%).} 
%\chiara{THE NUMBERS ARE FULLY CONSISTENT AND UN-DISTINGUISHABLE. ALSO FOR DELTAV =10 BUT THERE AT LEAST THE MEAN IS RATHER DIFFERENT. COMMENT ON THAT INSTEAD OF 40? OR REMOVE THE COMMENT AT ALL?}

%\st{By limiting the stream candidates to the cases where \texttt{Fract}$>0.5$, we considerably reduce the number of allowed set-up and the contamination is also reduced}. 
%NRN-8: Massi quantifica un po' qui sopra.
%\st{Overall, the $\Delta_v=40$  velocity errors are totally inappropriate to find streams in cold systems and $\Delta_v\leq20$ are needed.}
%Focusing on the case with $N_{\rm part}$=2000, 
Finally, as seen in Fig.~\ref{fig:gSa_prograde_parameter_space}, at least for the retrograde case, COSTA performs generally better when the velocity errors are smaller. Here, the algorithm reveals the stream in more set-ups.

An interesting difference between the gE0-dE0 case and the gSa-dS0 is that, for the latter, the number of particles makes a much larger difference in terms of number of set-ups with high reliability. And this \chiara{valid both for the prograde and for the retrograde case. Going from 2000 particles to 500, the $F_{\rm N}$ is between 3 and 5 times smaller while the CF increases.}

\nic{We also note that, for the gSa-dS0 cases, and in particular for the prograde encounters, the configurations for which we correctly detect the streams are often embedded in low-$F_{\rm N}$ areas. This is different from what happens in the  gE0-dE0 interaction and it is because, since COSTA finds more spurious, the configurations that allow to find the stream also find some spurious, at least with the change of other parameters. This means that, even if the stream is found, this has a general lower reliability in cold systems. 
We need to stress here that this conclusion is not general, as this applies to the case of a mass ratio 10:1, i.e. with a small contrast between the dispersion of the stream and the dispersion of the background velocity field (see below).}

%is especially appreciable in the retrograde case.
%\chiara{WHY FOR THE PROGRADE CASE, THE CASE WITH DELTAV = 40 IS BETTER THAN THE CASE WITH 10?}
 %\chiara{At the same time, it performs worse when decreasing the number of particles. In fact, }   especially for the retrograde encounter, 
%for both the prograde and 
%even for 
%for the 500 and the 1000 particle cases,  
%although, in this case,
%\massi{the stream has been detected only in very few cases ($<5\%$), and the contaminant fraction is generally larger ($\ge0.50$), as one can read from Tab.~\ref{tab:galmer_statistic_spiral}} . 
%\chiara{THE MAJORITY OF THE RED POINTS ARE IN BLUE REGIONS. WHY? COMMENT/JUSTIFY?}

%\st{we have detected false positive also in the area with high \texttt{Fract}.} 
%\st{In the prograde configuration, the number of set-ups where COSTA finds the stream is similar to the number of them which gives spurious structures (but they are a few in total, see also top right panel of figure 8, while in retrograde one most of the set-ups found only the stream (see top right panel of figure 8), and the contaminants percentage decreases to 10\%}. %NRN7 Aggiungerei: 
%\st{In all } 

\subsubsection{Stream kinematics}
\label{sec:kinem_gSa_dS0}
The $\sigma_{\rm cut}$ ($\le40$\kms) is an upper limit beyond which COSTA does not detect the stream anymore. The median of the velocity dispersion, using only set-ups with $F_{\rm N}$ > 0.5 in the $n-N_{\rm min}$ plot in the retrograde encounter, 
%\st{is consistent with the one of the stream} 
gives a velocity dispersion that is lower than the one of the parent dwarf galaxy ($\sigma_{\rm dwarf}=74$ \kms).  \chiara{In fact, we obtain as median of the velocity dispersion distributions:} 24 $\pm$ 13\kms, 31 $\pm$ 7\kms, 33 $\pm$ 2\kms, 37 $\pm$ 6\kms for the 2000(40), 2000(10), 1000(40), 500(20) cases. 

\nic{Here the worse performance of COSTA with respect to the gE0-dE0 is due to the little contrast of the dwarf velocity field (which is rather hot in the specific Galmer simulation, i.e. $\sigma_{\rm dwarf}\sim74$ \kms), with respect to the gSa ($\sigma_{\rm giant}\sim81$ \kms). } \chiara{Thus, the exercise we carried on here has therefore to be interpreted as an ``extreme case'' to set a guideline for the methodology to follow in ``real'' cases, where the difference between the velocity dispersion of the dwarf and that of the giant is larger. }%\nic{we shown that, even in very challenging configurations, COSTA can still find the stream.}  
%\nic{TBD: Need a comment after having checked the dispersion of the stream}
%\massi{As most of the set-ups with high $\sigma_{\rm cut}$ have a reliability below the selected threshold, that region is not sampled by COSTA, so it did not recover the real velocity stream dispersion, but a lower value.}
%Given that one can derive the kinematical properties of the stream from the collected particles, one understands that the velocity dispersion is possibly a function of the parameter configuration for that particular pick. Hence we need to check what is the  

%strengthen the view that we could constraint the internal kinematics of a stream by looking at highest value of $\sigma_{\rm cut}$.\\
%With a lower number of total particles COSTA has more difficult to detect the stream, just as the gE0-dE0 configuration.\\
%In synthesis, refining the measurements on the particle velocities definitively improves the ability of COSTA in detecting the stream and inferring their kinematics.
%\st{Finally, also for this case we re-scaled the dwarf face space in order to simulate a different mass ratio between the colliding galaxies in both encounters. We discuss it in the Appendix.}

%\subsection{COSTA dependence from the reliability threshold}
\subsection{The dependence of the performance of COSTA on the reliability threshold}
\label{sec:different thresholds}

In this section we explore how a different reliability threshold can affect the completeness and purity of COSTA. 
Overall, \chiara{logically,} a lower threshold allows us to rise the probability to find a stream, at the expense of a greater contamination, while an higher limit has the opposite effect.

In the case of the gE0-dE0 interaction, we increase the lower limit of the reliability and we use only combinations of free parameters above 90\%.
We run COSTA only on the $N_{\rm giant}$ = 2000, $N_{\rm dwarf}$ = 150
and $\Delta$v = 40 \kms\ configuration.
Of course, the number of parameter combinations which overcome the threshold are reduced with respect to the previously used threshold of 90\%. This can be seen in %In the top left panel of figure 
left panel of Fig.~\ref{fig:parameter_space_different_threshold} (left panel). %, where %we plot the parameter space using a reliability cut-off of 90\%  and it is possible to notice a slight decrease of the fraction of good set-ups with respect to the case of a threshold set to 70\%.
%\chiara{WHY THE CASE WITH 40kms IF YOU SAY IT IS NOT SUITABLE FOR SPIRAL AND 10 IS BETTER? 20 WOULD REPRODUCE OBSERVATIONS, AS YOU CLAIM SOMEWHERE.. .SO WHY 40?}
%\massi{answer: We wanted just to probe the dependence of the parameter space from the reliability threshold. Thus we used only 1 case speculating that the trend is similar also to the other cases. The case 2000-150 particles with $\Delta_v$ = 40 \kms is similar to what we have in Fornax.}
However, the stream is still detected in many set-ups, and they show almost the same distribution in the parameter space as for the lower threshold.

%\st{Furthermore, rising the accepted reliability has the effect to reduce the percentage of set-ups where COSTA finds spurious, from 39\% to 30\%.
%Overall, in this particular case an higher reliability threshold leads to a lower odds of finding spurious structures, while the stream is still detected (see also fig.8)}.

For the gSa-dSa interaction the situation is reversed, as we lower the acceptable value of reliability to 50\%. Indeed, in both the prograde and retrograde encounters, COSTA could not find streams using a higher threshold, so we check a lower one. We use, in both encounters $N_{\rm giant}$ = 2000 and $N_{\rm dwarf}$ = 150 and $\Delta$v = 40 \kms\. % and as acceptable value of reliability 50\%.
In the prograde case (central panel of Fig.~\ref{fig:parameter_space_different_threshold}), COSTA finds the stream in \chiara{a slightly larger number of %different 
set-ups, with } respect to the very few ones with a reliability cut-off of 70\%.
In the retrograde case (right panel of the same figure) \chiara{the improvement is even higher as} the number of set-ups where COSTA detects the stream increases by 50\% with respect to the previous case. Thus, \chiara{we conclude that} for the late-type case, the 70\% reliability threshold is too conservative and a lower reliability threshold would give more chances to identify streams.

%This stream will be the target to test the ability of the algorithm to detect cold structures coherent in the position and in the velocity.
%Then we launched the procedure on the 2130 particles of this cosmological simulation (2070 particles belong to the giant and 60 ones are those stripped from the body of the dwarf).
%We excluded again, for the same reasons stated in previous section, the inner part of the gE0, but because of the smaller scale of this sample we removed from the research of cold structures the region within $R = 1.8 kpc$, in order to have, in this area, a similar number of objects respect than Fornax cluster.
%We verified if, varying free parameters, some set up allowed the algorithm to find the target, made of 11 particles and having a velocity dispersion of 54 kms$^{-1}$, and no any other spurious structure.
%In table \ref{tab:galmer} we listed all these parameters, along with the numbers of the recovered particles of the target and those found together with stream but actual don't belong to.

%Following the same procedure for different times of this encounter we obtained similar results, hence we can speculate that, if it is not passed enough time, stripped tidal particles could be detected as coherent structures both in the position and in the velocity, moving with low velocity dispersion and surrounded by a huge number of other particles that move following the potential of the cluster, typically of the order of hundreds kilometer per seconds.

\section{The case of the Fornax cluster core}
\label{sec:fornax}
\massi{We test now COSTA on a cluster environment, where streams are produced in a more complex situation with many large coexisting galaxies. In such environment, multiple low surface brightness streams from a larger population of dwarf galaxies, with a given luminosity function and different kinematics properties, can be produced.}
%\st{Having demonstrated that COSTA is a viable tool to detect cold structure in the reduced phase space of kinematical tracers in typical minor encounters between a large galaxy and a smaller sized system with low velocity dispersion, now we want to make the case of a cluster environment, where streams are eventually produced in a more complex situation where many large galaxies are coexisting and might host multiple low surface brightness streams from a larger population of dwarf galaxies with a given luminosity function. }

\nic{In particular, we present here the case of the Fornax cluster, for which there are GCs (from P+18) and PNe (from S+18) available for stream search, that we will present in forthcoming analyses. }
%will be then followed up in a forthcoming paper of this series where we will use COSTA to detect real streams using GCs (from P+18) and PNe (from S+18).
%In particular, we want to show as a test case the one of the Fornax cluster because this can be used for real stream detection using GCs (from P+18) and PNe (from S+18), in a forthcoming paper. 

The aim of this section is to show that also for a more complex case, as the Fornax cluster core, COSTA can be set to detect cold streams of small number of particles, as done for the Galmer simulations. 

For the Fornax cluster, unfortunately, we do not possess a simulation realistic enough to produce the same large structure distribution of particles as reported in  GCs and PNe studies. We thus decide to build up Montecarlo realizations of the kinematical tracer distribution in the 3D phase space (i.e. 2D positions and radial velocity) \nic{over which we can obtain a reliability map for COSTA and test its stream detection performances.}
%we want to apply COSTA.   

Indeed, following the approach adopted for the Galmer simulations, the Montecarlo realizations of the Fornax core are needed, first, to have a smooth cluster background with no streams (i.e. the WNS). This allows us to explore the parameter space and assess the reliability function of COSTA as a function of the different parameters. Second, we add a number of artificial streams (hence generating different DSs) and run COSTA to recover them and to calculate the \texttt{OC} and \texttt{CF}.

\subsection{Montecarlo simulations of the Fornax cluster core}
\label{sec:simulation of the cluster}

To produce COSTA reliability maps, we perform a suite of Montecarlo simulations, which reassemble the Fornax core as close as possible, in terms of spatial distribution, local density and radial velocity distribution of the kinematical tracers (WNS).   

We simulate only the region covered by the current discrete tracer surveys (FVSS, P+18 and S+18), \chiara{covering }%that is the regions of Fornax of 
about $1.8$ deg$^2$ around the cD, NGC~1399. 
In this area, there are two other bright early-type galaxies: NGC~1404, located just below the cD, in the S-E direction at about $9$ arcmin; and NGC~1387, at a distance of $\sim$ $19$ arcmin in the West direction from NGC 1399. \chiara{A third relatively massive galaxy, }\nic{NGC~1379, located at $\sim60'$ toward W, }\chiara{has been observed with one FORS2 pointing in S+18. However, this system } %but it 
\nic{is excluded by this analysis because we do not have continuity with the rest of the Fornax core area, hence it is useless for stream finding. } 

We generate simulated GCs and PNe in a number that is as close as possible to what has been observed in S18 and P18. In the following, we assume that  both GCs and PNs trace the same underlying population of old stars
%We generate simulated GCs and PNe, which in the following we assume to both trace the same underlying population of old stars
\footnote{For a discussion about the statistical similarity between the two tracers, see Napolitano et al. in preparation.}, at the equilibrium in the gravitational potential of these three galaxies,\chiara{assumed} to be the superposition of the individual galaxy potentials with spherical symmetry.
Following \citet[][]{napolitano-2001}, we produce the 3D position starting from a 3D spherical density profile and projected them on the 2D sky plane (X-Y in our case). For each particle, we determine the 3D velocity vector, according to the hydrostatic equilibrium equations (see below), which we project along the line of sight to derive the intrinsic radial velocity\chiara{. We } %and 
finally simulate a radial velocity measurement by randomly extracting the measured velocity from a Gaussian having the truth radial velocity as mean and standard deviation equal to the measurement errors. 

In order to produce these Montecarlo realizations of particles sampling the total potential in the Fornax core, we assume for the cluster a total mass of about $10^{14}$ $M_\odot$ and a Hernquist \citep{hernquist-1990} density distribution of the stellar-like tracers. 
This is a good approximation for elliptical galaxies following a de Vaucouleurs law (1948)\chiara{. Indeed, for NGC~1399, which gathers most of the light in the cluster core, \citet[][]{iodice-2016} found a Sersic index $n = 4.5$, i.e. very close to $n = 4$, describing } the de Vaucouleurs law. 

The luminous mass density is expressed by the formula
\begin{equation}
	\rho(r) = C\frac{M_\textup{l}a}{2\pi}\frac{1}{r(r+a)^3}
\end{equation}
where $M_\textup{l}$ is the total luminous mass, \emph{a} is a distance scale
($R_\textup{e} = 1.81534$ \emph{a}) and \emph{C} is a normalization constant. We make the same assumption for all other galaxies in the area, with the adopted parameters as in Table~\ref{tab:parameters of the simulation}.

In addition to the stellar mass density, we also consider a dark halo following a Navarro-Frenk-White profile (NFW) \citep{Navarro1997}, to define a realistic internal kinematics for the simulated particles. 
Hence, the potential of the system at equilibrium is provided by the total mass:
\begin{equation}
	M_\textup{tot} = M_\textup{l} + M_\textup{dm}
\end{equation}

We assume no-rotation \footnote{Although some rotation of blue GCs in the radial range between 4 and 8 arcminutes was measured, the kinematics of the outskirt of NGC~1399 is dominated by the random motion \citep{schuberth-2010,coccato-2013}}, 
and an isotropic velocity dispersion tensor, and solve the radial Jeans equation
\begin{equation}
	\frac{d(\rho\sigma^2)}{dr} = -G\frac{M_\textup{tot}(r)\rho(r)}{r^2}
\end{equation}
to derive the 3D velocity dispersion $\sigma^2$ along the three directions in the velocity space, and generate a full 3D phase space.

As briefly anticipated above, we simulate an observed phase-space by projecting the tracer distribution on the sky plane and derive the line-of-sight (LOS) velocity of the individual particles. In particular, we use the X-Y plane as sky plane and Z-axis as the LOS. However, due to the full spherical symmetry of the model, the particular projection is irrelevant. 

Finally, in order to simulate a velocity measurement, we use the same approach as for the Galmer simulations: we adopt a Gaussian error distribution and re-sample the radial velocities produced by the Montecarlo simulations with a $\Delta_v=37$ \kms, consistent with typical measurement errors from P+18 and S+18. 
In this case we do not vary the errors, as this test is meant to demonstrate that COSTA can be applied on a real dataset and provide a series of reliable set-up to find stream candidates from real datasets in hands\chiara{, as we will do in the second paper of this serie. }
%with  took into account the typical velocity error in order to randomly extract the observed velocities (we refer to the original \citealt{napolitano-2001} paper for more details about the Montecarlo simulation). With the projected X and Y position and the observed radial velocity, we could mimic different mock observations to test our stream detection procedure. We have been careful to check that the mock catalogs of positions and radial velocities closely resemble the real one and also that the simulated cluster was consistent with the mean observables quantities of the galaxies in the area as reported in Table \ref{tab:parameters of the simulation}. 

\begin{table}
	\centering
    \caption{Parameters of the simulated galaxies. The effective radii are taken from the literature, unless differently specified. The velocity and velocity dispersion values are retrieved from the Nasa Extragalactic Database (NED), unless differently specified.}
	\label{tab:parameters of the simulation}
	\begin{tabular}{l c c c c}
	\hline
	Galaxy & Number & $R_{e}$ & Vel. & $\sigma_{p}$  \\
	& of Points & (arcsec) & (kms$^{-1}$) & (kms$^{-1}$) \\
	\hline
	NGC~1399 & 1855 & 138 & 1425 & 320\\
	NGC~1387 & 90 & 42 & 1302$^a$ & 160 \\
	NGC~1404 & 40 & 100$^a$ & 1947 & 247 \\
	\hline
	\end{tabular}

\medskip
\begin{minipage}{90mm}
%a: Values taken from \citet{caon-1994}\\
a: Values adopted to obtain a  more realistic reproduction (see text). %These values have been adapted to obtain a  more realistic reproduction (see below in the text)
\end{minipage}
\end{table}

%The parameters of the three simulated galaxies are reported in table \ref{tab:parameters of the simulation}.
\chiara{We include in the simulation 1985 particles, to reproduce as much as possible the number of observed PNe and GCs selected in the area.}
%From the 1985 particles included in the simulations, i.e. similarly to the number of PNe and GCs selected in the area, 

The number of points for each satellite galaxy is then obtained with a cross-match with the real data\chiara{,  %(i.e. 
counting the number of plausible PNe and GCs bound to the galaxies}, while both effective radii and velocity dispersion are taken from literature (see Table \ref{tab:parameters of the simulation}). 

%We used a mean measurement error on the velocities of 37 kms$^{-1}$, very similar to the velocity uncertainties in the real sample, which have an average of 37.5 kms$^{-1}$.

%NRN-10: fino a qui

%In Figure~\ref{fig:sim_ra_vs_dec} we show the simulated points with positions computed with respect to the simulated cluster center\chiara{. The position of the three galaxies are highlighted as black points}.  
%Overall, the 2D distribution of the particle positions looks fairly similar to the real one 
%\st{in Fig.} 
%, despite it cannot catch all the inhomogeneities of the sample. However the large scale features are well reproduced as we will discuss here below. 
%, while in the bottom panel we show the real data.

\begin{figure}
	\centering
	\includegraphics[width=8.3cm]
    {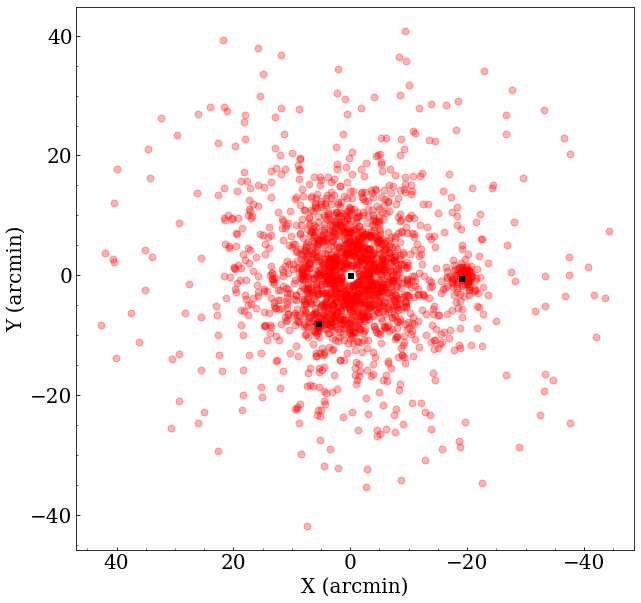}
	\caption{
	\chiara{Simulated datapoints for one of the Montecarlo realizations, }%Points of the simulation 
	with NGC~1399 at the origin of coordinates. NGC~1404 is just below the cD ($X \sim -5$, $Y \sim -5$) and NGC~1387 is at $X \sim -20$ arcmin. The positions of the three galaxies are indicated by  black squares.}
%NRNfin: forse puoi aggiungere una croce nella posizione delle galassie che citi nella caption
%NRNfin: attento che nella figura NGC1387 divrebbe andare spostato leggermente piu' a sud, perche' non si trova sull'orizzontale di NGC1399, potresti provare poi a sovrapporre i punti su una immagine BW del core di Fornax per fare vedere che c'e' una verosoimiglianza con la posizione delle galassie reali
	\label{fig:sim_ra_vs_dec}
\end{figure}

To obtain a realistic reproduction of the PN and GC systems around NGC~1404 (in terms of number and radial abundance), we need to adopt an effective radius (i.e. the radius enclosing half of the total light of the galaxy), $R_e\sim100''$, slightly larger than the one estimated by \citet{corwin-1985} ($R_e\sim80''$). 
%(i.e. 26\arcsec) from literature from 
%\citet{caon-1994}, i.e. 26\arcsec, but 
%used a $R_e\sim100''$, 
%&more consistent with $R_e\sim80''$ found from \citet{corwin-1985}.

\begin{figure}
	\centering
	\hspace{-0.5cm}
	\includegraphics[width=8.6cm]	
    {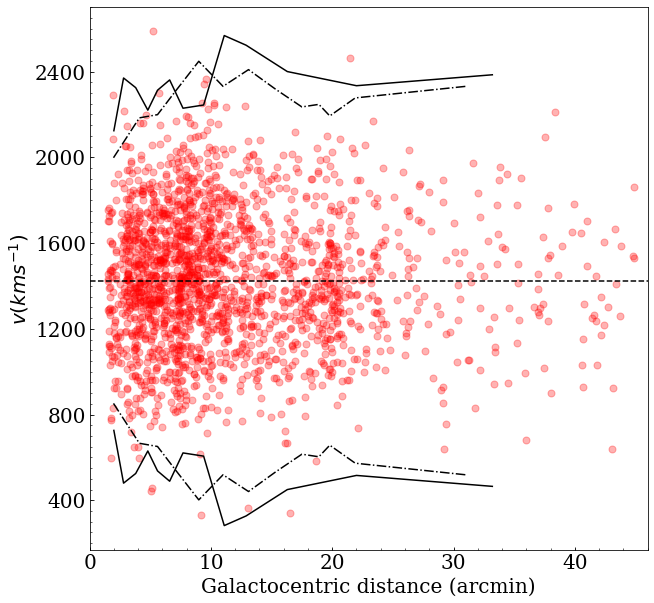}
    %\hspace{-0.5cm}
	%\includegraphics[width=8.4cm]
    %{histo_sim_vs_real.jpg}
	\caption{Phase-space of one of the Montecarlo simulations. \chiara{We plot} on the x-axis the distance from NGC~1399 in arcminutes and on the y-axis the velocities of the points in \kms.  The continuous solid and dashed-dotted lines represent the $\pm 3\sigma_P$ profiles of the GCs and PNe, respectively, extracted from P+18 (their Fig.~9). The dashed black \chiara{horizontal} line represents the systemic velocity of NGC1399 (1425 \kms)}.
	%{\em Bottom}:Histogram of the number of particles in function of the distance from the center of NGC 1399. Simulation is shown in black, whereas real data are red.}
%NRNfin: qui togli l'istogramma sotto e togli i dati reali nel phase space, mentre aggiungi le curcve a +/-3sigma di GCs e PNe che prendi da Pota et al. 2018 Fig. 9 (curva grigia e interpoli i punti fucsia delle PNe)
	\label{fig:sim_vs_real}
\end{figure}

%We found another inconsistency between the simulations and the real sample in the points surrounding NGC 1387; they were shifted downward in the phase-space respect to real particles. 
%We explain this difference with an inferred 
\chiara{For NGC~1387,} we %also 
take into account the velocity offset of PNe reported by S+18 (i.e. a mean velocity higher by $\sim 100$ kms$^{-1}$ than the systemic velocity of the galaxy reported in literature).
%which is tabulated in the literature (see \st{section ... and} S+18). 
Indeed in this area, we have a larger number of PNe than GCs, respectively 117 (88\%) and 16 (12\%) within three effective radii from NGC~1387; thus the offset of the PN velocities might generate an overall velocity excess of 100 kms$^{-1}$, that we thus artificially add to all simulated points around NGC~1387 in order to match the real objects, see also Table~\ref{tab:parameters of the simulation}. 

The final result of all these fine-tuning calibrations for the simulated sample gives the distribution of the simulated particle as shown in Fig.~\ref{fig:sim_ra_vs_dec}. \chiara{Here, we plot the simulated 
points for one of the mock realizations} with positions computed with respect to the simulated cluster center. We observe a fair spatial correspondence between the main galaxies in the \chiara{field of view (whose positions are highlighted as black squares)}  %F.O.V. 
and the simulated particles (red points). 
%\st{where we can see that there is a good correspondence of the simulated positions with respect to the real tracers in the surrounding regions of these two galaxies (see figure 10 for a comparison with real data),
%around 10$'$ for NGC 1404 and $\sim 19'$ for NGC 1387}.

In Fig.~\ref{fig:sim_vs_real}, we show the phase-space distribution of the same simulated particles together with the $\pm 3\sigma_p$ profiles (where $\sigma_p$ is obtained as in Eq.~\ref{eq:std}) of the GCs from P+18 and the PNe from S+18 (curves are extracted from Fig.~9 in P+18). Once again the similarities are quite evident \chiara{between} the overall kinematics of the simulated particles and of the observed ones. 

Once optimized the Montecarlo simulation set-ups to best reproduce the observed GC+PN dataset, we finally produce 100 realizations of the system, \chiara{that represent the WNS from which we obtain the reliability maps for COSTA. }
%which will represent the sample from which we need to obtain the reliability maps for COSTA. 

Differently from the Galmer simulations, where the statistical variation of the parameters have been obtained only perturbing the velocities of the particles, for the Fornax-like case we re-sample the full parameter space. Hence we add more statistical noise to the simulated sample, coming from different spatial configurations of the same physical streams. 
%The reason for the different approaches are that the test on simulated streams was meant to demonstrate that, given a real stream, it is possible to find a series of COSTA configuration to find a specific streams traced by a well defined group of particles. Now that we have demonstrated that the COSTA concept work, we want to demonstrate that in a system like the Fornax cluster: 1) what and how many configuration are reliable to search for streams; 2) if COSTA can find small groups of cold particles that can trace a cold stream.
%; 3) to make this as general as possible for Fornax.

\begin{figure}
%\vspace{-0.cm}
%\hspace{-0.4cm}
\centering
\includegraphics[width=0.5\textwidth]{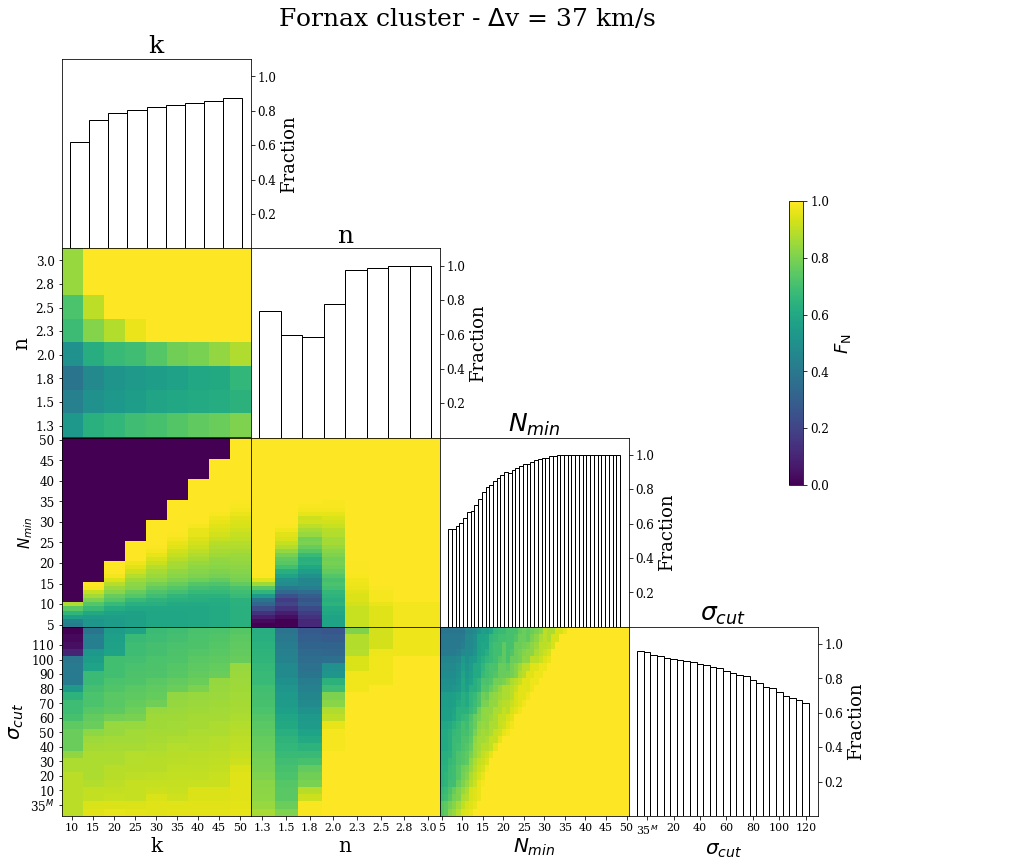}
    \vspace{-0.3cm}
    \caption{Reliability map for the Fornax cluster obtained with a reliability threshold of 70\%.}
    \label{fig:my_label}
\end{figure}{}

\subsection{COSTA set-up and reliability map}
To obtain the reliability map, we follow the same steps as in the Galmer simulations (\S\ref{sec:gE0-dE0} and \ref{sec:gSa-dS0}).  
\chiara{To begin, we} run COSTA for each parameter combination over the 100 Montecarlo realizations of the WNS and count the number of configurations for which COSTA finds no spurious streams. 

In this case we uniformly sample the \emph{k, n, N\textsubscript{\rm min}} parameters space, for different $\sigma_{\rm{cut}}$
%NRN6: forse dovremmo dire i limiti qui?
and run COSTA with all the possible combinations of the free parameters selected in the following ranges:
\begin{itemize}
    \item \emph{k}: from 10 to 50 with steps of 5
    \item \emph{n}: from 1.3 to 3 with steps of 0.2-0.3
    \item \emph{N\textsubscript{\rm min}}: all values from 5 up to \emph{k}
    \item $\sigma_{\rm cut}$: from 10 to 120 \kms with steps of 5 \kms.
\end{itemize}{}

The difference with respect to the gE0-dE0 case (i.e. the  case of another hot systems), is the adoption of larger $\sigma_{\rm cut}$, $k$, and a larger $N_{\rm min}$ range. \chiara{This is motivated by the fact that, } %because, 
being the Fornax environment hotter that the Galmer gE0, we can detect higher velocity substructure (if any). 
Similarly to the Galmer simulations, we use also $\sigma_{\rm cut}$ values below the instrumental errors, considering in these cases $\sigma_{\rm mea} \sim \sigma_{\rm obs}$ (see discussion in \S\ref{sec:gE0-dE0}). 

Fig.~\ref{fig:my_label} 
shows the reliability map, color coded by the fraction of the number of set-ups with a reliability $\geq$ 70\%. %, as usual. 
\massi{The case of the Fornax-like system is fairly different with respect to the configurations tested with the Galmer simulation. 
Indeed, the intrinsically higher velocity dispersion provides a much smaller chance to have a correlated group of particles characterized by a small dispersion, due to statistical fluctuation in the parameter space.}
%\st{It seems evident that the case of the Fornax-like system, due to the intrinsically higher velocity dispersion, has a much smaller chance to have a correlated group of particles characterized by a small dispersion, due to statistical fluctuation in the parameter space.} 
For this reason, COSTA has a quite large range of parameters that find a spurious stream in less than 30\% of the extractions. One would argue that maybe for this case 70\% is a too loose threshold \chiara{and higher values might be used too. } %(e.g. oppositely to the gSa/dS0 case) and one could also use a higher threshold. 
However the Montecarlo simulations are %currently 
only partially catching the full statistical fluctuations, \chiara{and they might be too} %or maybe they are just too
smooth with respect to the real data. 
\chiara{In conclusion, also taking into account that the higher is the threshold the lower is the chance to find a stream, we keep 70\% as reliability threshold, in line with the previous tests.}
%Moreover, also considering that the higher is the threshold the lower is the chance to find a stream (however with a higher contamination, see discussion at the end of \S\ref{sec:gSa-dS0}), we keep 70\% as a fair threshold for the Fornax-like case too.  

As a second step, we rau COSTA on the DS where the artificial streams have been added, to assess the effectiveness of COSTA detection. %the stream. 

\subsection{Recovering simulated substructures}
\label{sec:simulated_streams}
When applying COSTA to real cases, detection is the minimal goal we want to achieve (completeness), while we can compromise with the full recovery of stream particles vs. contaminants (purity), as we reasonably expect that we loose some particles and also get some contaminant (non-stream particle) as part of a correctly detected stream (see also the discussion in \S\ref{sec:gE0-dE0} and \S\ref{sec:gSa-dS0}). 
%, nor that a stream, with the characteristics we expect to find in the real data, is actually detectable by our algorithm.
%or real data have more complicated random substructures that our simulations could not reproduce.

To check the ability of COSTA \chiara{in recovering }%to recover 
known streams in the Fornax-like environment, and %possibly 
assess completeness and purity, we  add \chiara{three} artificial streams to our Montecarlo simulations. 
%A first stream, hereafter \emph{stream 1}, is constituted by 20 particles and has the size of 1$'\times$2$'$, having an intrinsic velocity  dispersion  $\sigma=35$ kms$^{-1}$ (i.e. very close to the velocity errors assumed in our simulations).
Since we cannot reproduce the full dynamics of a stream in our Montecarlo simulations, and we want to test COSTA into more observational situations, we choose typical stream sizes and kinematics that can be realistically found in real data. 
As shown in the case of Galmer simulations (see e.g. Fig.~\ref{fig:galmer_found}),
%NRN-8: Massi metti il riferimento a queste figure
despite a dwarf galaxy spreads a large number of particles along its encounter orbit, COSTA can identify only the closer %(last stripped) 
ones, that were the last to be stripped (of the order of a few tens, depending on the surface brightness of a stream), spread over $\sim$5--15 kpc, i.e. $1'-3'$ at the distance of Fornax.

%We chose these properties after having blindly ran some of the higher reliability configurations seen above on the real data to check what where the typical number of particles and sizes of the substructures found in the data.
%This artificial stream is similar to some of the streams found in the real data also in size but it is about twice in numbers, in order to check how many of the particles belonging to a real stream are actually recovered. Indeed, we cannot expect that a stream mixed to a relaxed component can have the totality of the particles recovered, hence we want to measure also the completeness of the algorithm ``on stream''.

A first stream (\emph{stream 1}, hereafter), is %constituted by 20 particles
\chiara{made of 20 particles, has the size of 1$'\times$2$'$ and an intrinsic velocity dispersion of $\sigma=35$ kms$^{-1}$.}% and has the size of 1$'\times$2$'$, having an intrinsic velocity  dispersion of $\sigma=35$ kms$^{-1}$.
Other two streams are extracted by randomly sampling particles from the tail of the Galmer gE0-dE0 case discussed in \S\ref{sec:gE0-dE0}.
%Although the bulk of the dwarfs in the core of Fornax is made of dE0, we used a dS0 because this had the mass and effective radius more similar to the dwarfs in the inner regions of Fornax.
%We selected an encounter starting with the dE0 at 100 Kpc and falling toward the gE0 in a prograde orbit with an inclination of 33 degrees and a pericentral distance of 16 Kpc.
%The parameters of the simulated galaxies are listed in the table \ref{tab:galmer parameters}. we have selected a simulation of an encounter between , 
We isolate a group of 30 particles, distributed over an area of about $3'\times1.5$' in one case (\emph{galmer 1}, hereafter) and $6'\times3'$ in a second case (\emph{galmer 2}), with an intrinsic velocity  dispersion of  $\sigma=45$ kms$^{-1}$ and $\sigma=62$ kms$^{-1}$, respectively. 
These \chiara{two} streams 
%which we have named \emph{galmer stream} 
have the advantage of being more realistic (in shape and density) as based on a simulated encounter, although the dynamics of the Galmer simulation adopted is not really close to the one of the Fornax core, in particular \chiara{because of }%for 
the smaller mass of the main galaxy as compared to NGC~1399. 
Also, we take larger sized streams (3$'$ roughly correspond to 30 kpc), in order to explore the ability of COSTA to find larger and more diffuse streams. 
%However, for the test we want to perform here this choice is the least arbitrary we can make and, as a sanity check, we have also performed a full self-consistent test of COSTA on the Galmer hydrodynamical simulations, as described in Appendix \ref{sec:galmer simulations}.

\begin{table}
    \centering
    \caption{Properties of the simulated streams}
    \begin{tabular}{lcccc}
    \hline
     & N & Size (arcmin) & $\sigma_{I}$ (kms$^{-1}$) & $\sigma_{M}$ (kms$^{-1}$)  \\
     \hline
     \emph{stream 1} & 20 & 2 X 1 & 35 & 42\\
     \emph{galmer 1} & 30 & 3 X 1.5 & 45 & 58\\
     \emph{galmer 2} & 30 & 6 X 3 & 62 & 69\\
     \hline
    \end{tabular}
    \label{tab:simulated streams}
\end{table}

%NRN5: Massimiliano aggiusta questa section per il fatto che abbiamo aggiunto lo stream Galmer...commenta che alla fine la contaminant fraction e la completeness che si ottengono per gli stream galmer sono simili indipendentemente dalle dimensioni...

The final properties of the artificial streams are summarized in  Table~\ref{tab:simulated streams}. 

In order to simulate a real measurement of the particle redshift, we randomly re-extract their ``measured'' velocities from a Gaussian having the central velocity equal to the intrinsic radial velocity, and standard deviation of 37 kms$^{-1}$. We stress here that the three streams have a velocity dispersion within the range 
%\st{we are considered} 
\massi{expected} for dwarf galaxies \citep{coma_dw_FJ}. 
%\ACTIONITEM{WHAT IS THE VELOCITY DISPERSION MID SIZE?}

Also, the number of particles is not arbitrary: indeed, assuming that we split the particles in the same number of PNe and GCs, and assuming a typical (bolometric) PN specific number density of $50\times10^{-9}$ PN$/L_\odot$ \citep[see e.g.][]{feldmeier2004}, the luminosity in $g$-band corresponding to 10-15 PN-like particles is of the order of $10^8L_\odot$. The corresponding surface brightness of streams\chiara{, with sizes as in Table~\ref{tab:simulated streams}, } %(taking into accounts the sizes in Table~\ref{tab:simulated streams})
is of the order of 28-29 mag$/$arcsec$^2$, which is close to the typical low surface brightness levels expected for these substructures \citep[see e.g.][]{cooper-2010}.  

%\begin{figure*}
 %   \centering
  %  \hspace{-0.9cm}
   % \includegraphics[width=9.2cm]{contaminants_frac_streams_fig_v2.pdf}
    %\includegraphics[width=9.2cm]{completeness_streams_fig_v2.pdf}
    %\caption{Contaminant Fraction as a function of the set-up parameter for the $R=5'-7'$ shell (1st column), and the $R=7'-12'$ shell (2nd column): red points are for the Galmer stream 1, green points for the Galmer stream 2 and blue points for stream 1. Observed completeness (full dots) and True completeness (transparent dots) as a function of the set-up parameter for the $R=5'-7'$ shell (3rd column), and the $R=7'-12'$ shell (4th column). Stream 1 is shown only for the $R=5'-7'$ shell, while Galmr stream 2 only for $R=7'-12'$ shell, as these are more representative for the typical structure we have found in real data (see \S\ref{Sec:Results}).}
   % \label{fig:streams_parameters}
%\end{figure*}

\begin{figure*}
%    \centering
    \includegraphics[scale = 0.2]{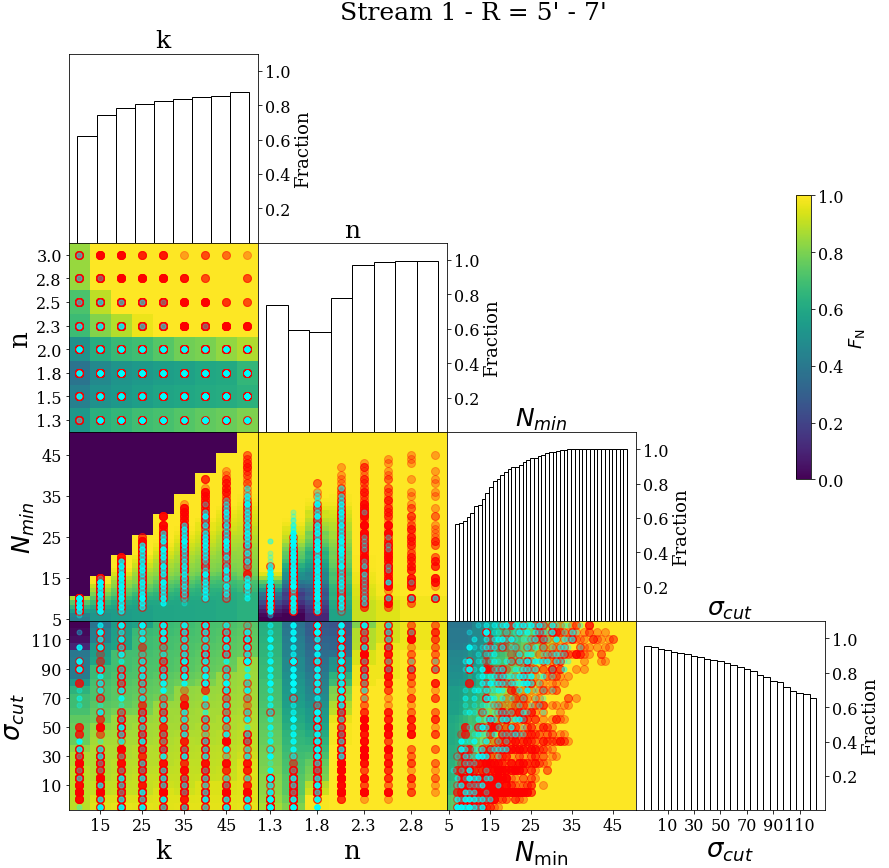}
    \includegraphics[scale = 0.2]{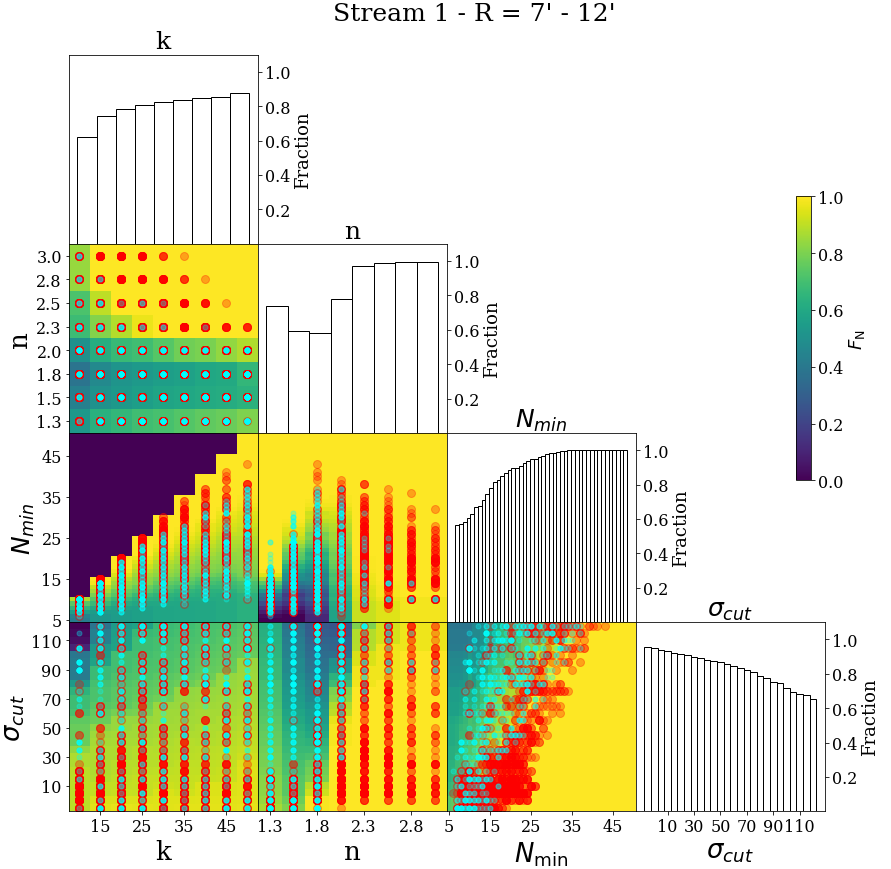}
    \includegraphics[scale = 0.2]{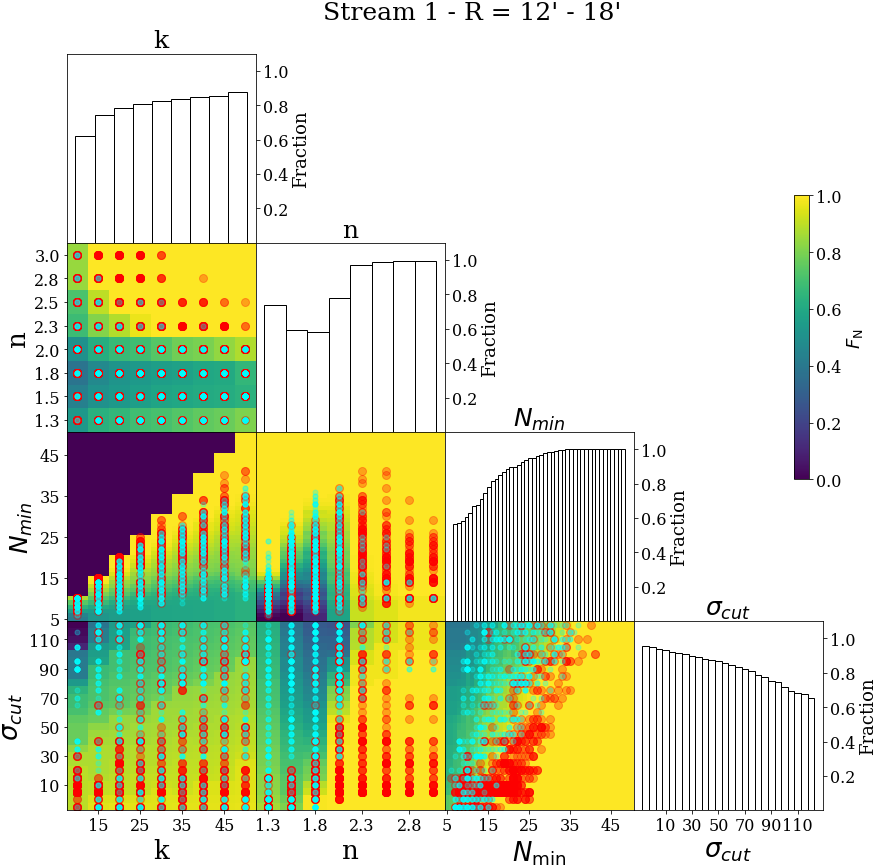}\\
    \includegraphics[scale = 0.2]{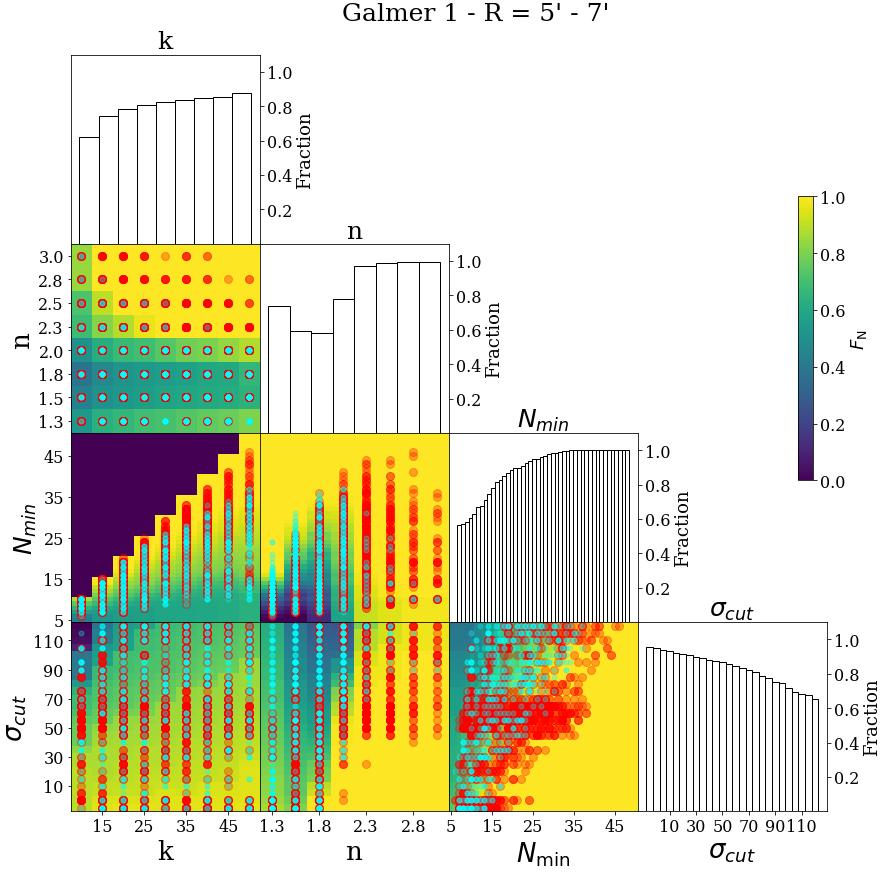}
    \includegraphics[scale = 0.2]{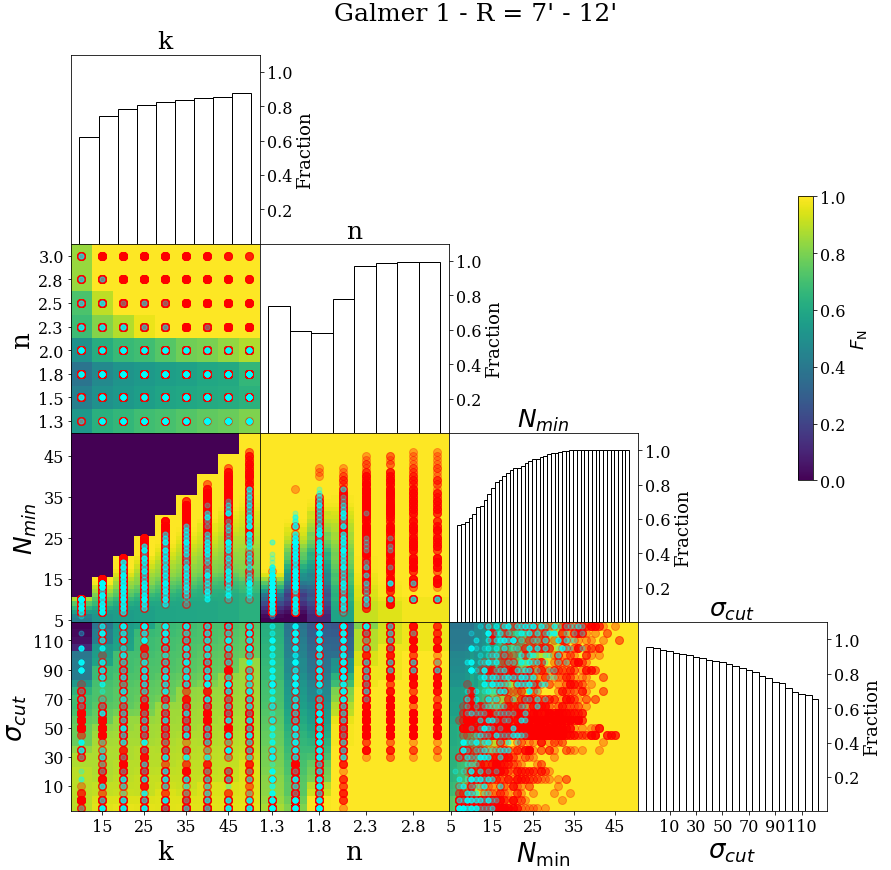}
    \includegraphics[scale = 0.2]{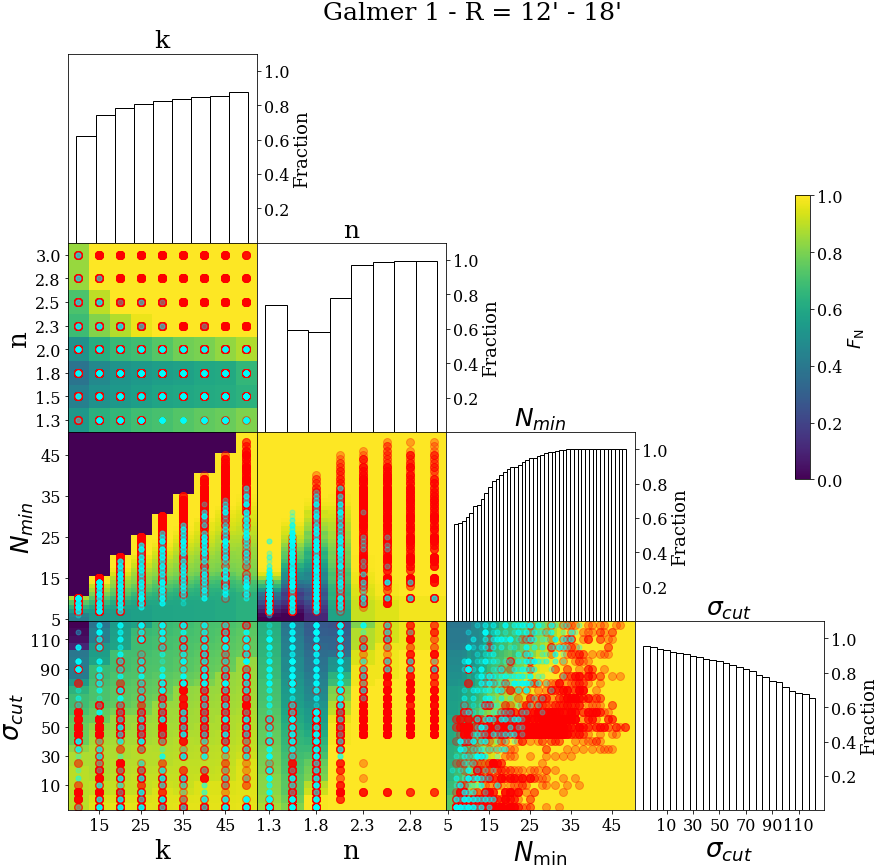}\\
    
\hspace{6.5cm}    \includegraphics[scale = 0.2]{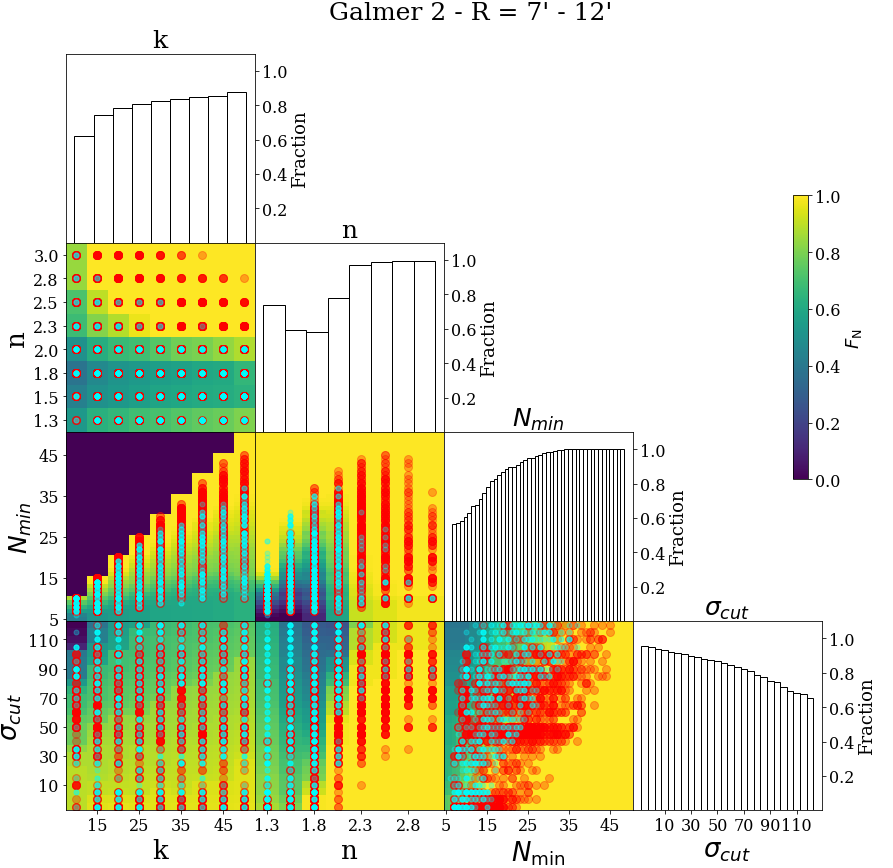}
    \includegraphics[scale = 0.2]{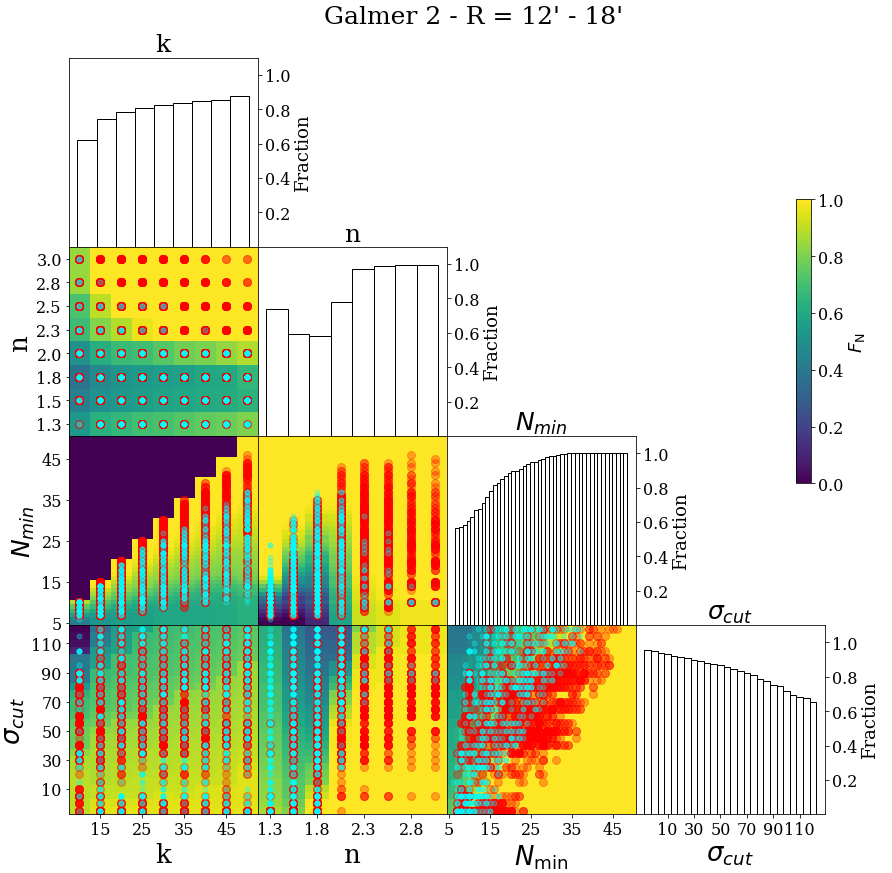}\\
    \caption{Stream (red points) and spurious structures (cyan points) overlapped on the Fornax cluster reliability maps in the different shells (left column R $= 5\arcmin-7\arcmin$, middle columns R $= 7\arcmin-12\arcmin$, right column R $= 12\arcmin-18\arcmin$) and for the three different simulated streams (top: \emph{stream 1}, middle: \emph{galmer1} and bottom \emph{galmer2}.}
    \label{fig:artificial_streams}
\end{figure*}{}

At this point, we build a set of Montecarlo simulations to which we add the particles of the three  simulated  streams  each time varying their mean
%The three simulated streams are added to the Montecarlo simulations, varying their mean 
velocity in a random way, using a Gaussian with a 0 kms$^{-1}$ mean, assuming the cD at rest in the cluster center, and with a standard deviation of 300 kms$^{-1}$, similar to the velocity dispersion of the Fornax cluster (see P+18 and S+18 for a discussion). 
We also randomize every time their positions to recover them in three different ranges of cluster-centric radii, in order to verify a dependence of the recovery rate with the distance (and hence the local particle density). We choose three shells: $R=5'-7'$, $R=7'-12'$, $R=12'-18'$, while we exclude the region within $5'$, \chiara{where the light is too dominated by the cD and it would be very hard to detect low surface brightness structures}. %, as usual. 
We do not put the streams at distances larger than $18'$ in order to be conservative, as in less dense regions the frequency with which the stream is recovered could be overestimated because of a lower background. Furthermore, the stream \emph{galmer 2} has been inserted only in the two external shell because this diffuse stream is typical of longer lived passages happening far from the cluster center (see e.g. the tail of the encounter in Fig.~\ref{fig:gE0-dE0 interaction}).

As for the Galmer systems, we run COSTA on 10 over 100 of the Montecarlo simulations from the reliability run, to evaluate the stream detection performances. 
%This is done to obtain more variance over the possible configurations that allow to recover the stream. We remind that we want to assess the effectiveness of COSTA to find real streams in the Fornax cluster, after having defined the parameter space where COSTA is likely to find real cold substructures.
%NRNfin: da qui
The results for the three streams at different radii and their reliability maps %for the three streams 
are displayed in Fig.~\ref{fig:artificial_streams} and listed in Tab.~\ref{tab:artificial_streams}. 

\begin{table}
    \centering
    \caption{Same of Tab.~\ref{tab:galmer_statistic} but in the case of artificial streams in a Fornax-like cluster. }
    \begin{tabular}{l|c|c}
    \hline
   \textbf{Stream}  & \textbf{F$_N$} & \textbf{CF}\\ 
   \hline
    \multicolumn{3}{c}{R = 5' - 7'}\\
    \hline
 %   Configuration & (\%) & & &    
 \emph{stream 1} & 0.69 $\pm$ 0.07 & 0.26 $\pm$ 0.18\\
    \emph{galmer 1} & 0.66 $\pm$ 0.22 & 0.24 $\pm$ 0.12\\
    \hline
    \multicolumn{3}{c}{R = 7' - 12'}\\
    \hline
    \emph{stream 1} & 0.69 $\pm$ 0.06 & 0.18 $\pm$ 0.15\\
    \emph{galmer 1} & 0.78 $\pm$ 0.09 & 0.22 $\pm$ 0.18\\
    \emph{galmer 2} & 0.68 $\pm$ 0.15 & 0.20 $\pm$ 0.15\\
    \hline
    \multicolumn{3}{c}{R = 12' - 18'}\\
    \hline
    \emph{stream 1} & 0.60 $\pm$ 0.19 & 0.17 $\pm$ 0.15\\
    \emph{galmer 1} & 0.77 $\pm$ 0.08 & 0.18 $\pm$ 0.18\\
    \emph{galmer 2} & 0.74 $\pm$ 0.10 & 0.15 $\pm$ 0.14\\
    \hline
    \end{tabular}
    \label{tab:artificial_streams}
\end{table}{}

%as in Table \ref{tab:reliability}; results for the three radial shells are reported in Tables  \ref{tab:sim streams 5-7}, \ref{tab:sim streams 7-12} and \ref{tab:sim streams 12-18}.
%In particular we report: 1) the set-up which allowed to correctly detect the simulated stream, together with reliability; 2) the frequency with which the stream is recovered over the 100 simulations 3) the number of recovered particles belonging to the stream and the number of contaminants, 4) the contaminant fraction defined below and the estimate of the velocity dispersion for each parameter set-up.

The general result is that COSTA is able to recover all streams with a 
%\st{variegate} 
\massi{broad} number of set-up parameters. On average,  
\chiara{for the \emph{galmer} streams, the recovery slightly increases at larger distances }%the recovery increases at larger distances, 
mainly because the signal from the particles belonging to the stream is higher with respect to the noise of the hot environment (with lower local density). \chiara{This does not happen for the \emph{stream 1}, which is smaller and more compact. Specularly, the contaminant fraction decreases with increasing radius, and this is true for all the streams. }
%This is also demonstrated by the lower number of contaminants, which is generally found in the more distant shells for any given set-up. 
%\st{Specularly, the number of correctly recovered particles increases with the distance from the center because, due to the lower density of non stream particles at larger distances, the chance to find close particles with similar velocity decreases.} 

%\massi{It is worth noting that the mean reliability slightly increases from the inner regions towards larger cluster-centric distances, even if they are within the uncertainties. What this would suggest is actually that the highest density of the central area may rise the possibility to random generate a group of coherent particles, meaning that the odds to find a spurious cold structure are highest at small distances from the cluster center. This supports our decision to do not run COSTA in the innermost regions.\\

\subsection{Stream kinematics}
\label{sec:fornax_streams_kin}
\chiara{The only aspect of the algorithm we are interested to discuss in more detail at this level is how much (and what kind) of physical information COSTA can provide, besides the stream detection. In fact, despite the detection is important {\emph per se}, as it provides candidates for follow-up observations (e.g. deep imaging, higher resolution spectroscopy), to have some predictions about relevant intrinsic properties, e.g. surface brightness, a trustable estimate of the velocity dispersion and, possibly, the membership of particles, is fundamental to plan such follow-up programs.

As introduced in \S\ref{sec:simulated_streams}, the estimation of the true kinematics of the stream 
%\st{this} 
is equivalent to estimate what is the real number of true particles belonging to the stream and the fraction of contaminants from the background system as a function of the set-ups. In principle, one can think that this should depends on the structure of the stream: compact, well populated and very cold streams should produce almost no contaminants, while very diffuse warmer streams %(despite being decoupled by the parent system)
would easier be contaminated by particles of the galaxy halo, \chiara{having similar velocities to that of the central galaxy.}}
%, mimicking similar velocity (unless the systemic velocity of the stream is very different from the one of the central galaxy). }
However, as we will discuss in the next \S\ref{sec:fornax_contam}, \massi{contaminants do not have a huge effect on the estimation of the velocity dispersion of the stream, since by construction COSTA picks up particles with similar velocity.} 

\begin{figure}
\centering
%\hspace*{-0.6cm}
    \includegraphics[scale = 0.26]{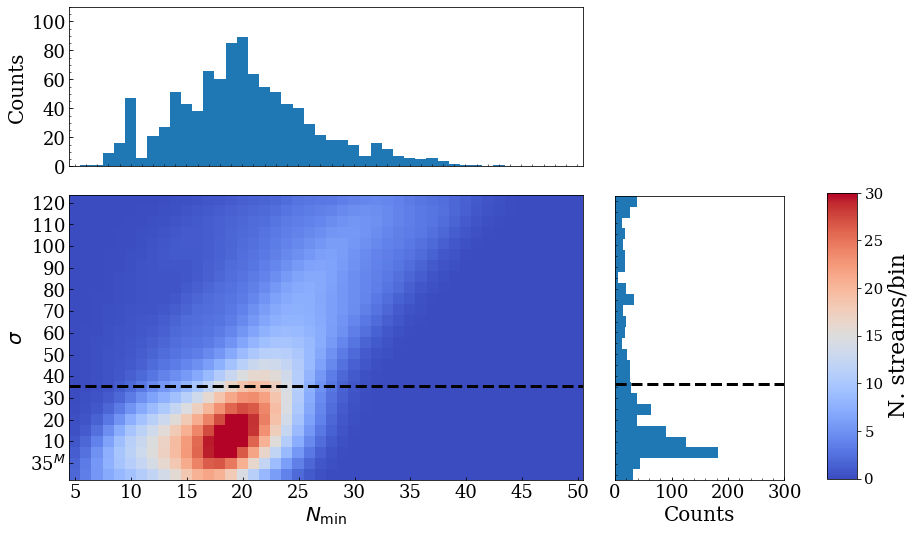}\\
%\hspace*{-0.6cm}
    \includegraphics[scale = 0.26]{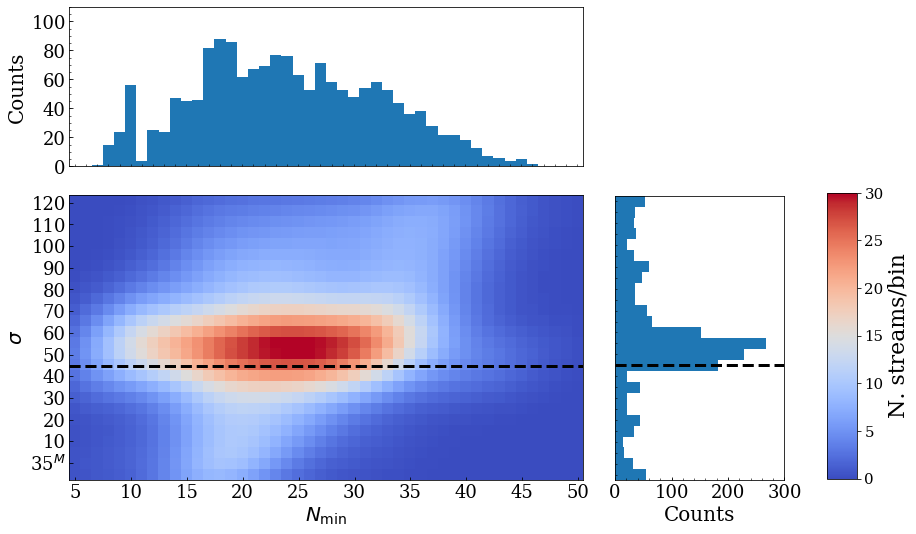}\\
%\hspace*{-0.6cm}
    \includegraphics[scale = 0.26]{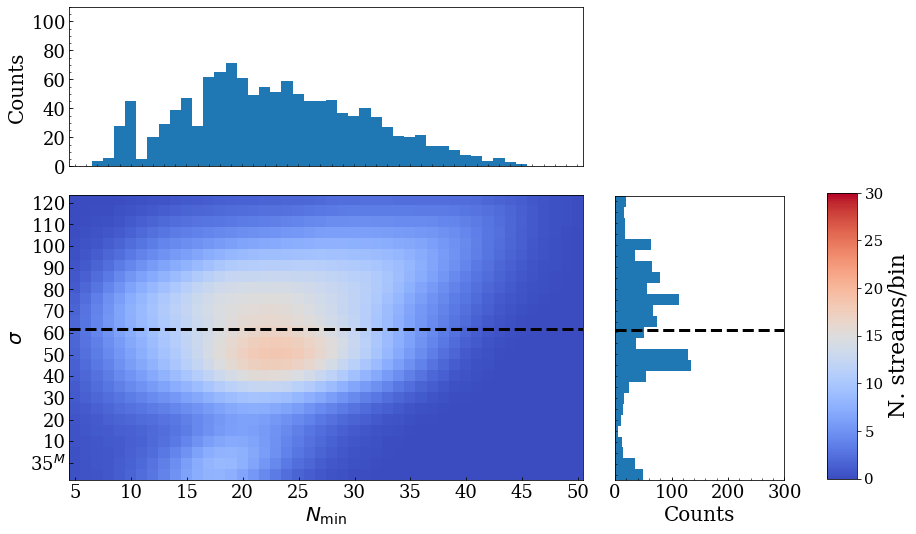}
    
    \caption{Density plot \chiara{in the $N_{\rm min}-\sigma$ diagram, and number counts} of the recovered artificial streams for the \chiara{middle}%central 
    shell (i.e. 7\arcmin-12\arcmin) for {\it stream 1} (top), {\it galmer 1} (middle) and {\it galmer 2} (bottom). \chiara{The horizontal dashed line show the true velocity dispersion of each stream.} The sample %of the detection 
    has been cleaned by using \texttt{Frac} = 0.5 as threshold in the $n-N_{min}$ space. Data have been smoothed with a Gaussian kernel having a bandwidth equal to 3.} %\chiara{MAYBE ADD WHAT IS WHAT? Stream 1 Galmer 1/2}}
    \label{fig:disp_vs_sigma_cut_artificial_streams}
\end{figure}

In Fig.~\ref{fig:disp_vs_sigma_cut_artificial_streams} we show the density plot of the recovered streams in the $N_{\rm min}-\sigma$ diagram, cleaned by using a threshold of $F_{\rm N}$ = 0.5 in the $n-N_{\rm min}$ panel, %(as done in S\ref{sec:kinem_gE_dE} and S\ref{sec:kinem_gSa_dS0}), 
in the $R = 7\arcmin-12\arcmin$ shell, as indicative of the ability to recover the stream kinematics.  \massi{The results for the other two annuli are similar to these obtained for the central one.}%, hence for the sake of simplicity we plot and focus only on the latter.} 
We notice that, differently from the Galmer simulations, we construct the streams with a given velocity dispersion, \chiara{and thus we can check if the 'real value' is recovered. }  %that we can check whether recovered. 
From the figure, it is evident that the set-ups where the stream is detected accumulate around the true velocity dispersion of the streams (35 \kms, 45\kms , 62\kms for {\it stream 1, galmer 1 } and {\it galmer 2}, respectively), indicated by the dashed black line. In particular, the median of the $\sigma_{mea}$ distribution is found to be 16 $\pm$ 24, 52$\pm$ 15, 62 $\pm$ 28, i.e.  \chiara{consistent within $1\sigma$ uncertainties.} %for the two {\it galmer} streams}. 

\massi{We notice that the difference between the median values and the true velocity dispersion values are of the same order of magnitudes than normal measurement errors for mid-resolution spectroscopy.}
For \emph{stream 1}, however, the mean value is below the accuracy allowed by the velocity, hence the median value %\st{($=10\pm15$)} 
%\chiara{WAHT IS THE DIFFERENCE BETWEEN THIS NUMBER AND $16.4\pm24.3$ ?}
just reveals that the assumed precision does not allow to recover the true kinematics, \chiara{for which a higher velocity accuracy is required. }

%which should need a higher velocity accuracy. 

We also estimate the median number of particles recovered for each stream, to check whether COSTA allows us to infer the total ``luminosity'' associated to the stream, and find %\st{for the same case} \chiara{WHICH SAME CASE?} 
a median of $ N_{\rm recov} = 22, 31, 29 $ for {\it stream 1, galmer 1} and {\it galmer 2}, respectively.  \massi{These median values are very similar to the true number of particles belonging to the stream (20, 30, 30, see Tab.~\ref{tab:simulated streams}),  even if we expect that not all of these particles are truly belonging to the stream (see discussion in the next session).} %\st{This median values, though, have to be read with caution because of the fraction of contaminants, as discussed in the next section.} 
%Thus, even if COSTA reveals the stream with a numerous collection of free-parameters, they are mostly confined around the intrinsic velocity dispersion of the stream.
%For larger $\sigma_{cut}$ values the number of set-ups that catch the stream sharply decreases, telling us that the stream velocity dispersion represents roughly an upper limit beyond which the stream is hardly detected. Thus, if a stream is disclosed in many different configurations the mode/median? of the $\sigma_{cut}$ could provides a hint of the true kinematics of the original stream.

{}

%We can define the recovery rate as the ratio of the frequency of the recovery of a stream for a given set-up, which is equivalent to a (100$\times$recovery probability)\%. Although we cannot generalise the results of the stream adopted because the recovery rate depends on the properties of a stream, from the Tables  \ref{tab:sim streams 5-7}--\ref{tab:sim streams 12-18}, we can derive some dependence of completeness and purity of the recovered streams, assuming these are close to average real streams. First, the recovery rate (or probability) ranges from $\sim30\%$ to $100\%$.
%MG: Actual, the recovery rate in some set-ups is smaller than 30%
%On average the recovery increases at larger distances, mainly because the signal from the particles belonging to the stream is higher with respect to the noise of the hot environment (with lower local density).
%This is also demonstrated by the lower number of contaminants, which is generally found in the more distant shells for any given set-up. Specularly, the number of correctly recovered particles increases with the distance from the center as, due to the lower density of non stream particles at larger distances, the chance to find close particles with similar velocity decreases. 

\subsection{Contaminants and fraction of recovered particles}
\label{sec:fornax_contam}
%We are finally interested on evaluating 
\chiara{We evaluate here} the fraction of contaminant ({\tt CF}, defined in \S\ref{sec:stream_detection}) particles within the different streams as a function of the different parameters. We use the Fornax simulations to figure how the contamination can change depending on %a larger variety of 
stream compactness (from the compact {\it stream} 1, to the very diffuse {\it galmer} 2), internal velocity dispersion and position within the central potential. 
\chiara{We stress that this is a test that one can do only {\emph a posteriori}, by placing stream  candidates in the Montecarlo simulations at the right distance and with the right geometry to perform a ``contamination run''. Nevertheless this is useful in order to assess what is the realistic contamination for a given stream. }
%We stress here that in order to assess what is the realistic contamination for a given stream, this is a test that one can do {\emph a posteriori}, by placing a stream candidates in the Montecarlo simulations at the right distance and with the right geometry to perform a ``contamination run''. 
%Here we want to provide a general example to show the impact of contaminants 
%that the fraction of contaminants one might reasonably expect from COSTA does not alter significantly 

\chiara{Our main goal is to provide a general example to show the impact of contaminants} on the inference one can derive from COSTA candidates on the intrinsic properties of the streams.

%We define the contaminant fraction as 
%\begin{equation*}
%\texttt{CF} = \frac{\texttt{\# recovered non-stream particles}}{\texttt{total recovered (stream and non-stream)}}
%\end{equation*}

\begin{figure*}
    \centering
    %\hspace{-0.9cm}
    \includegraphics[scale = 0.35]{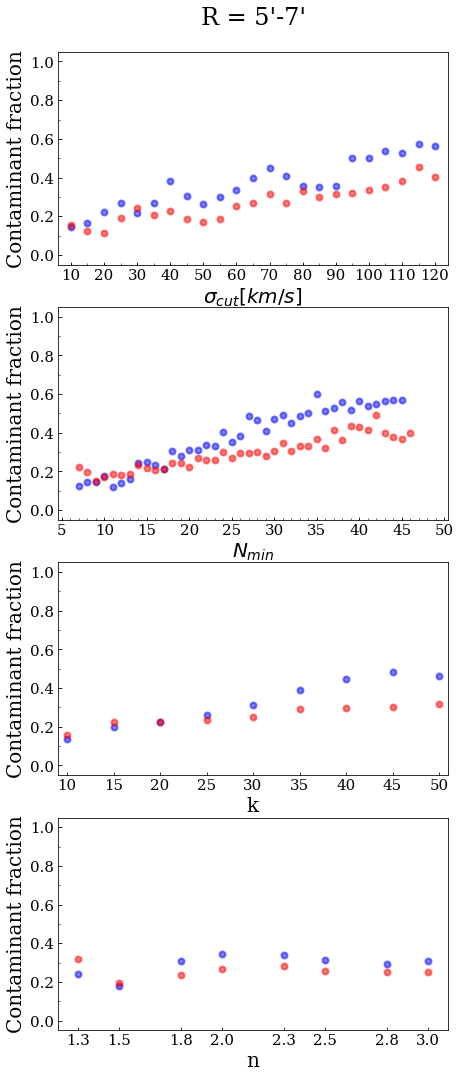}
    \includegraphics[scale = 0.35]{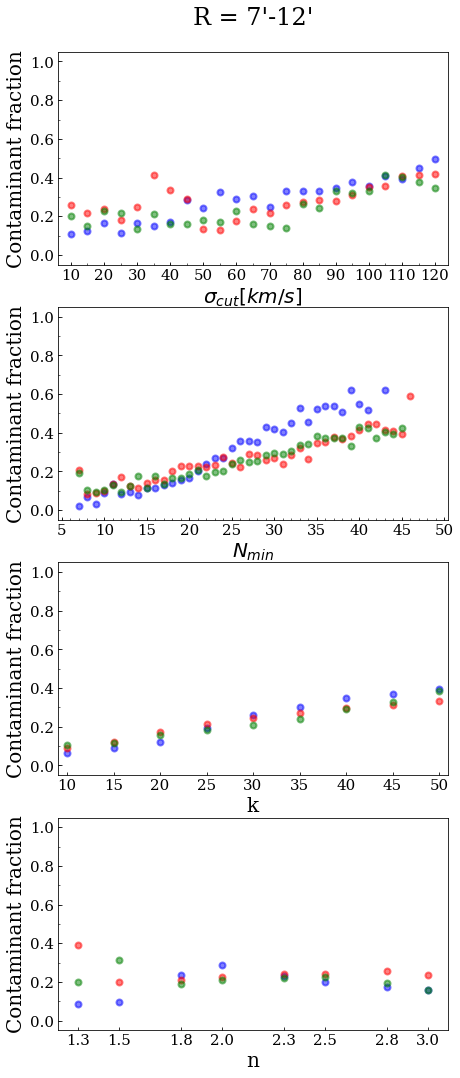}
    \includegraphics[scale = 0.35]{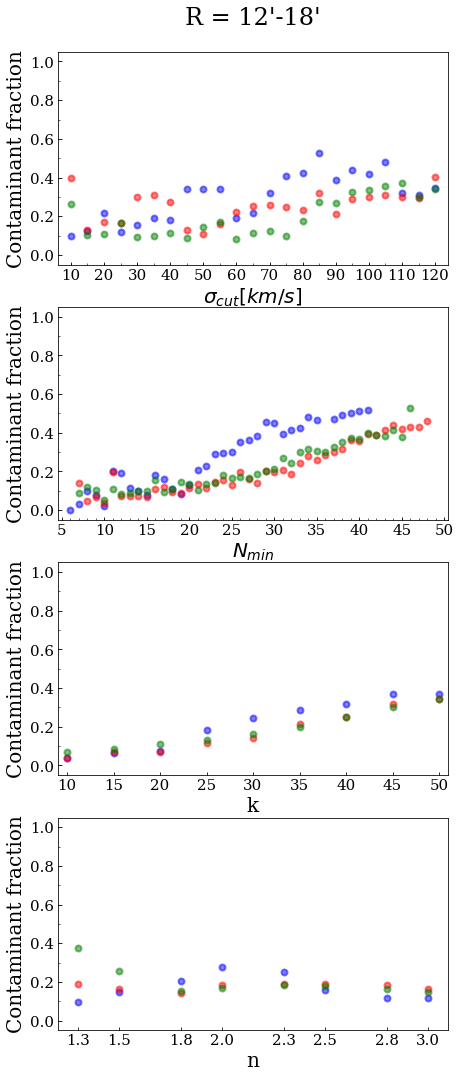}
    \caption{Contaminant Fraction as a function of the set-up parameters for the $R=5'-7'$ shell (left column), the $R=7'-12'$ shell (middle column) and the $R=12'-18'$ shell (right column). Red points are for the \emph{galmer 1} stream, green points for the \emph{galmer 2} stream and blue points for \emph{stream 1}.}
    
    %Observed completeness (full dots) and True completeness (transparent dots) as a function of the set-up parameter for the $R=5'-7'$ shell (3rd column), and the $R=7'-12'$ shell (4th column). Stream 1 is shown only for the $R=5'-7'$ shell, while Galmr stream 2 only for $R=7'-12'$ shell, as these are more representative for the typical structure we have found in real data (see \S\ref{Sec:Results}).}
    \label{fig:artificial_streams_parameters}
\end{figure*}

We show in Fig.~\ref{fig:artificial_streams_parameters} the trend of 
%\st{this} 
{\tt CF} as a function of the different set-up parameters, for every shell for the three streams.  
We see a clear dependence of the {\tt CF} on $N_{\rm min}$ and on $k$: i.e. the larger $N_{\rm min}$ and $k$ are, the higher the number of contaminants is. 
\chiara{This is valid for all the streams, in all the shell, but the slope of \emph{stream 1} is steeper and reaches larger {\tt CF}, especially for $N_{\rm min}$. A similar trend is also present for the velocity dispersion cut-off, although it is more noisy and dependent on the shell one considers (more evident at lower radii. %\st{looks less severe}
For $n$, instead, the {\tt CF} is constant around {\tt CF} $\sim0.2-0.3$.}

%\chiara{I PARTIALLY DISAGREE WITH WHAT WRITTEN ABOVE. THE TREND WITH SIGMA IS MORE NOISY AND DEPENDENT ON THE SHELL ONE CONSIDERS.}

%Interestingly, for $N_{\rm min}$, the correlation becomes less evident for the outermost radial bin.

%Being the number of the true recovered particles quite constant, this implies a larger contaminant fraction for larger cut-off. \st{The trend is not dramatic, however} 
%from a 15\% on average for $\sigma_{\rm cut}=10$ kms$^{-1}$ 
In the worst cases we see an increase up to 60\% for {\it stream 1} and $\sim40\%$ for the two {\it galmer} streams, 
%on average 
for $\sigma_{\rm cut}>100$ kms$^{-1}$ and $N_{\rm min}>35$.  
%for the inner shell and outer shells, respectively,
The main reason for the larger contamination is that, while the number of recovered true particles remain almost constant, the larger  $\sigma_{\rm cut}$ and $N_{\rm min}$ make COSTA select more particles with compatible velocities, but not belonging to the stream. 
\chiara{For $k=50$, the CF reaches the highest level of 50\% for the inner radial bin, while it stays below 40\% for the outer ones. For any value of $n$, the CF is always below 40\% and, also in this case, it is on average lower for the outer bins with respect to the inner one.}
%and f or the \emph{stream 1}, and 
%10\% on average for $\sigma_{\rm cut}=10$ kms$^{-1}$ 
%and  up to 40\% on average for $\sigma_{\rm cut}=120$ kms$^{-1}$ for the outer shells (see Fig. \ref{fig:artificial_streams_parameters})}. 
%\st{We will also define purity as the complement of the contaminant fraction, i.e. \texttt{PU}=1-\texttt{CF}}. 
%This fractions, though, do not reflect the true contamination, i.e. the number of the non-stream particles we select for every stream with respect to the real number of stream particles.

%the observed completeness
%MG: OC has the same definition of PU (purity), maybe we can call this two quantities in unique manner
%(\texttt{OC}) as
%\begin{equation*}
%\texttt{OC} = \frac{\texttt{\# recovered stream particles}}{\texttt{total recovered (stream and non-stream)}}
%\end{equation*}
%which is clearly complementary to the \texttt{CF} (i.e. \texttt{OC}$=1-$\texttt{CF}) and 
%Along the same line, i

%CS: MOVED
The number of particles recovered has a strong dependence on the number of $k$ (neighbors), as we can expect, and it also has a small dependence on the value $n$ of the sigma clipping: higher values of $n$ give better results. Completeness and contaminant fraction are very similar in the case of the two {\it galmer} streams, even if they have different sizes. This suggests us that COSTA is able to recover even diffuse streams, e.g. the frequency of the recovering is independent of the dimension of the structure; although we speculate that it could be harder to detect these kind of streams in the denser region, where the noise due to the hot component of the cluster increases.

It is also instructive to verify the correlation between the number of the recovered particles and the set-up parameters. 
To do that, we define 
the true completeness 
(\texttt{TC}) as
\begin{equation*}
\texttt{TC} = \frac{\texttt{\# recovered stream particles}}{\texttt{\# true stream particles}}
\end{equation*}
where the true stream particle numbers are given in Table ~\ref{tab:simulated streams} for the three simulated streams. 
The \texttt{TC} is used %\st{needed}
here to figure what is the \chiara{range of parameter set-ups} that maximize the number of recovered particles over the real ones, by contemporary \chiara{implying a low CF.} %cross-checking with the contaminant fraction. 

\begin{figure*}
    \centering
    %\hspace{-0.9cm}
    \includegraphics[scale = 0.35]{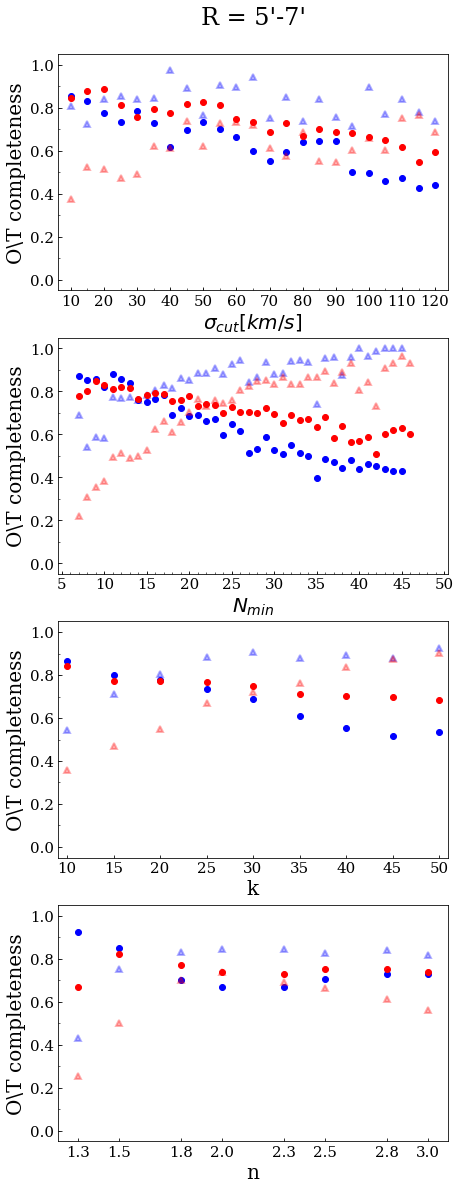}
    \includegraphics[scale = 0.35]{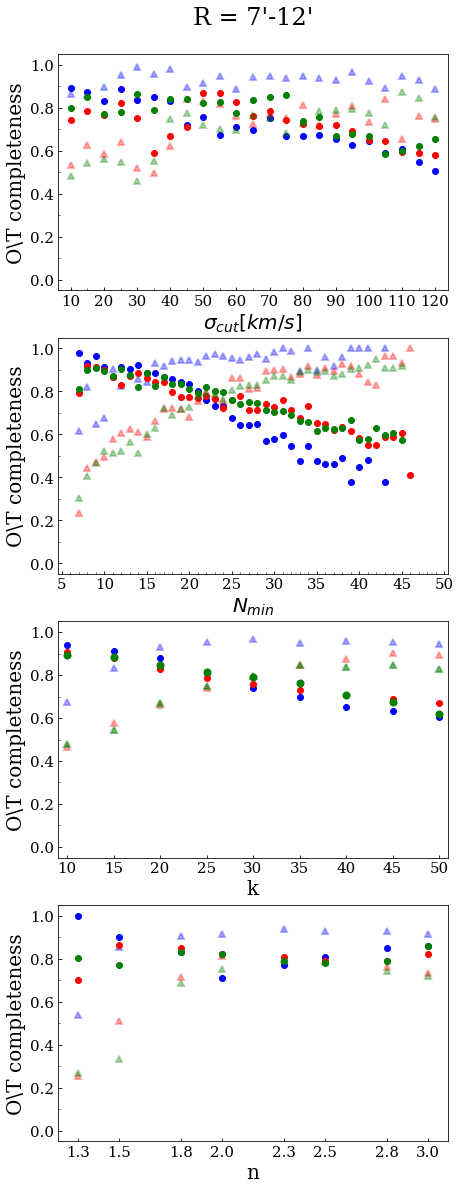}
    \includegraphics[scale = 0.35]{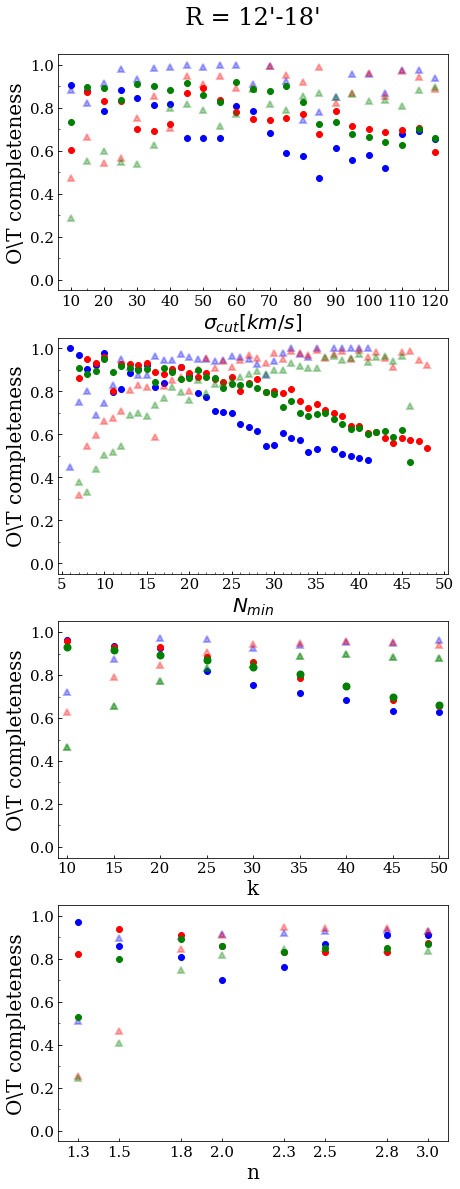}
    \caption{Observed completeness (full dots) and True completeness (transparent triangles) as a function of the set-up parameters for the $R=5'-7'$ shell (left column), the $R=7'-12'$ shell (middle column) and the $R=12'-18'$ shell (right column). \chiara{The color code is the same as in Fig.~\ref{fig:artificial_streams_parameters}: red points for {\it galmer} 1, green points for {\it galmer} 2 and blue points for {\it stream} 1.}}
    %Observed completeness (full dots) and True completeness (transparent dots) as a function of the set-up parameter for the $R=5'-7'$ shell (3rd column), and the $R=7'-12'$ shell (4th column). Stream 1 is shown only for the $R=5'-7'$ shell, while Galmr stream 2 only for $R=7'-12'$ shell, as these are more representative for the typical structure we have found in real data (see \S\ref{Sec:Results}).}
    \label{fig:artificial_streams_OTC}
\end{figure*}

All quantities \chiara{listed} above are not known in observations and depend on the specific data-set and on the distance (i.e. the local particle density). 
To investigate the effect of \chiara{distance, we plot}, %the latter parameter, we have plotted 
the values of the  \texttt{TC} (transparent triangles) parameter and of the \texttt{OC} (full dots) parameter, defined in \S\ref{sec:stream_detection},  for  \chiara{the three simulated streams in the three different shells.} 
%all streams in Fig.~\ref{fig:artificial_streams_OTC} in the three different shells.

%\st{defined them for different streams and radial shells according to the results reported in Tables.}  
%We see that generally the \texttt{OC} has a dependence on the four free parameters, this is expected since \texttt{OC} is the complement of \texttt{CF}.  
%\st{is quite
%constant as a function of the parameter ($\sigma_{\rm cut}$, $k$, $n$, $N_{\rm min}$), at a mean value of $\sim85-90\%$ slightly decreasing at $N_{\rm min}>15$.} \st{Despite the high nominal value, the \texttt{OP} is just an upper limit of} 

The \texttt{TC} has \chiara{overall an }%a clear 
increasing trend with increasing values of all the set-up parameters. \chiara{This indicates that COSTA recovers }
%, as seen in Fig.~\ref{fig:artificial_streams_OTC}, 
 %especially with $N_{min}$ and $k$, which tend to 
%increasingly 
more particles by using stronger constraints (i.e. larger number of neighbors, larger number of particles). However, this also causes a larger number of contaminants, as seen in Fig.~\ref{fig:artificial_streams_parameters}. %\st{lower sigma clipping and larger cut-off)} 
%at the cost of larger contaminants (see discussion about the figure \ref{fig:artificial_streams_parameters}). 
We also note that the behavior of the \texttt{TC} is more 'extreme' in the case of the {\it galmer} streams (red and green points). 
\chiara{We speculate that this can be due to the fact that {\it stream 1} is much more compact than the other two. 

The observed completeness (full dots) instead, decreases for larger $\sigma_{\rm cut}$, $N_{min}$ and $k$ while it stays roughly constant.}

A compromise rule-of-thumb we can derive cross-checking Fig.~\ref{fig:disp_vs_sigma_cut_artificial_streams},  Fig.~\ref{fig:artificial_streams_parameters} and Fig.~\ref{fig:artificial_streams_OTC} is that if a stream is found with different set-ups with similar reliability, one should first of all select the parameters that better reproduce its kinematics: 
the median of the distributions shown in Fig.~\ref{fig:disp_vs_sigma_cut_artificial_streams} for $N_{\rm min}$ and $\sigma_{\rm cut}$. Then, one should try to select values for $n $ and $k$ for which the \texttt{OB} and \texttt{TC} are similar, while the \texttt{CF} is the minimum possible. For the streams simulated here, this region is around the following values: $k\sim 25-30$ and $n\ge 1.8$. This should %in turns 
allow for a reasonable completeness ($\sim$80\%), with a contamination that is $\sim50$\% in the worst case (i.e. high $\sigma_{\rm cut}$), or lower than that. It is important to stress, however, that these numbers depends on the  system one considers and on the morpho-kinematic characteristics of the streams one aims at recovering. Thus, each run of COSTA needs to be properly trained with ad-hoc Montecarlo simulations before the algorithm can be used with real data.

%one should choose the larger  cut-off which present a 
%$N_{\rm min}\sim 25$ and $n\sim1.8-2.0$. This should allow a reasonable completeness ($\sim$80\%) with a contamination of the order of 20\% or less (purity $>80\%$).  
%\chiara{I DO NOT UNDERSTAND THIS! Do you mean the following: "if a stream is found in different set-ups with similar reliability, one should choose the set-up with the combination of parameters where the OB and TC are similar while the CF is the minimum possible, this regions is around the following values: $N_{\rm min}\sim 25$, $k\sim30$ and $n\ge 1.8$ . This should allow a reasonable completeness ($\sim$80\%) \st{with a contamination of the order of 20\% or less (purity $>80\%$)}.\massi{with a contamination that in the worst case (i.e. high $\sigma_{\rm cut}$ is $\sim50$ \% or lower.}

%We have already commented that generally COSTA cannot find all particles and, in order to minimize the contaminants, one can likely recollect up to 80\% of the true particles plus up to 20\% of contaminants. 
%%In general the number of recovered particles is smaller than the real number 

Finally, we check how accurate \chiara{COSTA} %the method 
can reconstruct the intrinsic properties of the streams. 
The presence of contaminants \chiara{(estimated to be of the order of 20\% of the total number of particles, for low $\sigma_{\rm cut}$, if the rule-of-thumb holds)} is expected to alter the properties of the recovered stream. \chiara{In fact, }%(e.g. 
these should have a hotter kinematics and \chiara{be characterized by a }%produce a overall 
higher velocity dispersion. However, to be selected as friend-of-friend particles, they likely have velocities reasonably compatible with the bulk of the stream, so the effect should not be dramatic. 
From Fig.~\ref{fig:disp_vs_sigma_cut_artificial_streams}, we see that 
%generally the final dispersion estimates are smaller than the intrinsic ones of the streams. T
the smaller the cut-off is, the smaller the measured stream dispersion is, although, if the adopted cut-off is larger than the real stream dispersion, the dispersion estimates tend to saturate around the true dispersion value. This implies that, if on one hand, using a too large cut-off does not produce a dramatic overestimate of the true velocity dispersion, on the other hand too small cut-offs might produce \chiara{an underestimation }%underestimates 
of the stream velocity dispersion.  \chiara{ Thus, adopting a larger cut-off would be the safest choice, } %}This might suggest to adopt a large cut-off as a safe choice, 
to have a more realistic estimate of the stream dispersion, but, as discussed before, this is at the cost of a larger fraction of contaminants.  
%Maybe not surprisingly, around the same cut-off values (i.e. the ones close to the real intrize nsic dispersion) we find also concentrated the higher reliability values: this is because, when searching for streams with a given intrinsic kinematics, the right cut-off minimizes the number of contaminants and  th

Overall, both the number of particles of the stream and the stream velocity dispersion are underestimated \chiara{by COSTA}.  These two quantities are the major parameters we want to retrieve for our stream candidates, because they can give information about the amount of dispersed stellar mass per event and the dispersion of the parent dwarf galaxy, and likely its virial mass. However, the optimization of the set-up to recover the best estimates is beyond the current goals of %this 
\chiara{the }preliminary test \chiara{carried on in this paper}, as we are primarily interested in the detection of the streams.  

\chiara{In conclusion,} for what we have discussed in this section we are confident that COSTA is able to find real streams in our data, if any.

\section{Conclusion}
\label{sec:conclusion}

In this work we have introduced COSTA (COld STream finder Algorithm): a new tool for the detection of cold kinematics substructures in the outer halo of massive galaxies, as probe of their recent and past merger history. 

As all massive galaxies built their halo through minor mergers \citep[e.g.][]{amorisco-2019}, it \chiara{is of enormous value}
%could be invaluable 
to reveal such tidal debris and infer their intrinsic properties in order to unveil the mechanisms playing a role in the mass assembly history of \chiara{massive} galaxies. 

Since these structures have a very low surface brightness, \chiara{it is incredibly difficult to detect them by means of photometric observations alone. } %it is unlikely that they can be detected by means of photometric observations. 
In the last years, \chiara{thanks to new instrumentation that allowed for}
%because of the 
more accurate spectroscopy, the research of merger signatures shifted to the exploration of the phase-space of kinematical tracers. 
So far, supported also by numerical simulations and analytical models, \chiara{the detection of streams has been limited to the search for}
%stream detection have  been based on the research of 
shells and narrow diagonal tracks, having a chevron-like shape in the position-velocity diagram \citep[e.g.][]{romanowsky-2012,longobardi-2015}. 
Unfortunately, these patterns \chiara{are not produced by }%do not represent 
low dispersion streams made of a handful of particles \chiara{(of the order of few tens)} and originated from dwarf galaxies in a recent encounter with a massive galaxy (see e.g. Fig.~\ref{fig:galmer_found}). 
\chiara{COSTA allows us to search for this low surface brightness streams in a {\it systematic way} in } 
%kind of tidal debris in a {\it systematic way} in} 
%With COSTA, we made the research of tidal debris systematic in 
the phase-space, being able to detect cold kinematics substructures moving in a warm/hot environment composed by relaxed particles. 

COSTA relies on a deep friend-of-friend \chiara{algorithm }%procedure 
that, through an iterative sigma clipping, detaches groups of neighbors particles with a cold kinematics (\chiara{with velocity dispersion of the order of } tens of \kms). 
%This procedure %allows one to \chiara{makes possible to identify smaller associations, }%to find out a small sample,  made of tens of particles, as expected for low surface brightness streams originated by dwarf galaxies, with a low velocity dispersion. 

The algorithm \chiara{ has }%depends on 
four free parameters that have to be set with  Montecarlo simulations,  mimicking the real system under exam. 

The final aim of this work has been the detection of streams with COSTA in simulated systems, and we are fully confident that our algorithm is able to detect them in real cases too, if any. 
In particular, in this work, we have
\begin{itemize}
    \item discussed the ability of COSTA in recovering cold substructures in different dynamical conditions, from cold giant spiral galaxies to giant ellipticals living in the core of large cluster of galaxies (sections \S\ref{sec:galmer simulations} and  \S\ref{sec:fornax}). 
    The general results is that, COSTA is able the detect real substructures with a variety of \chiara{combinations of the }%set-ups for its 
    four free parameters and with a limited number of spurious events, if one selects the regions in the parameter space that maximize the reliability of a stream detection and minimize the chance of false positive. 
    \item shown that the COSTA algorithm works in finding cold streams embedded in an hotter environment. The efficiency of COSTA in recovery streams heavily depends %mainly 
    on the ratio between the velocity dispersion of the stream and that of the host galaxy, favoring cold streams embedded in a hot surrounding. \chiara{In fact, the performance of COSTA are much worse when the difference between the velocity dispersion of the giant and that of the dwarf intruder is very small (see the case of a gSa-dSa interaction, \S\ref{sec:gSa-dS0}, versus the case of a gE0-dE0,\S\ref{sec:gE0-dE0},and that of the Fornax Core,\S\ref{sec:fornax}}). %see  \S\ref{sec:galmer simulations} and \S\ref{sec:fornax} versus section \S\ref{sec:gSa-dS0}) 
    \chiara{The algorithm also slightly depends on } the number of tracers of the stream (i.e. the mass ratio of the dwarf and the giant galaxy, see \S\ref{sec:fornax}). 
    \item  shown that the difference between the measured and the real velocity dispersion is similar to the uncertainties on velocities, hence the assessment of the stream kinematics would %greatly
    improve if higher spectroscopic resolution is available (section \S\ref{sec:kinem_gE_dE}).  
    \item demonstrated that in order to best estimate the reliability of the COSTA set-ups, we have first to construct a realistic realization in the phase space of the system under analysis. This is not an easy task and it also needs customized Montecarlo realizations (section~\S\ref{sec:fornax}).  
    Furthermore, the reliability map one can derive might depend on how accurate is the description of the underlying ``relaxed'' system with respect to the stream population. 
    However, we note that an over-detailed dynamical description of a given system, can also incorporate substructures as a part of the relaxed component, hence reducing the chance to be recognized as a true substructure. We believe that the Montecarlo approach, as in  \S\ref{sec:fornax}, is a reasonable start, but we cannot exclude that more sophisticated relaxed model based on N-body codes will be used in the future to refine the predictions on real systems applying COSTA to real data. 
    \item derived a rule-of-thumb to unveil the real kinematics of a given stream, and hence to get insights on the properties of the parent galaxy from which these particles have been stripped.  We note however that, overall, COSTA underestimates the true velocity dispersion of the stream (S\ref{tab:simulated streams}). 
%hence, to get insights on the parent galaxy these particles have been stripped from, but overall, COSTA underestimates the true velocity dispersion of the stream.
\end{itemize}

To conclude, we have proved here that COSTA is 
%To conlude, COSTA has proved to be 
a useful tool to detect stream candidates originated by close galaxy encounters in a 3D phase space, e.g. using \chiara{right ascension, declination } %R.A., Dec. 
and radial velocity of particles as input. 
Since its ability increases when the stream is considerably colder than the surrounding environment, its natural implementation shall be galaxy clusters, where the high velocity dispersion of the relaxed particles moving in the potential well of the cluster makes it easier to unveil a group of particles coherent both in position and in velocity.
%Anyway, 
\chiara{Nevertheless, we } demonstrated that COSTA is also suitable to other situations  like galaxy-galaxy encounters hence, in principle, it could be used everywhere a sample of kinematic tracers is available.

%Despite it is not its main aim, COSTA could also unveil the intrinsic properties of the stream, 

%To conclude, by using using R.A., dec and radial velocity of particles as input COSTA is able to detect stream candidates originated by close galaxy encounters in a 3-D phase space.

%we have presented a novel algorithm to find stream candidates originated by close galaxy encounters in a 3-D phase space, e.g. using R.A., dec and radial velocity of particles as input.
%We demonstrated by means of cosmological simulations that if the stream is "cold" enough with respect to the hotter environment, the algorithm is able to detect it.

%However, COSTA can be generalised to all problems of finding small substructure in the phase space of a limited sample of discrete tracers, provided that the algorithm is trained on realistic mock observations reproducing the specific dataset under exam. 

\medskip
\vspace{1.cm}
\begin{acknowledgements}
%\textbf{Acknowledgements:}
NRN acknowledges financial support from the "One hundred top talent program of Sun Yat-Sen University" grant N. 71000-18841229, and from the European Union Horizon 2020 research and innovation programme under the Marie Skodowska-Curie grant agreement n. 721463 to the SUNDIAL ITN network. 
MG acknowledges support from the INAF fund ``Funzionamento VST" (1.05.03.02.04).
CS is supported by a Hintze Fellowship at the Oxford Centre for Astrophysical Surveys, which is funded through generous support from the Hintze Family Charitable Foundation.  \\
\end{acknowledgements}

%%%%%%%%%%%%%%%%%%%% REFERENCES %%%%%%%%%%%%%%%%%%

% The best way to enter references is to use BibTeX:
\bibliographystyle{aa}
\bibliography{mybibliography} % if your bibtex file is called mybibliography.bib
\end{document}